\documentclass[reprint,amsmath,amssymb,aps,prl,superscriptaddress]{revtex4-1}

\usepackage{graphicx}
\usepackage{dcolumn}
\usepackage{bm}
\usepackage{hyperref}

\usepackage{xcolor}

\usepackage{subcaption}
\newcommand{\labelphantom}[1]{%
{\phantomsubcaption%
\label{#1}}%
}%

\begin{document}

\title[H-mode Access in Negative Triangularity]{H-mode inhibition in negative triangularity tokamak reactor plasmas} 

\author{A. O. Nelson}
\email[Corresponding author: ]{a.o.nelson@columbia.edu}
\affiliation{Department of Applied Physics and Applied Mathematics, Columbia University, New York, NY 10027, USA}

\author{C. Paz-Soldan}
\affiliation{Department of Applied Physics and Applied Mathematics, Columbia University, New York, NY 10027, USA}

\author{S. Saarelma}
\affiliation{UK Atomic Energy Authority, Culham Science Centre, Abingdon OX14 3DB, United Kingdom}

\date{\today}

\begin{abstract}
Instability to high toroidal mode number ($n$) ballooning modes has been proposed as the primary gradient-limiting mechanism for tokamak equilibria with negative triangularity ($\delta$) shaping, preventing access to strong H-mode regimes when $\delta\ll0$. To understand how this mechanism extrapolates to reactor conditions, we model the infinite-$n$ ballooning stability as a function of internal profiles and equilibrium shape using a combination of the CHEASE and BALOO codes. While the critical $\delta$ required for avoiding $2^\mathrm{nd}$ stability to high-$n$ modes is observed to depend in a complicated way on various shaping parameters, including the equilibrium aspect ratio, elongation and squareness, equilibria with negative triangularity are robustly prohibited from accessing the $2^\mathrm{nd}$ stability region, offering the prediction that that negative triangularity reactors should maintain L-mode-like operation. In order to access high-$n$ $2^\mathrm{nd}$ stability, the local shear over the entire bad curvature region must be sufficiently negative to overcome curvature destabilization on the low field side. Scalings of the ballooning-limited pedestal height are provided as a function of plasma parameters to aid future scenario design.
\end{abstract}

\maketitle

\section{Introduction}

The leading configuration for quick realization of a reactor-relevant fusion plasma is currently the positive triangularity (PT) tokamak H-mode \cite{Wagner1984}. In this regime, once a certain threshold heating power is reached, a strong edge transport barrier (ETB) is naturally formed in the outer portion of the confined plasma, allowing for significant gains in temperature, density and energy confinement throughout the entire device. While the H-mode ETB is fundamentally a product of strong $E\times B$ shear which reduces turbulent transport in the plasma edge \cite{Burrell1997}, the ``\textit{dee-shaped}'' positive triangularity arrangement of these configurations also offers an enhancement in edge magnetohydrodynamic (MHD) stability, raising the pressure gradient limit for high-$n$ modes, and in some cases allowing for second stability to high-$n$ modes and further increase in pressure gradients until they are limited by other mechanisms such as intermediate or low $n$ modes \cite{Snyder2002}. Due to the high performance achieved by these regimes, PT H-mode plasmas have been selected as the default operating regimes for many next generation fusion experiments \cite{ITER1999, Creely2020}.

However, the edge pressure gradients that build up in PT H-mode pedestals cannot increase indefinitely, and are eventually limited by global MHD instabilities called edge-localized modes (ELMs) \cite{Nelson2020, Leonard2014}. In current experiments, the intermittent bursts of energy and particle flux caused by ELMs are generally within the allowable margins for machine safety. However, it is well-established that the peak energy fluence of these events will increase with both plasma current and the pedestal pressure, extrapolating to high levels which could unacceptably limit component lifetimes in future fusion power plants \cite{Gunn2017}. As such, mitigation or elimination of these modes, which are caused by peeling-ballooning (PB) instabilities, is a subject of urgent interest. 

Several potential solutions to the ELM mitigation issue are currently under investigation in the community. Of these, plasma operation in a  ``\textit{reversed-dee}'' or negative triangularity (NT) configuration is unique in that it relies solely on simple plasma shaping to achieve an ELM-free regime. The triangularity ($\delta$) of a particular tokamak equilibrium is defined as the mean of the upper $(\delta_\mathrm{u})$ and lower $(\delta_\mathrm{l})$ triangularities, which are explicit functions of the plasma shape: $\delta_\mathrm{u,l} \equiv (R_\mathrm{geo}-R_\mathrm{u,l})/a$, where $R_\mathrm{geo}$ is the geometric major radius, $R_\mathrm{u,l}$ are the major radius of the highest and lowest points along the plasma separatrix, and $a$ is the minor radius of the plasma.
While strong positive triangularly ($\delta>0$) is known to enhance MHD stability and facilitate access to H-mode, recent experimental work has shown that plasmas with $\delta < 0$ are able to achieve high plasma performance while retaining an L-mode edge \cite{Austin2019}. If these NT regimes are shown to extrapolate well to reactor conditions, they could thus comprise an interesting set of alternative reactor scenarios with several tangible benefits over the more traditional PT approach, including operation in a robustly ELM-free regime (ELMs generally do not occur in L-mode) and elimination of the need to meet a $P_\mathrm{LH}$ power criterion to reach H-mode in a reactor \cite{Kikuchi,Kikuchi2019,Austin2019,medvedev_single_2016,coda_enhanced_2022}. 

To assess the potential of NT scenarios as an ELM-free regime, it is critical to understand why H-mode is prevented in NT plasmas and if this phenomenon will persist in the hot and dense conditions of a fusion reactor. Building off off early work on the subject \cite{medvedev_beta_2008, Medvedev2015, merle_pedestal_2017}, recent experimental analysis of NT discharges on the DIII-D tokamak has shown that ideal infinite-$n$ ballooning modes, while normally stabilized at high pressure gradients in PT plasmas, are unstable at much lower pressure gradients in plasmas with $\delta<0$, preventing the virtuous cycle of strong edge $E\times B$ shear which typically begets H-mode operation \cite{saarelma_ballooning_2021}. Further, it has been suggested that ballooning stability is prevented in these plasmas as a direct result of the plasma shape, as strong NT prohibits the magnetic shear in the pedestal region from dropping below the threshold necessary to stabilize the infinite-$n$ ballooning mode \cite{marinoni_brief_2021}.

In this study, we asses how infinite-$n$ ballooning stability is impacted by both plasma profiles and shaping in order to determine if the L-mode phenomenology of the NT edge will persist in reactor-relevant parameter spaces. To start, we give a more in-depth overview of the relationship between ballooning stability and H-mode access in section~\ref{sec:theory}, along with a description of the assumptions and tools applied throughout this study in section~\ref{sec:model}. In section~\ref{sec:tri} we demonstrate that strong negative triangularity is directly responsible for robustly preventing ballooning stability in the plasma edge before considering the effects of other shaping parameters in section~\ref{sec:shape}. Then, in section \ref{sec:disc}, a full database of over 1,500 equilibria is used to develop a rough scaling law for the ballooning-limited pedestal height in negative triangularly scenarios.


\section{Infinite-n ballooning stability theory}
\label{sec:theory}

The ballooning mode is a well-studied MHD instability that is driven by interaction of the pressure gradient and the magnetic curvature \cite{Freidberg2014}. In contrast to traditional interchange modes, which are approximately constant along field lines in order to minimize the stabilizing effects of field line bending, the infinite-$n$ ballooning perturbation responds to the fact that the magnetic curvature in a toroidal device can oscillate between favorable and unfavorable along a given field line. By including a small amount of field line bending, the ballooning perturbation then has the potential to localize in regions of unfavorable magnetic curvature, producing a net effect which leads to greater instability than experienced in the interchange case and allowing for instability in plasmas where the Mercier criterion is already satisfied \cite{mercier_necessary_1960}. Notably, the radial extent of these modes still tends towards zero as the toroidal mode number $n\rightarrow\infty$, allowing for the stability along each flux surface to be calculated individually from a 1D equation.

\begin{figure}
	\includegraphics[width=0.9\linewidth]{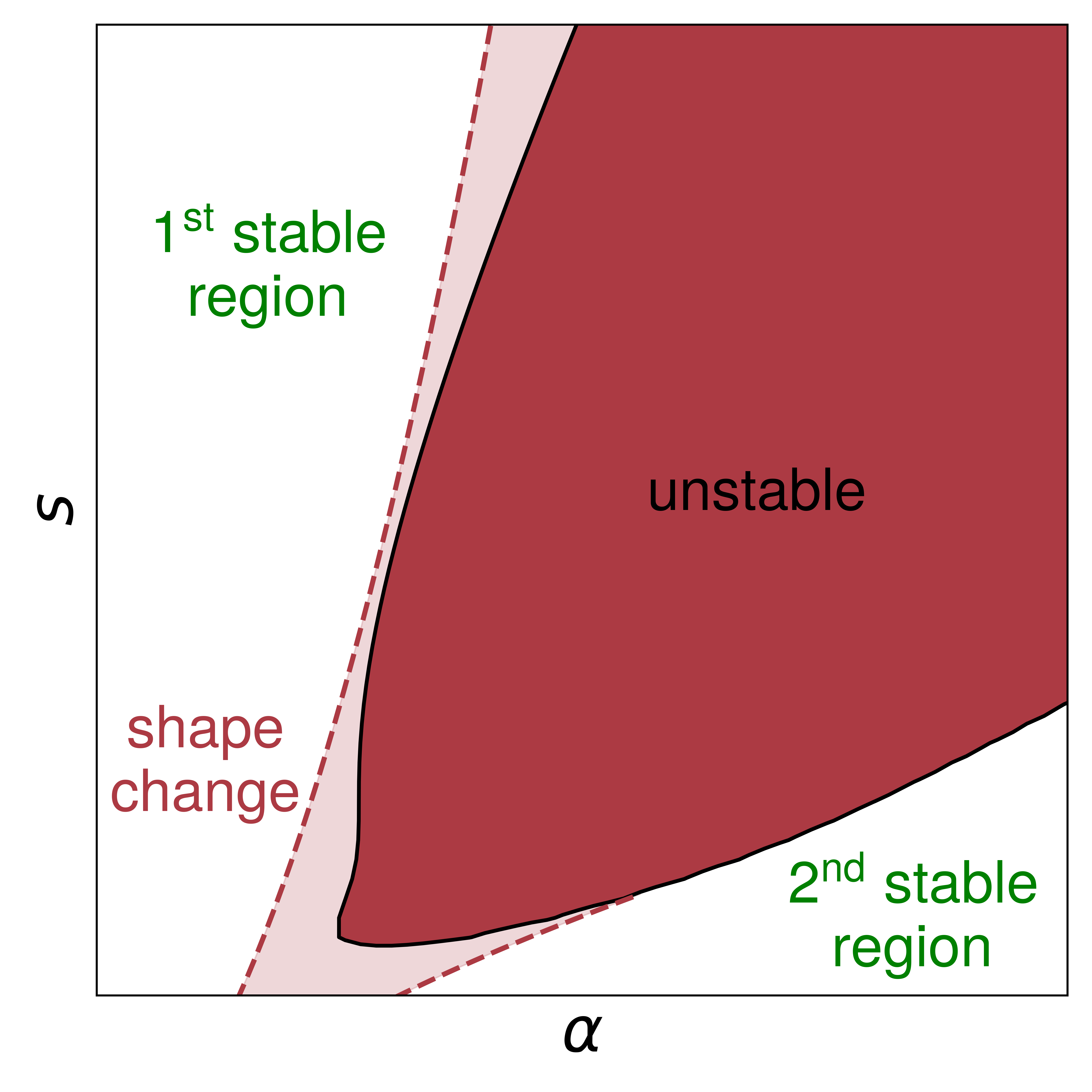}
	\caption{A cartoon (cylindrical equilibrium) marginal stability diagram for infinite-$n$ ballooning modes as a function of the normalized pressure gradient ($\alpha$) and magnetic shear ($s$). Two stability regions -- one at all $s$ and low $\alpha$; one at low $s$ and high $\alpha$ -- are separated by configurations unstable to ballooning modes. The boundary of the unstable region is strongly influenced by the plasma shape, and can potentially be modified to close the pathway to the second stability region, as explored below.}
	\label{fig:salphastab}
\end{figure}

This calculation can be competed routinely by solving the infinite-$n$ ballooning mode equation first derived by Connor \textit{et. al.} \cite{connor_shear_1978}. The main result of such a calculation is a depiction of marginal stability in $s-\alpha$ space, where the global magnetic shear $s$ and the normalized pressure gradient $\alpha$ are given by the expressions
\begin{equation}
    \label{eq:s}
    s = \frac{r}{q}\frac{dq}{dr}
\end{equation}
and
\begin{equation}
    \label{eq:alpha}
    \alpha = \frac{\mu_\mathrm{0}}{2\pi^2}\frac{\partial V}{\partial\psi}\bigg(\frac{V}{2\pi^2R}\bigg)^{1/2}\frac{dp}{d\psi},
\end{equation}
respectively. Here, as throughout this manuscript, $q$ is the plasma safety factor, $B_\mathrm{T}$ the toroidal magnetic field, $V$ the volume enclosed by each flux surface, $\psi$ the poloidal flux, $p$ the plasma pressure, $R$ the plasma major radius and $r$ the plasma minor radius. Note that an expression for the local magnetic shear, which is found provide a strong link between shape and ballooning stability, is provided in section~\ref{sec:shear}.

In the example of this solution space provided in figure~\ref{fig:salphastab}, which is calculated for cylindrical equilibria for illustrative purposes, two distinct stability regions can be observed. In the first region, the plasma is stable against ballooning modes up to some maximum $\alpha$, which increases with increasing shear. This dependence is commonly observed in core plasmas and can be understood intuitively by considering shear as a stabilizing influence for ballooning modes, increasing the energy needed to bend adjacent field lines as a perturbation balloons outward. This picture becomes more complicated in the edge, where the slope of the $1^\mathrm{st}$ stability boundary can be reversed such that the critical $\alpha$ increases as shear decreases. This edge phenomena is associated with feedback with the bootstrap current, and requires full consideration of the plasma shape (as is explored below.) Perhaps surprisingly, there is another stability region at low $s$ and high $\alpha$, commonly referred to as the ``$2^\mathrm{nd}$ stability region.'' In this space, as the pressure gradient is increased, the local magnetic shear shear in the unfavorable region becomes large and negative such that the total shear stabilization (which is proportional to $s^2$) produces a stabilizing effect \cite{greene_second_1981, Freidberg2014, gerver_access_1988}. Importantly, it is possible to open up a stable path between these two stability regions (as shown in figure~\ref{fig:salphastab}) by adjusting the plasma shape, shear and pressure, allowing for potential access to the $2^\mathrm{nd}$ stability region and the high $\beta$ that comes with it, though a maximum limit on $\alpha$ will sill be imposed by finite-$n$ PB modes \cite{wilson_access_1999}.

\subsection{Relationship between ballooning stability and H-mode access}

While access to the high-$n$ $2^\mathrm{nd}$ stability region could be achieved at any minor radius, in practice it is most often associated with increased $\alpha$ in the H-mode pedestal. Indeed, access to the $2^\mathrm{nd}$ stability region can be quite robust in the edge, especially in plasmas with strong PT shaping and low global shear. $2^\mathrm{nd}$ stability access allows pressure gradients to build, creating more sheared rotation through the diamagnetic drift effect, which leads to a positive feedback loop: more $E\times B$ shear suppresses turbulence and allows for even steeper pressure gradients. Additionally, increased pressure gradients raise the bootstrap current, lowering the magnetic shear and increasing the pressure limit, allowing for steeper pedestals to continue to form. This effect becomes particularly relevant in the plasma edge where steep pressure gradients and large bootstrap currents are present. Without access to the $2^\mathrm{nd}$ stability region, however, edge pressure gradients would instead lead to enhanced MHD turbulent transport instigated by the ballooning perturbation, preventing this virtuous cycle of gradient enhancement. As such, $2^\mathrm{nd}$ stability access is generally regarded as facilitating H-mode operation (though H-mode is observed in first stability limited regimes as well, these typically have higher pressure gradient limits than the negative triangularity cases discussed here). 

The relationship between access to the $2^\mathrm{nd}$ stability region and high pedestal H-modes has been observed experimentally on many machines, including on DIII-D \cite{ozeki_plasma_1990, lao_effects_1999,  ferron_modification_2000, saarelma_ballooning_2021}, JET \cite{korotkov_edge_2000, saibene_improved_2002}, JT-60U \cite{lao_dependence_2001}, NSTX \cite{canik_edge_2013} and Alcator C-Mod \cite{mossessian_h-mode_2002}. In these cases, pressure gradients in the edge have been found to exceed the first infinite-$n$ ballooning stability limit (calculated in the limit of high magnetic shear) by accessing the $2^\mathrm{nd}$ stable region. This leads to strong pedestal gradients and in some cases large Type-I ELMs, which are triggered by coupling between finite-$n$ ballooning and peeling modes and will pose a significant risk for vessel walls in reactor-like conditions \cite{Leonard2014, Gunn2017}. The foreseen need to operate a reactor in this regime has inspired a tremendous amount of research on ELM control and mitigation, from which various ELM control strategies have emerged.  Many of these control schemes focus on preventing the pedestal from encountering low to intermediate $n$ PB boundaries, which is naturally also achieved by operating in a plasma configuration unable to access the $2^\mathrm{nd}$ stability region for infinite-$n$ ballooning modes.

It should be noted here that $2^\mathrm{nd}$ stability access is not necessarily a hard-and-fast requirement for H-mode operation, especially at high shear and collisionality. For example, ELMing H-mode was achieved with and without access to the $2^\mathrm{nd}$ stability region during a squareness scan on DIII-D \cite{ferron_modification_2000}. However, a factor of $\sim10$ increase in the ELM frequency was reported with loss of access to the $2^\mathrm{nd}$ stability region, along with a corresponding decrease in the ELM amplitude and height of the H-mode pressure pedestal. Discharges in ASDEX and C-Mod have also been observed to achieve H-mode while remaining in the  $1^\mathrm{st}$ stability stability region by operating at high collisionality, which then suppresses the bootstrap current and prevents the edge shear from dropping below the critical value needed for ballooning stability. Again, $1^\mathrm{st}$ stable H-mode operation in these cases is associated with a strong reduction in the ELM amplitude. This picture is a bit more complicated on JT-60U, where large ELMs have been observed in discharges without access to the $2^\mathrm{nd}$ stability region \cite{lao_dependence_2001}. However, a significant reduction in ELM amplitude was reported after a shift of the unstable mode number from $n=8$ to $n>10$, suggesting that limiting the pedestal through infinite-$n$ ballooning modes could still be an effective method of ELM control. Since $1^\mathrm{st}$-stable H-modes are generally found to be weaker and less ELMy than their $2^\mathrm{nd}$-stable counterparts, focus in this work solely on the regime separation between shapes with and without access to the $2^\mathrm{nd}$ stability region, treating $1^\mathrm{st}$-stable operation as a potential solution to the ELM problem. 

\subsection{Effects of shape on infinite-n ballooning stability}

Previous studies have shown, both through modeling and experiment, that the simplified schematic in figure~\ref{fig:salphastab} can be significantly modified by plasma shape. For example, early models of bean-shaped equilibria demonstrated that increased outer midplane indentation is strongly beneficial for $2^\mathrm{nd}$ stability access \cite{chance_ballooning_1983, grimm_mhd_1985}. More recently on DIII-D, experimental results have indicated that large changes in the squareness of PT discharges closely coincide with the loss of access to the $2^\mathrm{nd}$ stability regime in the edge region \cite{ferron_modification_2000}. This comes as a result of an increase in field line connection lengths in the destabilizing bad curvature region ($\nabla B^2 \cdot \nabla p >0$), which increases the weight of the destabilizing toroidal field line curvature in the ballooning stability equation \cite{greene_second_1981}. Similarly, in PT discharges on JET, reduced triangularity was found to restrict the stable operational space by reintroducing infinite-$n$ ballooning instability in the edge region, resulting in strongly reduced pedestal pressures at low $\delta$ \cite{saibene_improved_2002}.

When operating at $\delta<0$, the plasma experiences a significant bulge outwards towards the low field side, dramatically increasing the region of the plasma in the bad curvature region, which contributes significantly towards the destabilization of infinite-$n$ ballooning modes. As such, the infinite-$n$ stability limit has been proposed as a primary instability in the NT edge \cite{medvedev_beta_2008, Medvedev2015, saarelma_ballooning_2021, marinoni_brief_2021, merle_pedestal_2017}. Recent experimental work on DIII-D has demonstrated that, while H-mode can be achieved at modest negative triangularity, it is quickly lost when $\delta$ is decreased below some critical value \cite{saarelma_ballooning_2021}. Similarly on TCV, a strong reduction of the pedestal height was observed when going from positive to negative top triangularity \cite{merle_pedestal_2017}. In both of these cases, infinite-$n$ ballooning stability analysis demonstrated that the edge gradient reduction in NT was closely correlated to loss of access to the $2^\mathrm{nd}$ stability region (and reduction of the critical gradient,) suggesting that prohibited access to the $2^\mathrm{nd}$ stability region at $\delta\ll0$ may forbid robust H-mode operation in NT configurations. This trend was highlighted in the recent NT review by Marinoni, \textit{et. al.}, where an increase in the edge shear in the bad curvature region in NT is credited with closing access to the $2^\mathrm{nd}$ stability region \cite{marinoni_brief_2021}. 

The direct effect of shape on $2^\mathrm{nd}$ stability access (and as a result on the critical pressure gradient) potentially offers a straight-forward mechanism with which to control H-mode access. In the following, we numerically investigate how changes in plasma shape can open or close access to the $2^\mathrm{nd}$ stability region, finding that, while the exact nature of the marginal stability boundary is a complicated function of plasma shape and composition, strong negative triangularity robustly destabilizes infinite-$n$ ballooning modes in the edge region and thereby limits the maximum achievable pedestal gradients.


\section{Ballooning stability modeling workflow}
\label{sec:model}

To quantify the ballooning stability of various model equilibria, we use a combination of the CHEASE \cite{lutjens_chease_1996} and BALOO \cite{miller_stable_1997} codes. CHEASE is a well-established equilibrium reconstruction code that was used to generate over $\sim10,000$ equilibria over the course of this study. For stability analysis of these equilibria, we used the BALOO code, which follows the local $n=\infty$ ideal MHD ballooning mode stability formalism and treats each flux surface individually to determine its stability. Throughout this work, we use the normalized pressure gradient $\alpha$ as a metric of the equilibrium and stability limits. To calculate the infinite-$n$ stability limit, BALOO artificially varies the local pressure gradient (holding all other equilibrium quantities constant) until marginal value of $\alpha$ corresponding to the $1^\mathrm{st}$ and $2^\mathrm{nd}$ stability limits are found. If no critical $\alpha$ is found for a particular flux surface, it is infinite-$n$ ballooning stable and has access to the $2^\mathrm{nd}$ stability regime. The BALOO code has previously been experimentally validated on various machines over a wide range of conditions \cite{saarelma_ballooning_2021, canik_edge_2013, ferron_modification_2000, mossessian_h-mode_2002, miller_stable_1997, lao_dependence_2001, lao_effects_1999}.

Throughout this work, we make several assumptions regarding the plasma profiles in order to most clearly capture the effect of shape on ballooning stability. First, we use the EPED $\texttt{mtanh}$ parameterization to generate density and temperature profiles \cite{Snyder2009, Snyder2011}. This profile shape was chosen to provide a consistent and realistic H-mode edge profile which can be smoothly varied to an L-mode-like profile by controlling the pedestal height \cite{Sauter2014}. Further, this assumption facilitates integration with existing integrated modeling tools (like STEP) that make use of the EPED pedestal formalization \cite{Meneghini2021}. To identify the critical gradient for infinite-$n$ ballooning instability, the pedestal height in the EPED parameterization is scaled, keeping all other aspects of the profile constant, until a marginally-stable value of $\alpha$ is attained. Effects of changes in the pedestal width are also considered in appendix~\ref{app:pedwidth}, though shown to have little impact on the overall results. It should be noted that the $1^\mathrm{st}$ stability boundary could be used to generate arbitrary profiles with marginal stability over the entire edge region, as was done in a recent study on DIII-D \cite{saarelma_ballooning_2021}. This approach was not taken here, as the edge profiles resulting from maximizing $\alpha$ at all radii often feature interesting features not observed in experiment. Additionally, since an accurate description of the bootstrap current is essential for the ballooning stability calculation, we use the Sauter model \cite{Sauter1999, Sauter2002} to determine the bootstrap current and the pressure profile from model density and temperature profiles specified with hyperbolic tangent shapes in the edge barrier region. Variations in the relative contributions of the density and temperature profiles to the pressure profile, which will effect the magnitude of the bootstrap current and shear even at constant $\alpha$, are also considered in sections~\ref{sec:boot} and \ref{sec:disc}. 


\section{Effect of triangularity on ballooning stability}
\label{sec:tri}

\begin{figure*}
	\includegraphics[width=1\linewidth]{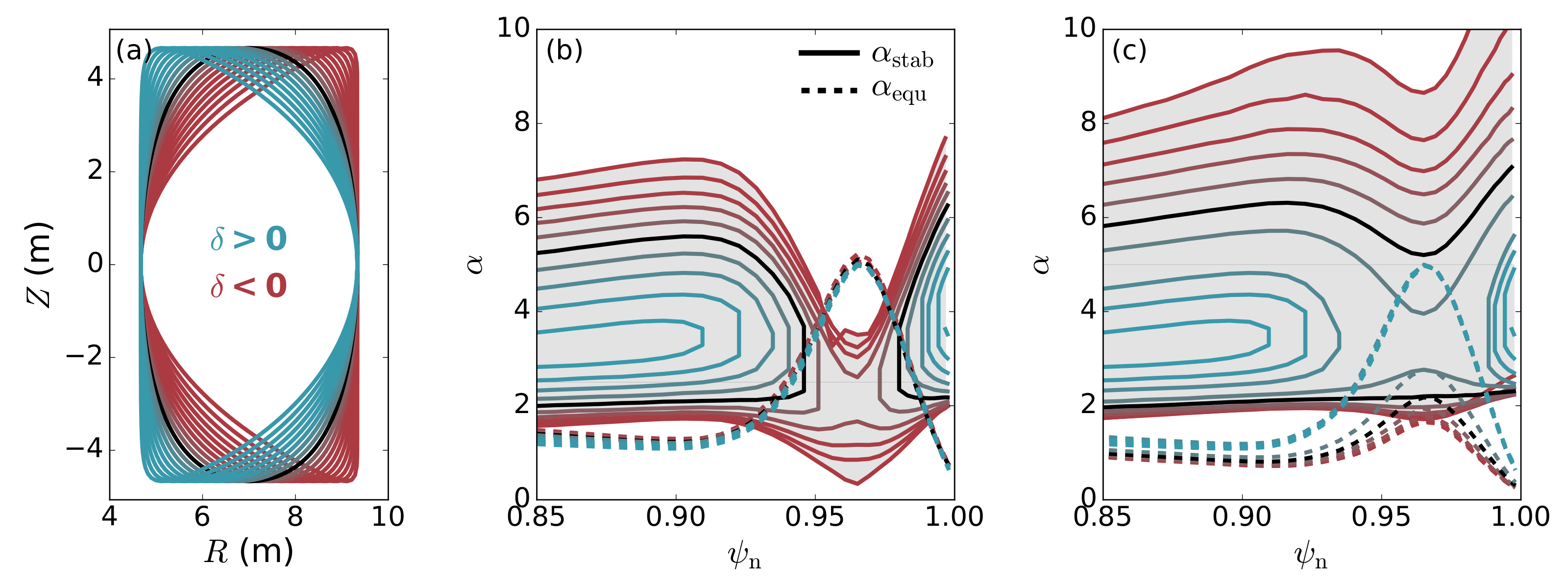}
    \labelphantom{fig:deltaEx-a}
	\labelphantom{fig:deltaEx-b}
	\labelphantom{fig:deltaEx-c}
	\caption{The effect of triangularity on infinite-$n$ ballooning stability is demonstrated by scanning $\delta$ for fixed equilibrium parameters. (a) The separatrix of each equilibria is plotted, providing a color scale for the rest of the figure. (b) In a scan with constant edge pressure gradient, the critical gradient required for instability -- $\alpha_\mathrm{stab}(\psi_\mathrm{N})$, solid lines -- is found to vary with $\delta$. The unstable region is indicated by a grey shading. In (c), any unstable pressure profiles are reduced to remain in the stable region, leading to dramatically reduced $\alpha_\mathrm{equ}$ at strong NT.}
	\label{fig:deltaEx}
\end{figure*}

We use an up-down symmetric base equilibrium loosely modeled off of the NT tokamak reactor targets developed in Medvedev, \textit{et. al.} (2015) as an initial condition for various parameter scans \cite{Medvedev2015}. This design is targeted at a plasma with major radius $R_0=7\,$m, minor radius $a=2.6\,$m, elongation $\kappa=2$, on-axis magnetic field $B_0=6\,$T, plasma current $I_p=15\,$MA and reactor-relevant normalized beta of $\beta_N=2.5$, though all of these parameters are scanned in the following sections to understand their role in setting $2^\mathrm{nd}$ stability access for infinite-$n$ ballooning modes. 

From this initial equilibrium, the impact of triangularity on controlling $2^\mathrm{nd}$ stability access is immediately evident. In figure~\ref{fig:deltaEx}, the triangularity is scanned from $\delta=-0.9$ to $\delta=0.9$. Then, using the BALOO code, the critical $\alpha$ needed to access the $1^\mathrm{st}$ and $2^\mathrm{nd}$ stable regions (solid lines, $\alpha_\mathrm{stab}$) are determined at each flux surface and plotted in figure~\ref{fig:deltaEx-b}, along with the modeled equilibrium gradient $\alpha_\mathrm{equ}$. For the strong positive triangularity cases in blue, an opening to the $2^\mathrm{nd}$ stable region exists in the edge near $\psi_\mathrm{N}\sim0.96$, where $\psi_\mathrm{N}$ is the normalized poloidal flux. However, this stability pathway quickly closes at negative enough triangularity, as demonstrated by the red contours in figure~\ref{fig:deltaEx-b}, showing the strong effect of $\delta<0$ on infinite-$n$ ballooning stability. 

\subsection{Pressure profile stability}
\label{sec:press}

While pedagogically informative, the results plotted in figure~\ref{fig:deltaEx-b} are unphysical at $\delta\lesssim0.1$ as portions of the equilibrium profiles reside within the infinite-$n$ unstable region. To address this, the pedestal height for each unstable equilibria is systematically decreased in figure~\ref{fig:deltaEx-c}, keeping the rest of the equilibrium parameters, including the pedestal width $\Delta_\mathrm{ped}=0.05$ (in $\psi_\mathrm{N}$ units,) constant. For triangularities in the range $-0.1<\delta<0.1$, this results in a closing of the instability boundary, as $\alpha$ is no longer strong enough to reduce the local shear below the critical value needed for ballooning stability. For shapes with strong negative triangularity that were already closed to $2^\mathrm{nd}$ stability, the lower bound of the instability region actually increases slightly, though access to the $2^\mathrm{nd}$ stable region becomes increasingly difficult due to a significant widening of the instability region. These equilibria with $\delta\lesssim0.1$ are expected to be strongly limited by the first ballooning stability limit at low pressure gradient and thereby constrained to L-mode operation.

\begin{figure}
	\includegraphics[width=1\linewidth]{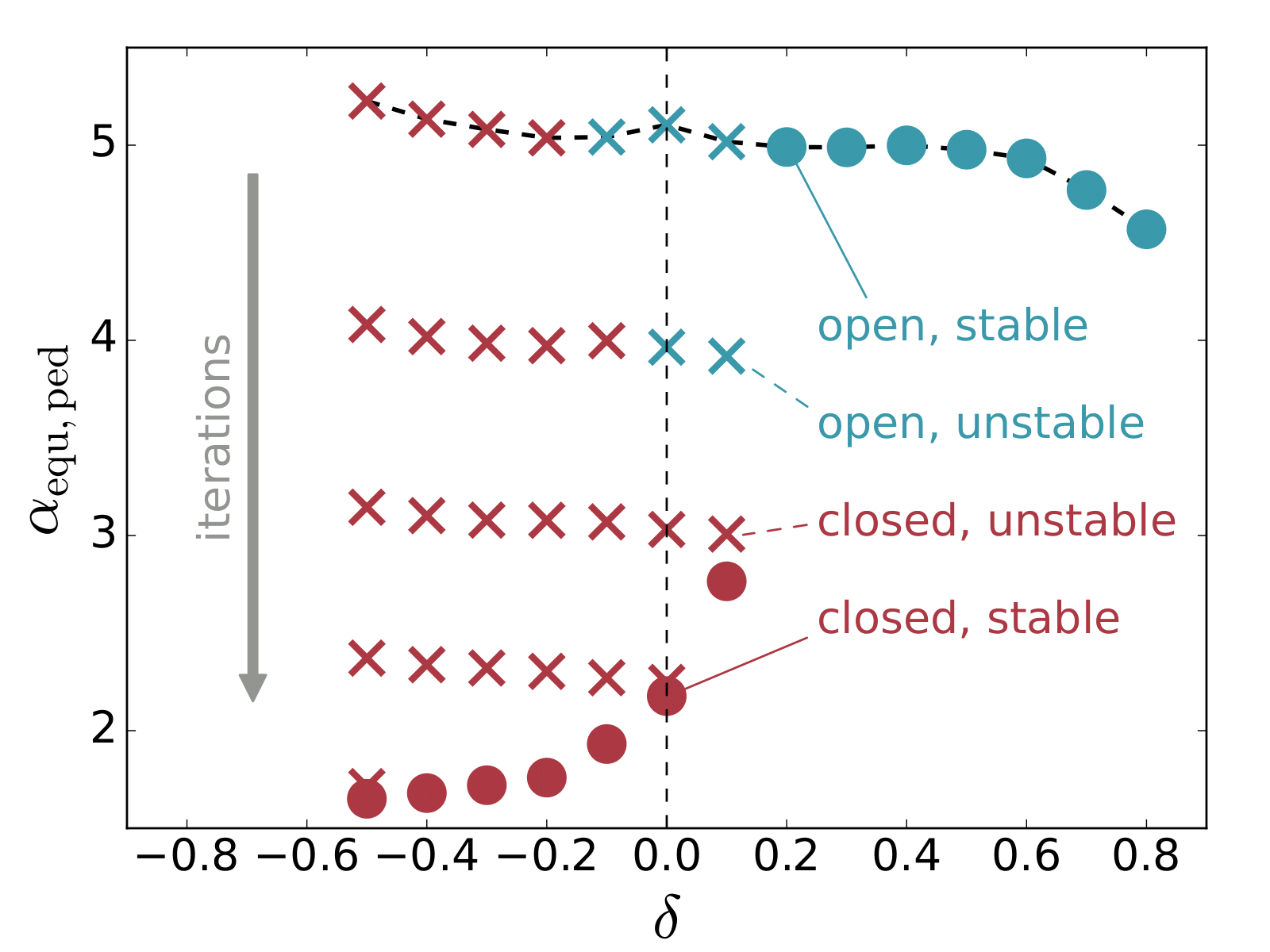}
	\caption{A one-dimensional representation of the pedestal height reduction applied in figure~\ref{fig:deltaEx-c} to retain stable pressure profiles at all $\delta$. Profiles stable to ballooning modes are marked with dots, and unstable profiles are marked with X's. Configurations with access to the $2^\mathrm{nd}$ stability region are colored blue, and configurations closed to $2^\mathrm{nd}$ stability are colored red. Due to closure of the $2^\mathrm{nd}$ stability region, a significant reduction of $\alpha_\mathrm{equ,ped}$ is needed to retain stability at triangularities $\delta\lesssim0.1$. }
	\label{fig:delta1D}
\end{figure}

The process of iteratively decreasing the pedestal height is depicted in figure~\ref{fig:delta1D}, where the normalized pressure gradient in the steepest edge gradient region ($\alpha_\mathrm{equ,ped}$) is plotted for the various equilibria in  figure~\ref{fig:deltaEx}. Profiles stable to ballooning modes are marked with dots, and unstable profiles are marked with X's. It is easily seen here that the requirement of ballooning stability begets a significantly reduced pedestal gradient at strong NT. The eventual closing of $2^\mathrm{nd}$ stability access as $\alpha_\mathrm{equ,ped}$ is reduced to stable values is also evident in equilibria with $\delta\sim0$. Note that $\alpha_\mathrm{equ,ped}$ is not modified from its initial value for equilibria that are already ballooning stable, which for this particular scan occurs at $\delta\geq0.2$, since there is no limit imposed on the pedestal gradient from infinite-$n$ ballooning modes in these cases. In experiment, these equilibria will instead be limited by finite-$n$ PB modes, which are outside the scope of this paper. These points are still included in the discussion to illustrate the $2^\mathrm{nd}$-stability-open region, but should not be used as a quantitative estimate of the pedestal height except that they feature greater $\alpha_\mathrm{equ,ped}$ than the ballooning-limited cases. 

\subsection{Collisionality and the bootstrap current}
\label{sec:boot}

\begin{figure}
	\includegraphics[width=1\linewidth]{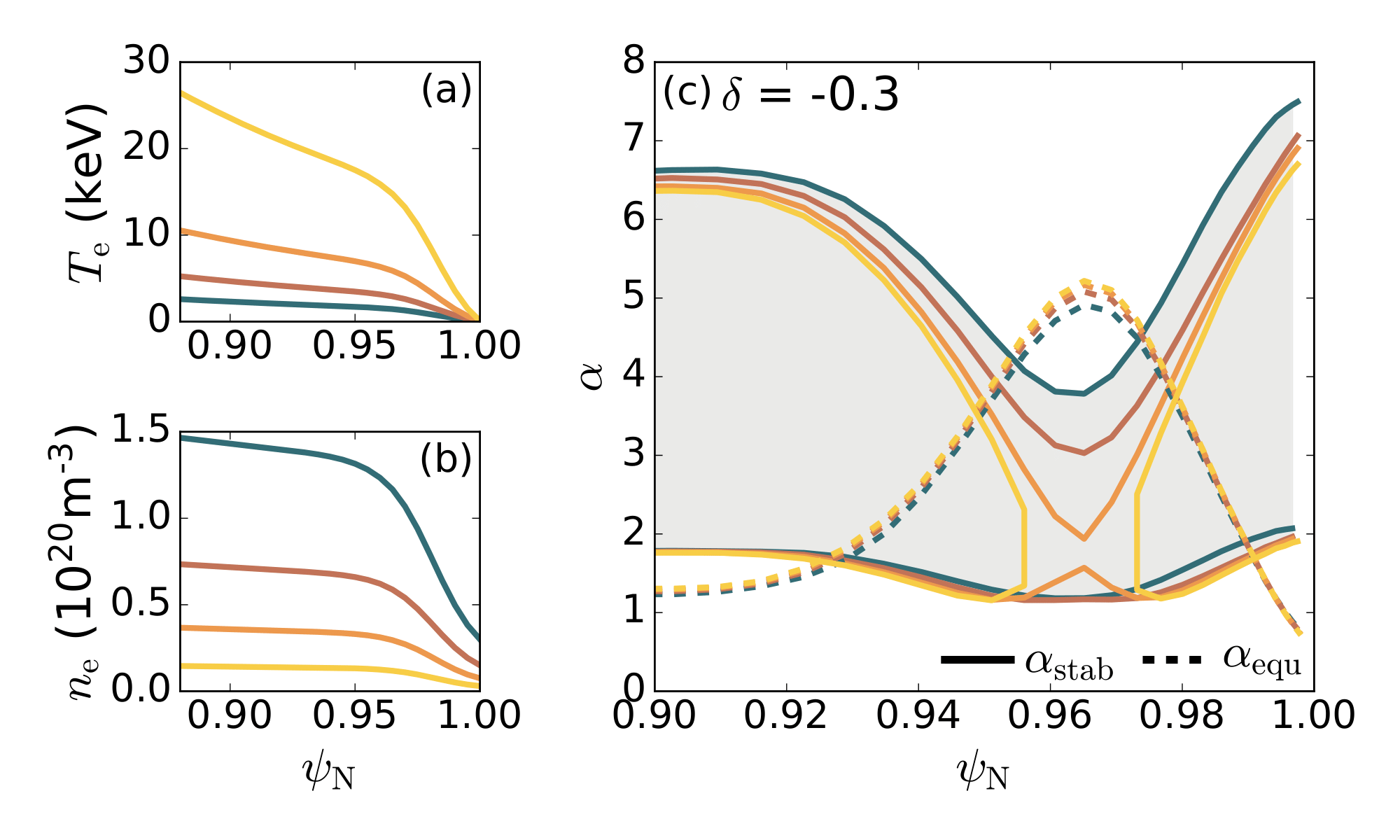}
	\labelphantom{fig:neteEx-a}
	\labelphantom{fig:neteEx-b}
	\labelphantom{fig:neteEx-c}
	\caption{At constant pressure, the edge temperature (a) and density (b) are varied to change the bootstrap current and edge shear. (c) At low collisionality, a small window to the $2^\mathrm{nd}$ stability region can be opened for this example equilibrium, which has $\delta=-0.3$.}
	\label{fig:neteEx}
\end{figure}

Another aspect of the equilibrium profiles that can play an important role in $2^\mathrm{nd}$ stability suppression is the edge collisionality. At constant pressure, if the temperature is increased while the density decreased, the collisionality will decrease and lead to a corresponding increase in the edge bootstrap current. In turn, the increased bootstrap current will then lower the global magnetic shear in the edge, potentially to a level below the minimum shear required for instability, at which point access to the $2^\mathrm{nd}$ stability region could be recovered. This process is demonstrated in figure~\ref{fig:neteEx}, where the electron temperature and density are inversely varied to modify the edge collisionality at constant pressure. The resulting stability boundaries for a representative equilibrium with $\delta=-0.3$ are depicted in figure~\ref{fig:neteEx-c}, showing a gradual reduction of the instability barrier with increasing edge temperature until a small window to the $2^\mathrm{nd}$ stable region is eventually recovered, albeit at an extremely high pedestal temperature of $T_\mathrm{e,ped}\sim15\,$keV (and low density) that may not be realizable on a reactor. 

\begin{figure}
	\includegraphics[width=1\linewidth]{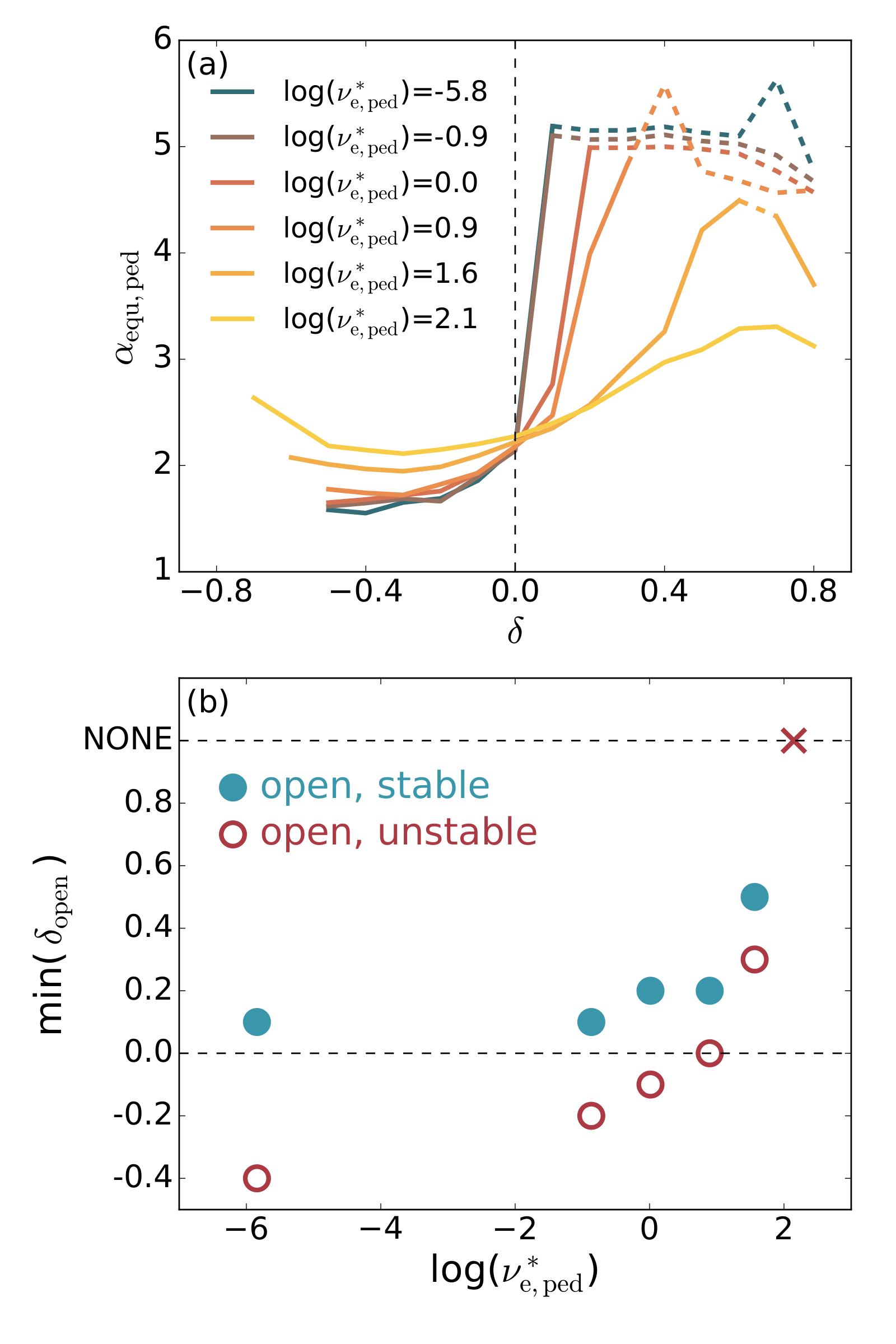}
	\labelphantom{fig:nustar-a}
	\labelphantom{fig:nustar-b}
	\caption{Changes in the collisionality impact (a) the maximum attainable gradient at each ballooning-unstable $\delta$, and (b) the minimum $\delta$ at which access to $2^\mathrm{nd}$ stability remains open. In (b), two equilibria are represented where applicable: equilibria with lowest $\delta$ featuring access to $2^\mathrm{nd}$ stability and stable profiles (blue dots) and equilibria with the lowest $\delta$ featuring access to $2^\mathrm{nd}$ stability and unstable profiles (red circles).}
	\label{fig:nustar}
\end{figure}

As before, this analysis can be repeated at all values of $\delta$ while decreasing the height of any unstable pedestals to uncover the effect of collisionality on the ballooning-limited $\alpha_\mathrm{ped}$ and the critical $\delta$ necessary for $2^\mathrm{nd}$ stability suppression. The results are shown in figure~\ref{fig:nustar-a}, which depicts the maximum attainable $\alpha_{ped}$ for various triangularities as a function of the normalized pedestal collisionality $\nu_\mathrm{e,ped}^*$. For experimentally relevant values of $\nu_\mathrm{e,ped}^*$ ($-1\lesssim\log(\nu_\mathrm{e,ped}^*)\lesssim1$), the maximum attainable $\alpha_\mathrm{ped}$ is not strongly impacted. This trend holds even for extremely hot plasmas with $\log(\nu_\mathrm{e,ped}^*)\sim-5$, which are tested here to explore the absolute limits of the underlying physics, while for extremely cold plasmas with $\log(\nu_\mathrm{e,ped}^*)\sim2$, the ballooning-limited $\alpha_\mathrm{ped}$ is slightly increased at $\delta<0$. 

\begin{figure*}
	\includegraphics[width=1\linewidth]{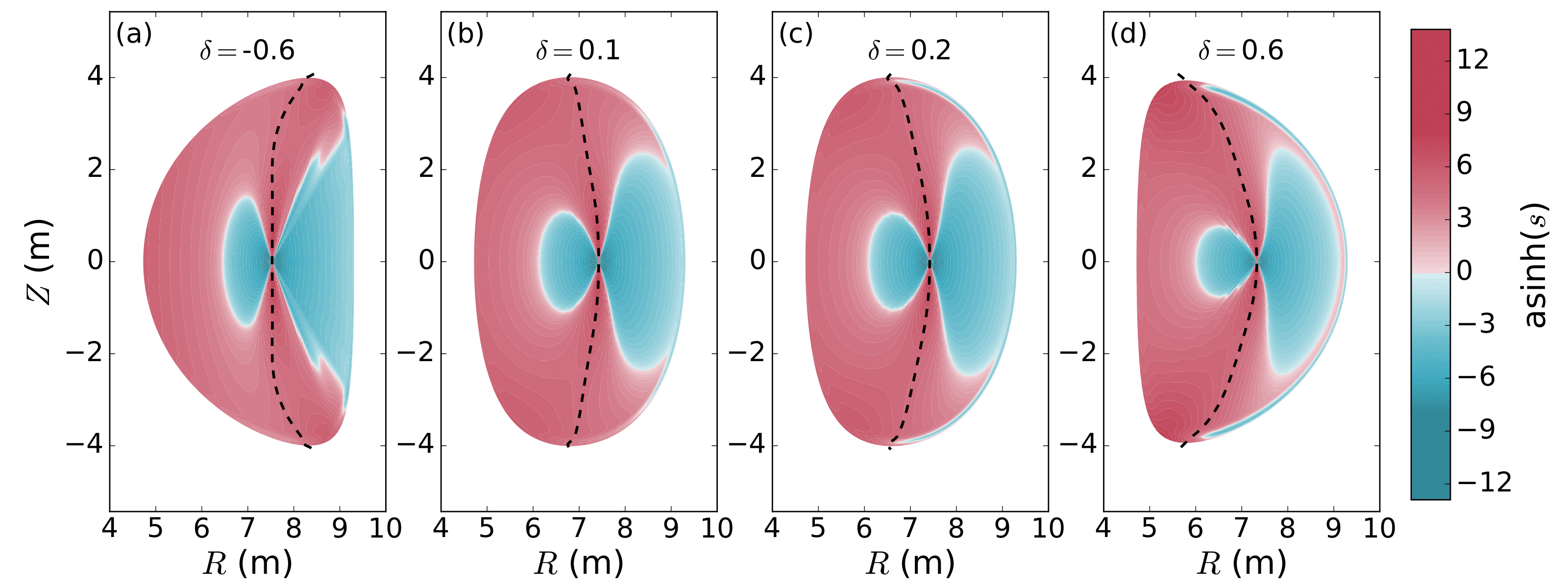}
	\labelphantom{fig:sheareg-a}
	\labelphantom{fig:sheareg-b}
	\labelphantom{fig:sheareg-c}
	\labelphantom{fig:sheareg-d}
	\caption{Local shear for various equilibria at (a) $\delta\ll0$, (b) just before $2^\mathrm{nd}$ stability access, (c) just after $2^\mathrm{nd}$ stability access and (c) $\delta\gg0$. The boundary between the good and bad curvature region is marked with a black dashed line.}
	\label{fig:sheareg}
\end{figure*}

The minimum triangularity at which the $2^\mathrm{nd}$ stability region remains open as a function of $\nu_\mathrm{e}^*$ can easily be extracted from this analysis and is shown in figure~\ref{fig:nustar-b}. For completeness, we show two values here -- the minimum $\delta$ with both open $2^\mathrm{nd}$ stability access and stable profiles, and the minimum open $\delta$ regardless of profile stability (as is true for the equilibrium shown in figure~\ref{fig:neteEx-c}.) While an extreme reduction in $\nu_\mathrm{e}^*$ can open up the $2^\mathrm{nd}$ stable region at modest negative triangularity if unstable configurations are considered, the minimum \textit{stable} $\delta_\mathrm{open}$ meets a hard limit near $\delta\sim0.1$ for this particular set of equilibrium parameters. The robustness of this closure even at extremely (unrealistically) small values of $\nu_\mathrm{e,ped}^*$ highlights the robustness of the shaping effect. Conversely, for extremely cold and dense plasmas, $2^\mathrm{nd}$ stability is lost even for strong positive triangularity, since plasmas with low bootstrap current and high shear are never able to open access to the $2^\mathrm{nd}$ stability region, regardless of the plasma triangularity. These results again highlight that specifics of the plasma profiles can have some effect on both the ballooning-limited $\alpha_\mathrm{ped}$ and the particular value of $\delta$ at which access to the $2^\mathrm{nd}$ stability regime closes, but that these effects are generally secondary to the shaping effects setting the ballooning instability limits. 

\subsection{Importance of local shear}
\label{sec:shear}

\begin{figure}
	\includegraphics[width=1\linewidth]{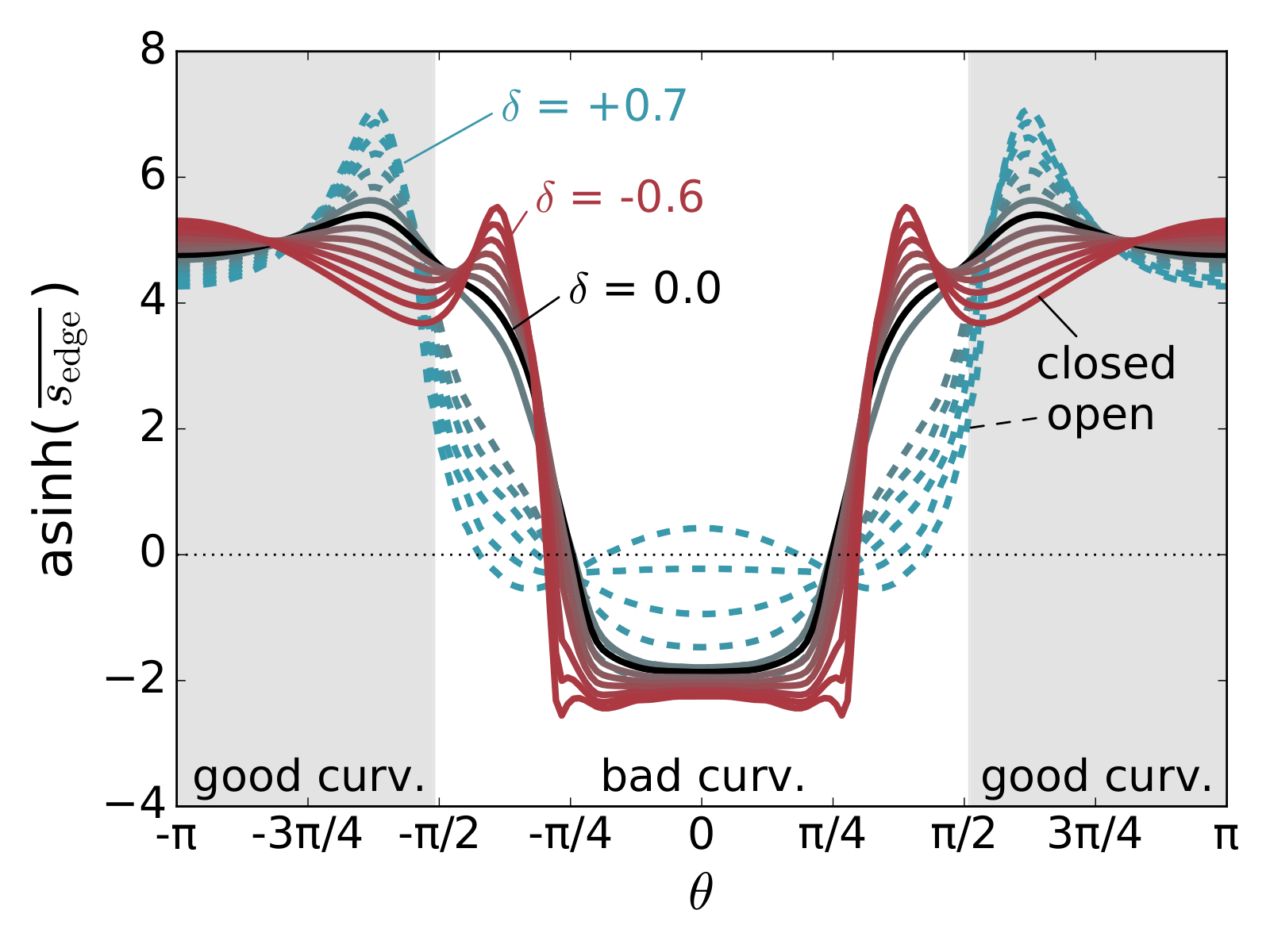}
	\caption{Average edge shear plotted as a function of poloidal angle $\theta$ (with $\theta=0$ corresponding to the outer midplane) for a series of equilibria with varying $\delta$. The transition from a open to closed regime is marked by a sudden rise in the edge shear near the top and bottom of the plasma, preventing low shear over the entire bad curvature region.}
	\label{fig:shearslice}
\end{figure}

To understand why modification of the plasma triangularity has such a strong impact on infinite-$n$ ballooning stability, it is informative to look at the local shear $s$ as a function of $\delta$. In the first stable region, increased local magnetic shear contributes to the stabilization of infinite-$n$ ballooning modes by increasing the amount of energy required to bend neighboring field lines as the mode balloons outward \cite{Freidberg2014, connor_shear_1978}. Since shear stabilization is proportional to $s^2$, a similar effect can be achieved for large enough values of negative local shear. This effect is most relevant on the outboard side of the torus, since the destabilizing normal curvature is maximum there. It is important to note that this process is aided by starting from an equilibrium with already low global shear, since then less pressure gradient is required to eventually achieve a large local shear on the outboard side \cite{greene_second_1981, marinoni_brief_2021}.

The effect of local shear on controlling access to the $2^\mathrm{nd}$ stability region is demonstrated as a function of $\delta$ in figure~\ref{fig:sheareg}. Here we calculate the local shear 
\begin{equation}
    s \equiv - \frac{\vec{B}\times\nabla\psi}{|\nabla\psi|^2} \cdot \nabla \times \bigg(\frac{\vec{B}\times\nabla\psi}{|\nabla\psi|^2}\bigg)
\end{equation}
for each equilibrium \cite{greene_second_1981}. Note that the quantity $\text{asinh}(s)$ is plotted for clarity. In figure~\ref{fig:sheareg}, four different equilibrium are shown with triangularities just below and above the critical $\delta$ for $2^\mathrm{nd}$ stability ($\delta_\mathrm{crit}\sim0.15$) and at the extremes of NT and PT tested in this scan, highlighting the contrast between NT and PT discharges. In each equilibria, regardless of $\delta$, a portion of the plasma near the outer midplane features strongly reversed local shear, which has a stabilizing influence on the ballooning mode. However, in the two equilibria with $\delta<\delta_\mathrm{crit}$, the region of the plasma with positive shear extends into the bad curvature region (which is marked with a black dashed line.) This prevents the local shear from being large and negative across the entire bad curvature region, and since the local shear must become zero somewhere inside the bad curvature region as a result, shear stabilization disappears and the $2^\mathrm{nd}$ stability region cannot be accessed \cite{marinoni_brief_2021}. In contrast, equilibria with $\delta>\delta_\mathrm{crit}$ exhibit a narrow but strong band of negative shear across the entire bad curvature region, which is caused by the strong pressure gradient in that region. Importantly, this band of negative shear pushes the local maximum of $s$ along each pedestal flux surface onto the inboard side where shear stabilization is not critical since the curvature is favorable. By allowing for $s\ll0$ across the entire outboard side, the equilibrium with $\delta>\delta_\mathrm{crit}$ can thereby provide access to the $2^\mathrm{nd}$ stability region at large enough $\alpha$. 

The movement of the local maximum in $s$ from the good curvature to bad curvature region is further demonstrated in figure~\ref{fig:shearslice}, which shows $\overline{s_\mathrm{edge}}$ -- the local shear in the edge region (averaged across flux surfaces with $\psi_\mathrm{N}>0.92$) -- as a function of poloidal angle $\theta$ for various triangularities. For equilibria with $\delta>\delta_\mathrm{crit}$, the local maximum in $\overline{s_\mathrm{edge}}$ appears on the good curvature side (indicated by shaded boxes,) allowing for low shear across the entire bad curvature region. However, as $\delta$ approaches $\delta_\mathrm{crit}$, the local shear in the bad curvature region quickly rises, reducing the stabilizing effect of low $s$ in the bad curvature region. For equilibrium with $\delta\ll0$, this effect is accentuated with the maximum of $\overline{s_\mathrm{edge}}$ appearing firmly in the bad curvature region. Notably, the transition from stabilizing to de-stabilizing local shear occurs very quickly at a critical $\delta$, explaining the sudden loss of H-mode with decreased triangularity observed in experiment \cite{saarelma_ballooning_2021}. The precise value of $\delta_\mathrm{crit}$ at which this phenomena occurs is a function of various parameters, as is explored below. 


\section{General shaping effects on ballooning stability and H-mode inhibition}
\label{sec:shape}

To better understand the full operating space for potential negative triangularity tokamak design, the effects of inverse aspect ratio $\epsilon$, elongation $\kappa$, squareness $\zeta$ and up-down asymmetry on infinite-$n$ ballooning stability are considered below for various values of $\delta$. In these simulations, the edge collisionality and initial profiles are kept constant. As such, the exact numerical limits of the stability boundary should not be taken as precise limits for all scenarios, though the general trends described in this section have been observed to hold true for all other initial conditions examined in section~\ref{sec:disc}. 

\subsection{Effect of aspect ratio}

\begin{figure*}
	\includegraphics[width=1\linewidth]{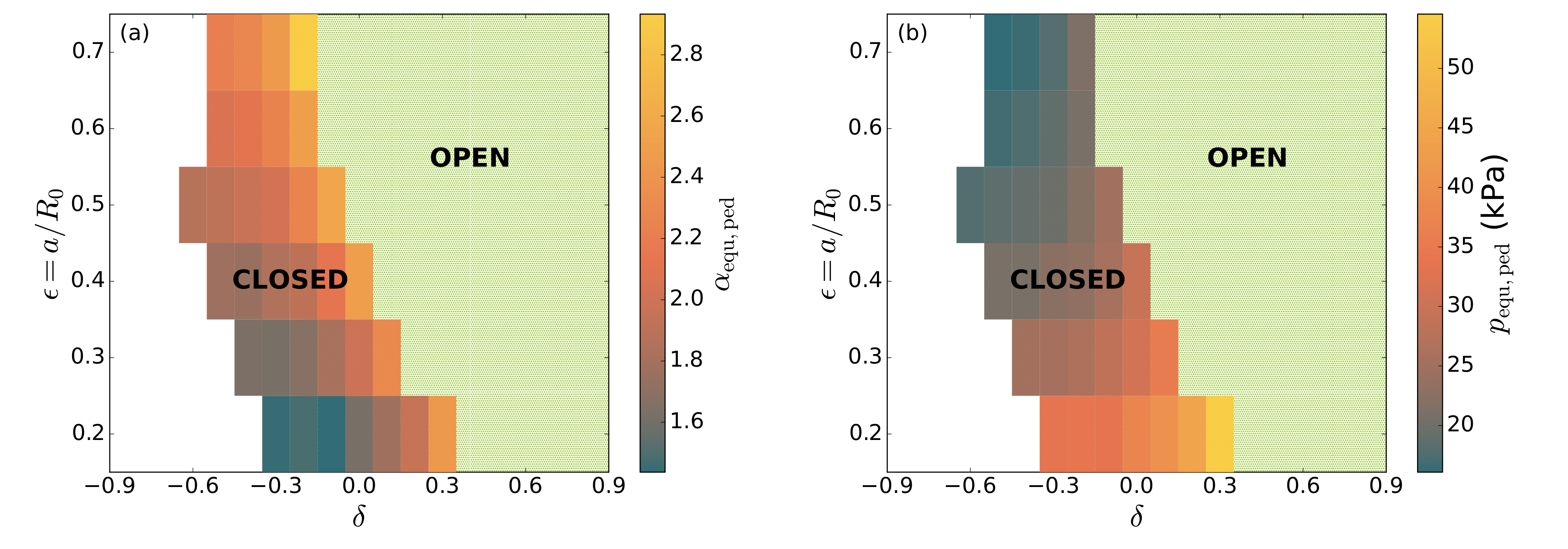}
	\labelphantom{fig:eps2D-a}
	\labelphantom{fig:eps2D-b}
	\caption{(a) The maximum stable edge pressure gradient and (b) the maximum stable pedestal top density are plotted as a function of inverse aspect ratio and $\delta$ for equilibria with constant $I_\mathrm{p}=15\,$MA, $B_{t}=6\,$T, $\kappa=2$ and minor radius $a_\mathrm{minor}=2.33\,$m. For $\delta\ll0$, the $2^\mathrm{nd}$ stability region cannot be accessed, and the edge gradients are thus limited by the infinite-$n$ ballooning mode. Configurations open to $2^\mathrm{nd}$ stability are shaded in green.}
	\label{fig:eps2D}
\end{figure*}

\begin{figure}[h!]
	\includegraphics[width=1\linewidth]{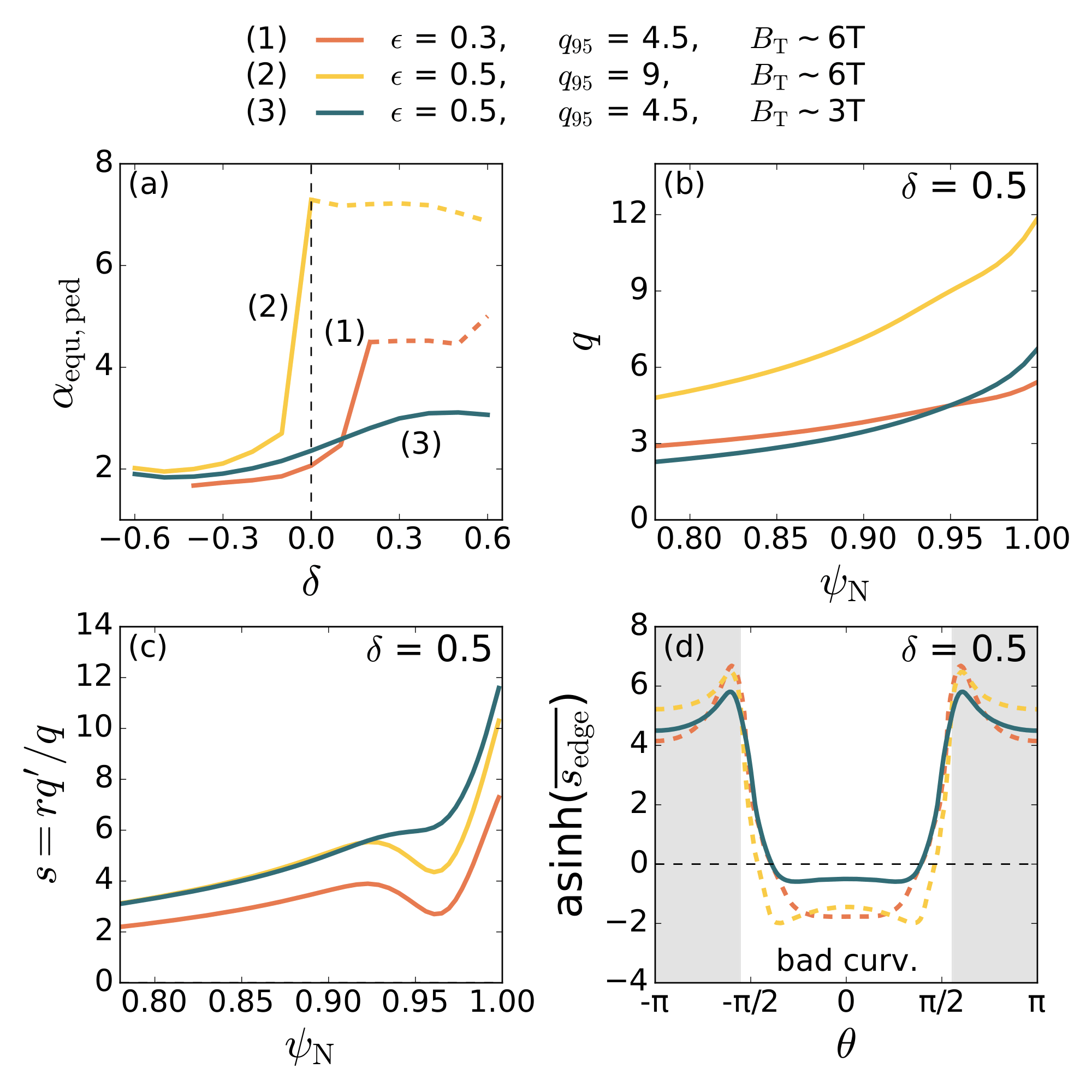}
	\labelphantom{fig:epsq95-a}
	\labelphantom{fig:epsq95-b}
	\labelphantom{fig:epsq95-c}
	\labelphantom{fig:epsq95-d}
	\caption{(a) Maximum achievable $\alpha_\mathrm{equ,ped}$ as a function of $\delta$ for three configurations. No $2^\mathrm{nd}$ stability access is available for low aspect ratio at low $q_\mathrm{95}$. Also shown are (b) the $q$-profile, (c) the global shear and (d) the average local shear for equilibria with $\delta = 0.5$.}
	\label{fig:epsq95}
\end{figure}

Through a dedicated scan of modeled equilibria, we find that decreased aspect ratio allows for $2^\mathrm{nd}$ stability access at lower $\delta$ but also leads to lower maximum pedestal pressures in ballooning-limited configurations. In figure~\ref{fig:eps2D}, the inverse aspect ratio $(\epsilon = a/R_\mathrm{0}$) and triangularity are varied at fixed minor radius to produce a two-dimensional image of the ballooning instability boundary in $\epsilon$ -- $\delta$ space. In this figure, and those following in this section, $\alpha_\mathrm{equ,ped}$ is not plotted for equilibria with access to the $2^\mathrm{nd}$ stability limit in order to highlight any trends in the closed region. Unsurprisingly, the $2^\mathrm{nd}$ stability region is robustly open for equilibria with $\delta\gg0$ and closed for negative triangularity cases. However, the critical delta at which access to $2^\mathrm{nd}$ stability is lost is described by a weak function of $\epsilon$ at fixed $I_\mathrm{p}$ and $B_\mathrm{T}$, with access to $2^\mathrm{nd}$ stability in the edge extended to equilibria with more negative triangularity (up to about $\delta\sim-0.1$) at smaller aspect ratio. Further, almost a factor of two variation in the maximum achievable normalized pressure gradient is observed between the lowest and highest aspect ratios included in figure~\ref{fig:eps2D-a}. At fixed $I_\mathrm{p}$, $B_\mathrm{T}$ and $a_\mathrm{minor}$, $\alpha$ is a strong function of volume (goes like $\alpha\sim1/R_\mathrm{0}$,) leading to higher values of $\alpha_\mathrm{ped}$ at smaller aspect ratio. The actual pedestal pressure is plotted in figure~\ref{fig:eps2D-b}, showing roughly two times lower pedestal top values at $\epsilon = 0.7$ than at $\epsilon = 0.2$. A more quantitative assessment of the role $\epsilon$ plays in reducing the maximum achievable pedestal top density is performed in section~\ref{sec:disc}. More negative $\delta$ at constant aspect ratio leads to smaller $\alpha_\mathrm{ped}$ and $p_\mathrm{ped}$ over the full scan. These results suggest both that $2^\mathrm{nd}$ stability suppression via NT may be more difficult at low aspect ratio and that, once $2^\mathrm{nd}$ stability is suppressed, ballooning-limited NT discharges will be more strongly impacted by the $1^\mathrm{st}$ stability boundary at lower aspect ratio, leading to strongly reduced pedestal pressures. 

When considering the changes in $2^\mathrm{nd}$ stability access with $\epsilon$, it is important to note that changes in the plasma geometry also impact the safety factor profile $q(\psi_\mathrm{N})$, which can in turn affect the ballooning stability calculations through the local shear. To demonstrate this effect, figure~\ref{fig:epsq95} highlights three triangularity scans with specified $q_\mathrm{95}$ that are generated by allowing $B_\mathrm{T}$ to vary while holding $I_\mathrm{p}$ constant. When moving from $\epsilon=0.3$ to $\epsilon=0.5$ at fixed toroidal field, $q_\mathrm{95}$ increases by a factor of two. Several effects are at play here that lead to the increased ballooning stability seen in figure~\ref{fig:epsq95-a}. First, the ``bad curvature connection length" -- defined as the distance along a magnetic field line within the bad curvature region -- shortens at low aspect ratio, leading to a stabilizing effect that diminishes ballooning drive due to the normal curvature. Second, there is a stability enhancement in the region of favorable magnetic curvature at high $q$ and low aspect ratio which results from toroidal modification of the magnetic curvature \cite{gerver_access_1988}. Together these effects allow $2^\mathrm{nd}$ stability access at lower $\delta$ for the low aspect ratio, high $q$ case. 

In contrast, when moving from $\epsilon=0.3$ to $\epsilon=0.5$ at $q_\mathrm{95}$, access to $2^\mathrm{nd}$ stability is lost even at strong positive triangularity, despite the reduced bad curvature connection length for the low aspect ratio configuration. In this case, the toroidal magnetic field must be decreased by a factor of $\sim2$ to match $q_\mathrm{95}$ at the edge. As a result, the magnitude of the negative local shear in the bad curvature edge is diminished and no longer enough to overcome the (enhanced) global shear to provide access to the $2^\mathrm{nd}$ stability region. This behavior is displayed in figures~\ref{fig:epsq95-c} and \ref{fig:epsq95-d}, which show the global and edge-averaged local shear, respectively, for equilibria with $\delta=0.5$ from all three cases. From figure~\ref{fig:epsq95-d}, it can be easily seen for all cases that $\overline{s_\mathrm{edge}}$ peaks in the good curvature region and is negative throughout most of the bad curvature region, as was described in section~\ref{sec:tri}. However, for the low aspect ratio, low $q$ equilibrium in blue, the absolute magnitude of $\overline{s_\mathrm{edge}}$ is reduced across the low field side, leading to a significant destabilizing effect since shear stabilization is proportional to $s^2$. This case study demonstrates that it is not sufficient to move the peak of $\overline{s_\mathrm{edge}}$ to the good curvature side to archive $2^\mathrm{nd}$ stability access, as the relationship between the local shear, global shear and path length can still play a decisive role. 

\subsection{Effect of elongation}

\begin{figure}
	\includegraphics[width=1\linewidth]{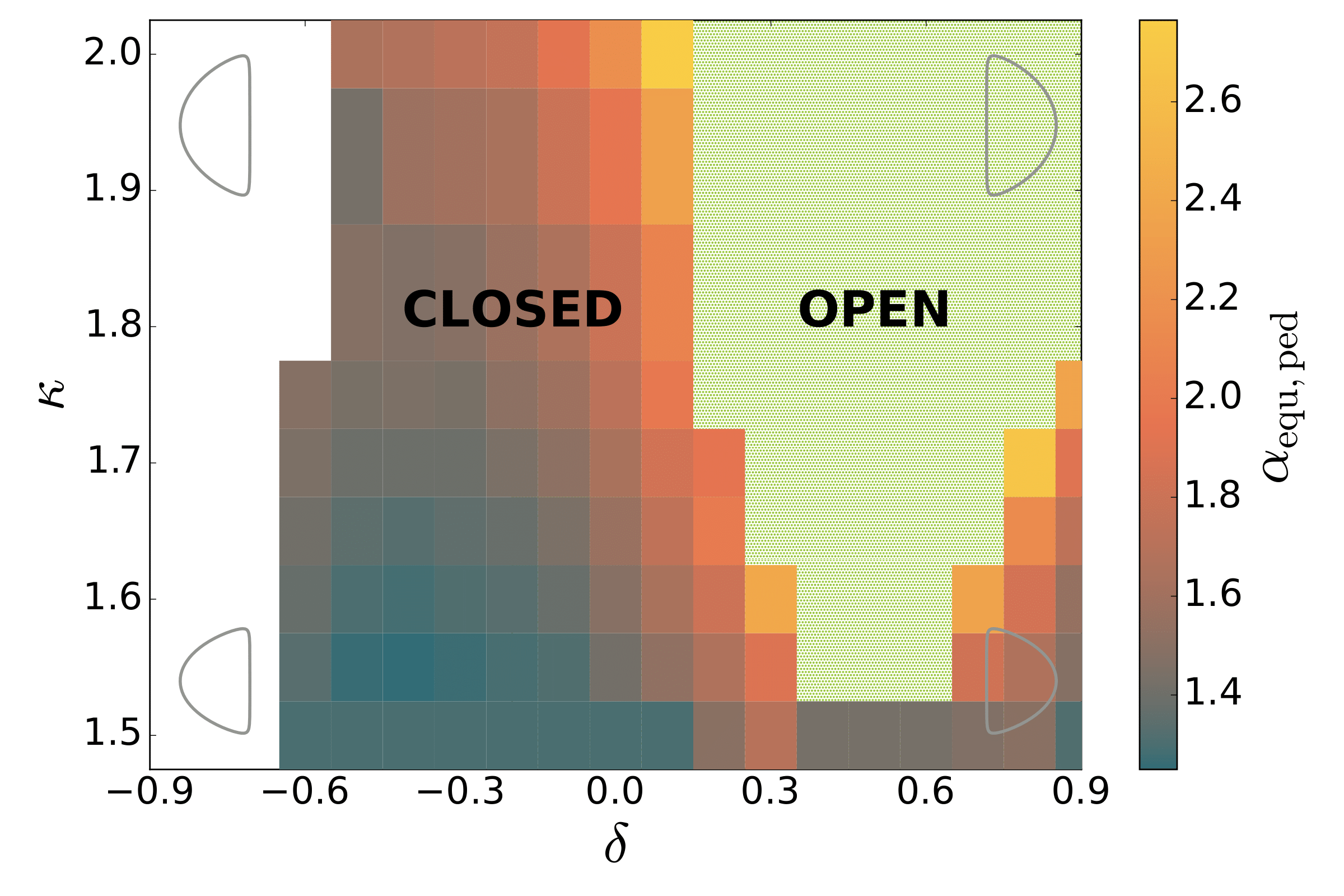}
	\caption{The maximum stable edge pressure gradient is plotted as a function of $\kappa$ and $\delta$ for equilibria with constant $I_\mathrm{p}=15\,$MA, $B_{t}=6\,$T and $\epsilon=1/3$. For $\delta<0$, the $2^\mathrm{nd}$ stability region cannot be accessed, and the edge gradients are thus limited by the infinite-$n$ ballooning mode. Configurations open to $2^\mathrm{nd}$ stability are shaded in green. Cartoons of equilibria at the extremes of the scan are shown in each quadrant.}
	\label{fig:kappa2D}
\end{figure}

In this section we show that, while high elongation ($\kappa$) is found to have relatively little effect on ballooning stability, low elongation can lead to robust closing of the $2^\mathrm{nd}$ stability region even in PT. As before, the ballooning stability as a function of $\kappa$ and $\delta$ is displayed in figure~\ref{fig:kappa2D} for plasmas with constant $I_\mathrm{p}=15\,$MA, $B_{t}=6\,$T and major radius $R_\mathrm{0}=7\,$m. The region of $\kappa$ -- $\delta$ space where the $2^\mathrm{nd}$ stable region is open is shaded in green, highlighting the improved plasma performance typically associated with strongly shaped, high-$\kappa$ discharges in PT configurations. The minimum triangularity needed to access $2^\mathrm{nd}$ stability is found to be roughly constant with elongation for $\kappa\gtrsim1.6$, with a slight trend towards earlier access at high $\kappa$. At low enough $\kappa$, the $2^\mathrm{nd}$ stability region is closed for all triangularity, an effect related to the decreasing absolute magnitude of the (negative) local shear in the bad curvature region with decreasing $\kappa$. This effect is strong enough to close $2^\mathrm{nd}$ stability region for plasmas with moderately low $\kappa$ and $\delta\sim1$ as well. For equilibria with $\delta<\delta_\mathrm{crit}(\kappa)$ where the pedestal gradients are limited by infinite-$n$ ballooning modes, slightly higher normalized pressure gradients are attainable at high $\kappa$. As with the aspect ratio scans above, we note that $\alpha$ is dependent on the plasma volume, which leads to larger equilibrium $\alpha$ at high $\kappa$ for equivalent pedestal pressures. The pedestal top pressure itself displays the opposite trend as $\alpha_\mathrm{equ,ped}$ in figure~\ref{fig:kappa2D}, with $\sim30\%$ lower ballooning-limited pedestals achieved at $\kappa=2.0$ than $\kappa=1.5$ for equilibria with $\delta<0$ in this scan. For all elongation, larger (more positive) values of triangularity lead to an increase in the maximum attainable pedestal pressure. Also evident in figure~\ref{fig:kappa2D} are singular equilibria just below the critical triangularity for $2^\mathrm{nd}$ stability access, which feature critical pressure gradients significantly above equilibria with slightly more unstable $\delta$, again emphasizing the abrupt nature of the critical stability boundary. 

\begin{figure}
	\includegraphics[width=1\linewidth]{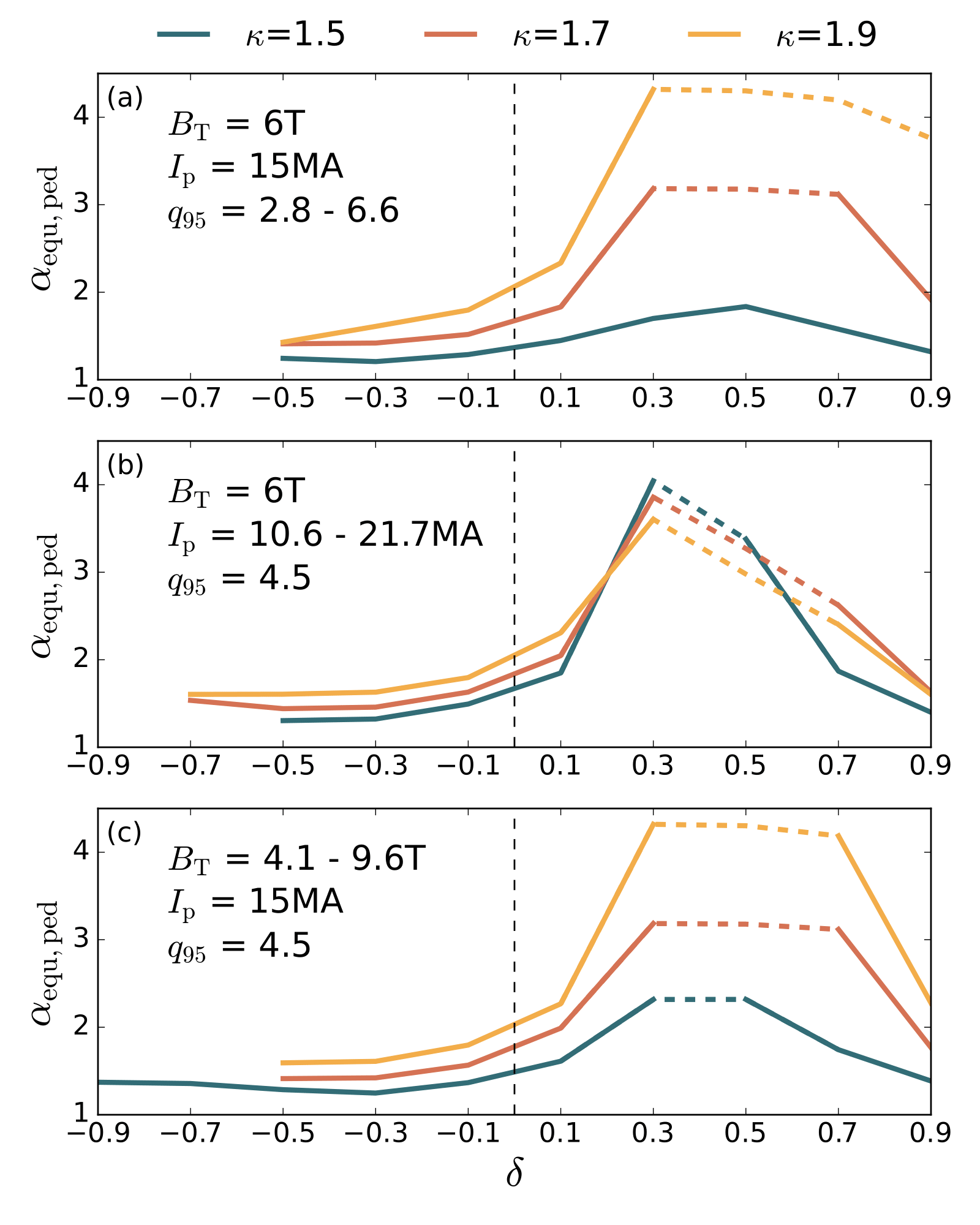}
	\labelphantom{fig:kappaq95-a}
	\labelphantom{fig:kappaq95-b}
	\labelphantom{fig:kappaq95-c}
	\caption{Maximum pedestal gradients for various $\kappa$ and $\delta$ at fixed (a) $I_\mathrm{p}$ and $B_\mathrm{T}$, (b) $B_\mathrm{T}$ and $q_\mathrm{95}$, and (c) $I_\mathrm{p}$ and $q_\mathrm{95}$. Regions open to $2^\mathrm{nd}$ stability are marked with a dashed line.}
	\label{fig:kappaq95}
\end{figure}

Modification of the plasma elongation changes the plasma volume, in turn influencing the edge safety factor and shear profiles. In figure~\ref{fig:kappa2D}, this effect is folded into the stability calculations, since the edge $q$ decreases significantly with decreasing $\kappa$ at constant $B_\mathrm{T}$ and $I_\mathrm{p}$. By allowing $B_\mathrm{T}$ and $I_\mathrm{p}$ to vary, a suite of equilibria with constant $q_{95}$ can be generated. An overview of these scans is given in figure~\ref{fig:kappaq95}, where the initial pedestal height guess was the same for all scans. Holding $q_\mathrm{95}$ constant while varying other parameters does not make a large impact on the maximum allowable $\alpha_\mathrm{ped}$ in configurations where access to the $2^\mathrm{nd}$ stable region is not available. This is especially evident in figure~\ref{fig:kappaq95-b}, where changing the plasma current to hold $q_\mathrm{95}$ constant across $\delta$ and $\kappa$ has a very large change in the starting value of $\alpha_\mathrm{equ, ped}$ in the open region due to compression of the flux surfaces $\psi$ at high $I_\mathrm{p}$. This effect almost completely vanishes in the closed region at $\delta<0$, however, as the gradient-limiting mechanisms of the infinite-$n$ ballooning mode come into effect. Also evident in figure~\ref{fig:kappaq95} is the requirement of high enough $q$ to access ballooning mode stabilization in $\delta>0$. In figure figure~\ref{fig:kappaq95-a}, the $\kappa=1.5$ simulation is unstable to ballooning modes at all $\delta$ due to the low value of $q$ in the edge. By raising $q$, either by increasing current or decreasing magnetic field, $2^\mathrm{nd}$ stability access at $\delta>0$ can be recovered. 

\subsection{Effect of squareness}

\begin{figure}
	\includegraphics[width=1\linewidth]{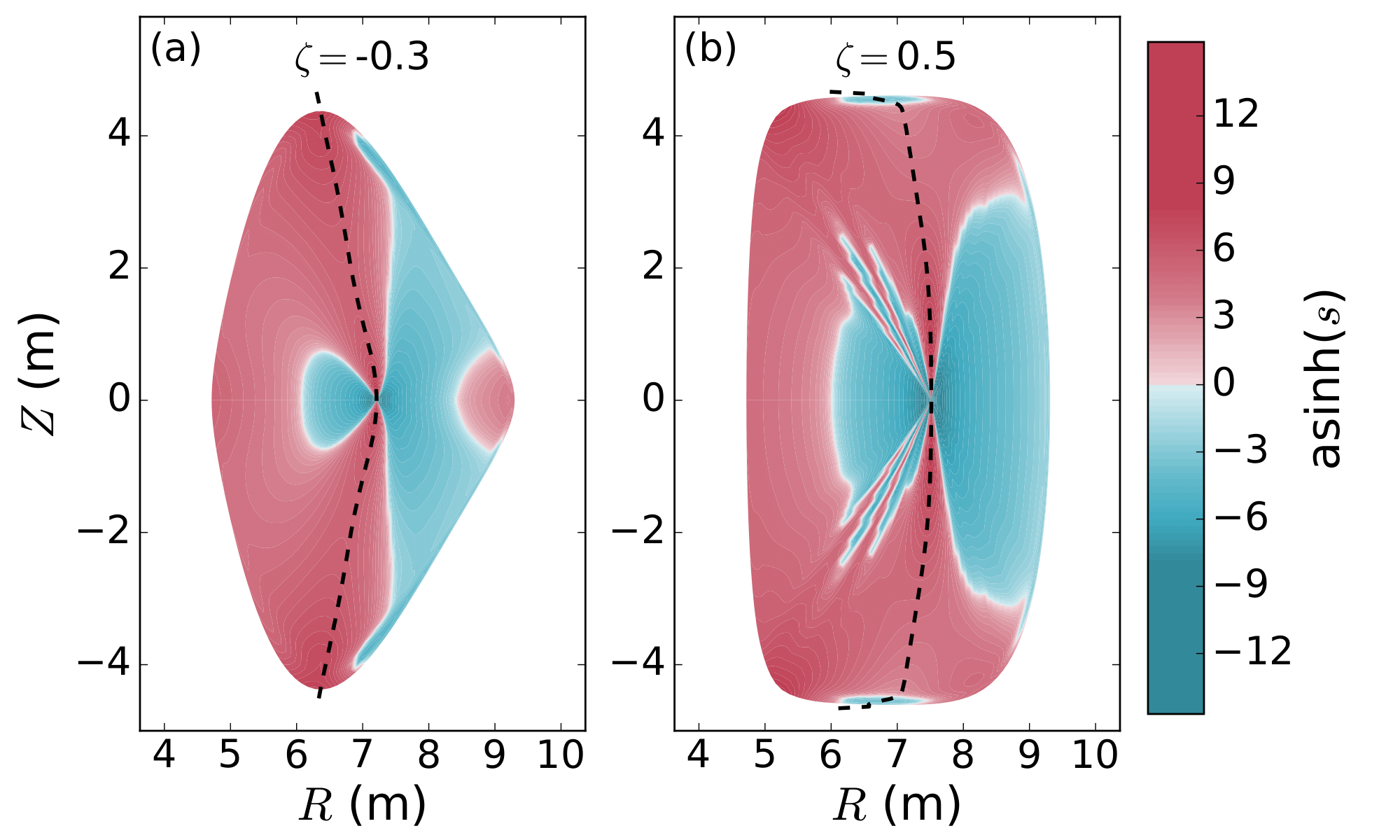}
	\labelphantom{fig:zetaeg-a}
	\labelphantom{fig:zetaeg-b}
	\caption{The local shear is plotted for two PT equilibria with (a) extremely low and (b) extremely high squareness $\zeta$. Due to the concentration of $s>0$ on the bad curvature side (right of the dashed line), both equilibria are closed to the $2^\mathrm{nd}$ stability regime.}
	\label{fig:zetaeg}
\end{figure}

The effect of the plasma squareness ($\zeta$) on $2^\mathrm{nd}$ stability access has been previously studied in experiments on the DIII-D tokamak, where an abrupt loss of access to the $2^\mathrm{nd}$ stability regime was observed at sufficiently large or sufficiently small $\zeta$ \cite{ferron_modification_2000}. In figure~\ref{fig:zetaeg}, we offer an explanation for these results by examining the local shear in reactor-like configurations with $I_\mathrm{p}=15\,$MA, $B_{t}=6\,$T and $\delta=0.3$. At an extremely low value of $\zeta$ (figure~\ref{fig:zetaeg-a},) the local shear in a portion of bad-curvature plasma edge near the outboard midplane switches from negative to positive, introducing a zero crossing into the low field side local shear that is destabilizing to ballooning modes. In figure~\ref{fig:zetaeg-b}, high squareness is responsible for pushing a local maximum in the local shear to the bad curvature side, much as negative triangularity did in the examples shown in figure~\ref{fig:sheareg}. As a result, the requirement of small $s$ over the entire bad curvature region is not met, preventing access to the $2^\mathrm{nd}$ stability regime. 

\begin{figure}
	\includegraphics[width=1\linewidth]{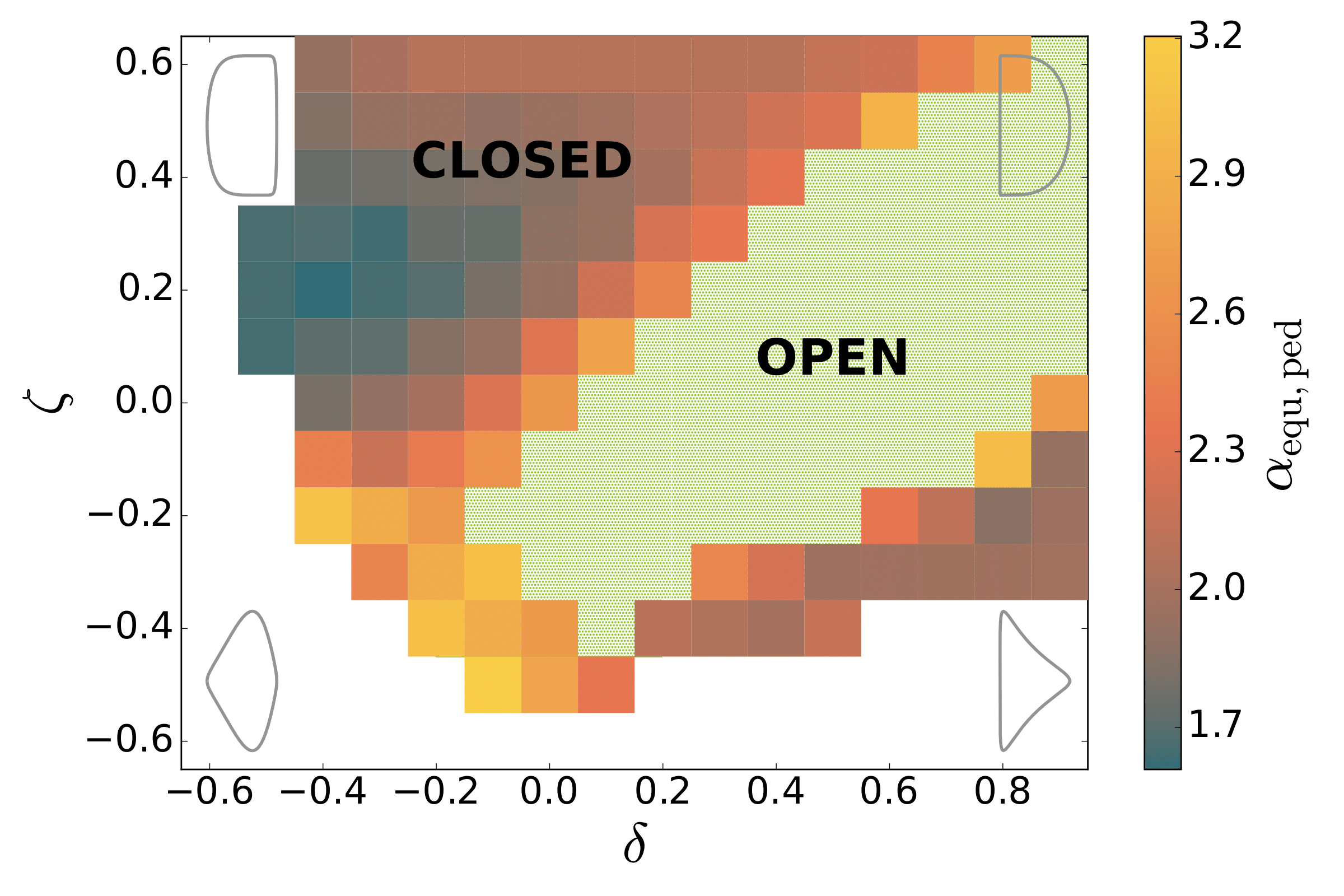}
	\caption{The maximum stable edge pressure gradient is plotted as a function of $\zeta$ and $\delta$ for equilibria with constant $I_\mathrm{p}=15\,$MA and $B_{t}=6\,$T. $2^\mathrm{nd}$ stability (shaded in green) can be accessed at $\delta>0$ and moderate squareness. Cartoons of equilibria at the extremes of the scan are shown in each quadrant.}
	\label{fig:zeta2D}
\end{figure}

A full picture of the impact of squareness on infinite-$n$ ballooning stability is provided in figure~\ref{fig:zeta2D}, which shows $\alpha_\mathrm{equ,ped}$ as a function of both $\zeta$ and $\delta$. The infinite-$n$ ballooning instability is found to vary relatively strongly with $\zeta$ across the full range scanned as the two effects illustrated in figure~\ref{fig:zetaeg} compete with the effects of triangularly to control local shear on the low field side. Generally, equilibria with more moderate values of $\zeta$ typical of reactor design $(-0.1\lesssim\zeta\lesssim0.1)$ behave as expected from previous studies, with increased ballooning stability at large $\delta>0$, suggesting that relatively extreme modification of $\zeta$ is needed to achieve the effects demonstrated in figure~\ref{fig:zetaeg}. Increasing the squareness above $\zeta=0$ results in an increase in $\delta_{crit}$. Interestingly, by reducing $\zeta$ slightly, the access window for $2^\mathrm{nd}$ stability can be extended significantly below $\delta=0$, though stability is eventually lost at $\delta\lesssim-0.2$. These results highlight the robustness of NT in preventing $2^\mathrm{nd}$ stability access, even through the the complex dependence of local shear on other aspects of the plasma geometry. 

\subsection{Effect of up-down asymmetry}

Throughout these scans, variations without up-down symmetry were also studied. The effect of asymmetric upper and lower triangularity is relatively simple to understand, since the access to the $2^\mathrm{nd}$ stability region is prevented by the minimum of $\delta_\mathrm{u}$ and $\delta_\mathrm{l}$. For example, if for a particular scenario $\delta_\mathrm{l}=-0.2$ is strong enough to push the positive region of the local shear on the bottom of the tokamak from the good curvature to the bad curvature region, $2^\mathrm{nd}$ stability access will be prevented no matter the value of $\delta_\mathrm{u}$. Similarly, earlier work on the effect of asymmetry has suggested that the minimum of $\delta_\mathrm{l}$ and $\delta_\mathrm{u}$ matters more for kinetic ballooning mode stability than the maximum $\delta$ by roughly a factor of 2:1, which leads to a 2/3 weighting of the minimum triangularly in the EPED code \cite{Snyder2011}. These results emphasize the importance of achieving low local shear across the entire bad curvature region in $2^\mathrm{nd}$ stability access, and are consistent with studies of varying upper triangularity on DIII-D and TCV \cite{saarelma_ballooning_2021, merle_pedestal_2017}. 



\section{Database analysis}
\label{sec:disc}

In the above discussion, emphasis was placed on key scans in order to demonstrate the specific relationship between individual shaping parameters and access to the $2^\mathrm{nd}$ stability region. In this section, we apply the same concepts to a database of randomized equilibria in order to understand the robustness of the above trends, and to more fully characterize the broader reactor operational space. From this database, scaling laws for the normalized pressure gradient and the pedestal pressure can be generated, providing a simple starting point for the inclusion of physics dominating the NT edge into integrated tokamak models \cite{Meneghini2021, Schwartz2022}. 

\subsection{Database generation}

To that end, over 1500 separate marginally-ballooning-stable equilibria have been generated with a simple loop of the CHEASE and BALOO codes. The shapes of these equilibria were chosen randomly from the parameter space including triangularities $-0.9<\delta<0.9$, inverse aspect ratios $0.1<\epsilon<0.9$, elongations $1.2<\kappa<2.2$, squareness $-0.1<\zeta<0.1$ and minor radii $2<a_\mathrm{minor}<4.5\,$m. Magnetic field and plasma current were constrained within $4 - 10\,$T and $8 - 20\,$MA, respectively. In line with the previous analysis, the EPED parameterization was used to define the electron temperature and density profiles, with $0.5\times10^{20}<n_\mathrm{e,ped}<2\times10^{20}\,\mathrm{m}^{-3}$, $1\times10^{20}<n_\mathrm{e,core}<3\times10^{20}\,\mathrm{m}^{-3}$, $1<T_\mathrm{e,ped}<2.5\,$keV and $15<T_\mathrm{e,core}<3\,$keV and the pedestal width varying between $0.03<\Delta_\mathrm{ped}<0.07$. The profiles were subject to an additional monotonicity constraint, forcing $n_\mathrm{e,core}>n_\mathrm{e,ped}$. For each equilibrium configuration, the bootstrap current was calculated self-consistently from the density and temperature profiles using the Sauter formalism \cite{Sauter1999, Sauter2002}. Configurations which required a bootstrap fraction ($f_\mathrm{BS}$) greater than unity were neglected. 

\begin{figure*}
	\includegraphics[width=0.75\linewidth]{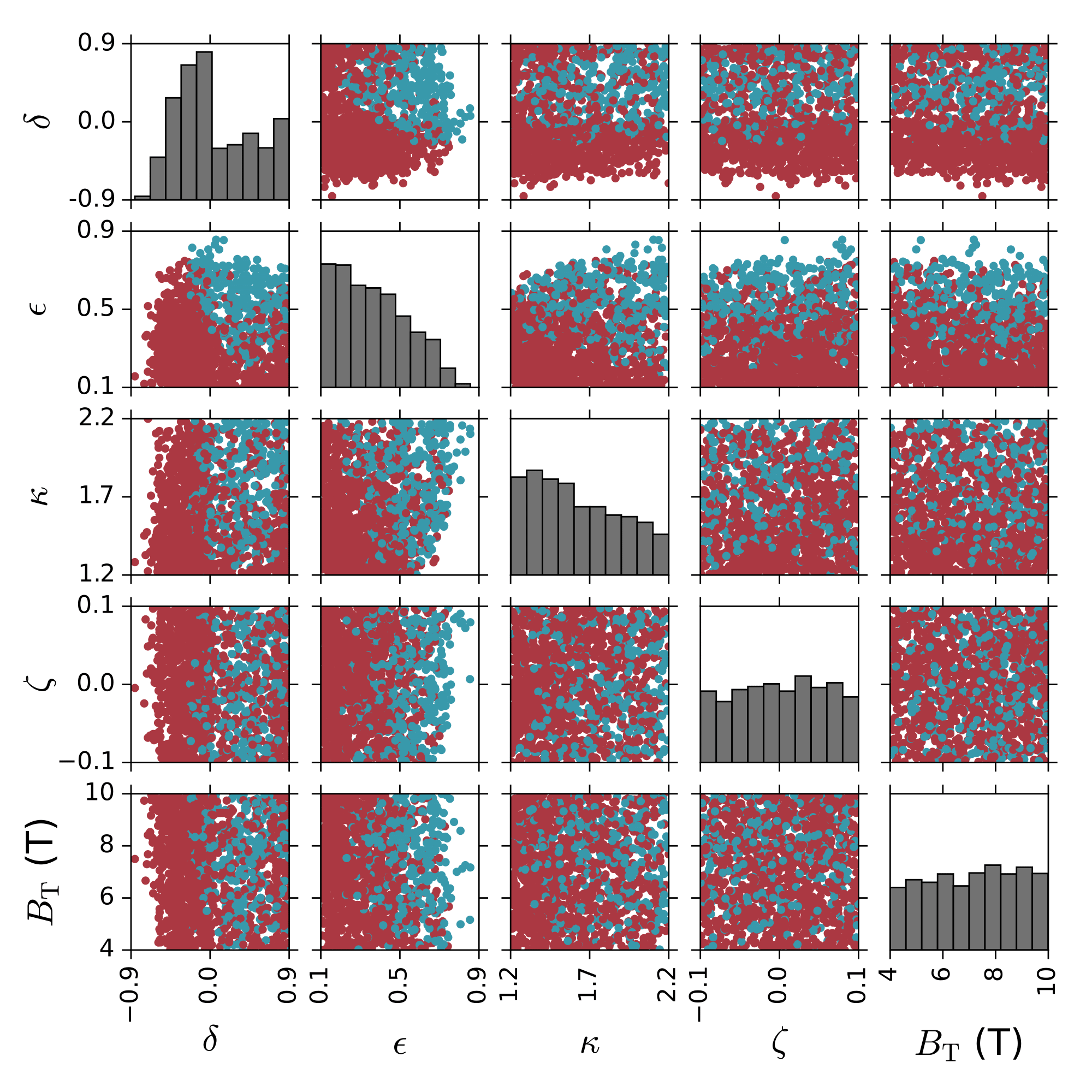}
	\caption{Distributions of the randomized equilibria across the five most impactful regression parameters: triangularity ($\delta$), inverse aspect ratio ($\epsilon$), elongation ($\kappa$), squareness ($\zeta$) and magnetic field ($B_\mathrm{T}$). Red dots represent equilibria that cannot access the $2^\mathrm{nd}$ stable region, and blue dots those that can. Single-parameter distributions are included along the diagonal to illustrate the shape of the database.}
	\label{fig:LDA}
\end{figure*}

After generation, the infinite-$n$ ballooning stability of each equilibria was analyzed using the BALOO code. In cases where the initial profile guess results in a configuration unstable to ballooning modes, regardless of access to the $2^\mathrm{nd}$ stability region, the pedestal top pressure was reduced until the equilibrium $\alpha$ was within 5\% of the $1^\mathrm{st}$ stability limit. Conversely, in cases were the initial profile is stable and access to the $2^\mathrm{nd}$ stability region was closed, the pedestal pressure was increased until either the $1^\mathrm{st}$ stability limit was reached, or an opening to the $2^\mathrm{nd}$ stability region was realized. If at any point a stable profile was achieved with access to the $2^\mathrm{nd}$ stability region, no further changes were made to the equilibrium. This iteration scheme produces equilibria with the maximum achievable $\alpha_\mathrm{ped}$, except in cases where the gradients are not limited by infinite-$n$ ballooning modes.

In order to best capture the dynamics of equilibria with negative triangularity, a slight weighting was placed on generating configurations with $\delta<0$. However, due to numerical difficulties, full iterations with extremely negative triangularities less that $\delta<-0.6$ were less likely to converge, resulting in inclusion of approximately equal numbers of PT and NT equilibrium in the database. Similar issues were encountered at large $\epsilon\gtrsim0.8$ and $\kappa\gtrsim2$, slightly weighting the resulting database towards high aspect ratio (low minor radius) and lower elongation. The requirement of bootstrap fraction $f_\mathrm{BS}<1$ also skewed the database towards more equilibria with higher $I_\mathrm{p}$. All other parameters are represented approximately equally in the final database. The full distributions of the most important shaping parameters are shown in figure~\ref{fig:LDA}.

\subsection{Predicting access to the $2^\mathrm{nd}$ stability region}

Using the full database of equilibria, a simple least discriminant analysis (LDA) is performed using singular value decomposition in order to characterize separation between configurations with and without access to the $2^\mathrm{nd}$ stability regime. The best access prediction is obtained using five input features: $\delta$, $\epsilon$, $\kappa$, $\zeta$ and $B_\mathrm{T}$, with an accuracy score of $\sim89\%$. The classification can be written as
\begin{equation} \label{eq:LDA}
\begin{split}
    x = &15.629 - 0.419B_\mathrm{T} - 13.140  \epsilon\\  
    &- 3.096 \kappa - 3.361 \delta + 3.450 \zeta,
\end{split}
\end{equation}
where a configuration is predicted to have access to the $2^\mathrm{nd}$ stability region if
\begin{equation} 
\begin{split}
    \frac{1}{1+\mathrm{exp}(x)} > 0.5.
\end{split}
\end{equation}
It is noteworthy that the majority of the separation between `closed' and `open' equilibria can be accounted for using only a combination of $\delta$ and $\epsilon$, which is already able to achieve an accuracy score of $\sim86\%$. 

The relative simplicity of this prediction is partially illustrated in figure~\ref{fig:LDA}, which shows the distributions of database equilibria over the input features found to yield the most successful predictive model. In this figure, closed equilibria are shown in red and the equilibria with access to $2^\mathrm{nd}$ stability are shown in blue. In the first column, a sharp cutoff in the data is seen when transitioning to negative triangularities, with almost all of the equilibria with $\delta<0$ closed to $2^\mathrm{nd}$ stability. In particular, inspection of the subplots corresponding to relationships $\delta$ vs. $\epsilon$ and $\delta$ vs. $\kappa$ reveal strong trends reminiscent of those identified in dedicated scans in figures~\ref{fig:eps2D} and \ref{fig:kappa2D}, respectively. This is especially evident when compared to the parameters with weaker influences on ballooning stability access, such as in the subplots comparing $\zeta$ and $B_\mathrm{T}$, which shows no clear parameter dependence on stability access. After the triangularity, the second most important parameter in this database is $\epsilon$: low $\epsilon$ closes access to the $2^\mathrm{nd}$ stability region even at $\delta>0$, leading to most of the false positives for this model occurring at large $\delta$ and low $\epsilon$, as can be inferred from figure~\ref{fig:LDA}.

Of course, there is still significant scatter in the boundary between closed and open equilibria shown in figure~\ref{fig:LDA} that is not captured by this simple model. The internal profiles of the database equilibria, which vary greatly ($0.05 \lesssim \nu_\mathrm{e, ped}^* \lesssim 36$) across the parameter space and impact ballooning stability through the effect of the bootstrap current and local shear, account for a large percentage of this scatter. To assess this effect, the LDA can be expanded to include the normalized pedestal top collisionality $\nu_\mathrm{e, ped}^*$, which impacts infinite-$n$ ballooning stability by decreasing the bootstrap current and raising magnetic shear in the edge. 
Interestingly, no statistically significant improvement is made by including $\nu_\mathrm{e, ped}^*$ into the model predictions, suggesting that access to the $2^\mathrm{nd}$ stability region is more strongly impacted by shape than collisionality.
It is likely that further improvements in the model can be made by exploring other classifiers, but this will provide marginal additional physics understanding and is not attempted here. 

\subsection{Prediction of ballooning-limited maximum $\alpha_\mathrm{ped}$}

In addition to exploring which configurations are and are not limited by infinite-$n$ ballooning modes, it is interesting to explore variations in the maximum achievable normalized pressure gradient $\alpha_\mathrm{equ,ped}$ (see equation~\ref{eq:alpha}) as a function of shape for equilibria where $\alpha$ is limited by the first ballooning instability boundary. For these cases, the infinite-$n$ ballooning mode sets a hard limit on the edge gradients, much in the same way that kinetic ballooning modes, calculated with the ballooning-critical-pedestal technique, do in the EPED model for H-mode pedestals \cite{Snyder2011}. It must be noted, however, that we do not include a secondary instability in this analysis as is done in the EPED model. In the case of EPED, inclusion of global peeling-ballooning mode physics allows for simultaneous determination of both the pedestal height and the pedestal width for an ELMy H-mode plasma.  

\begin{figure}
	\includegraphics[width=1\linewidth]{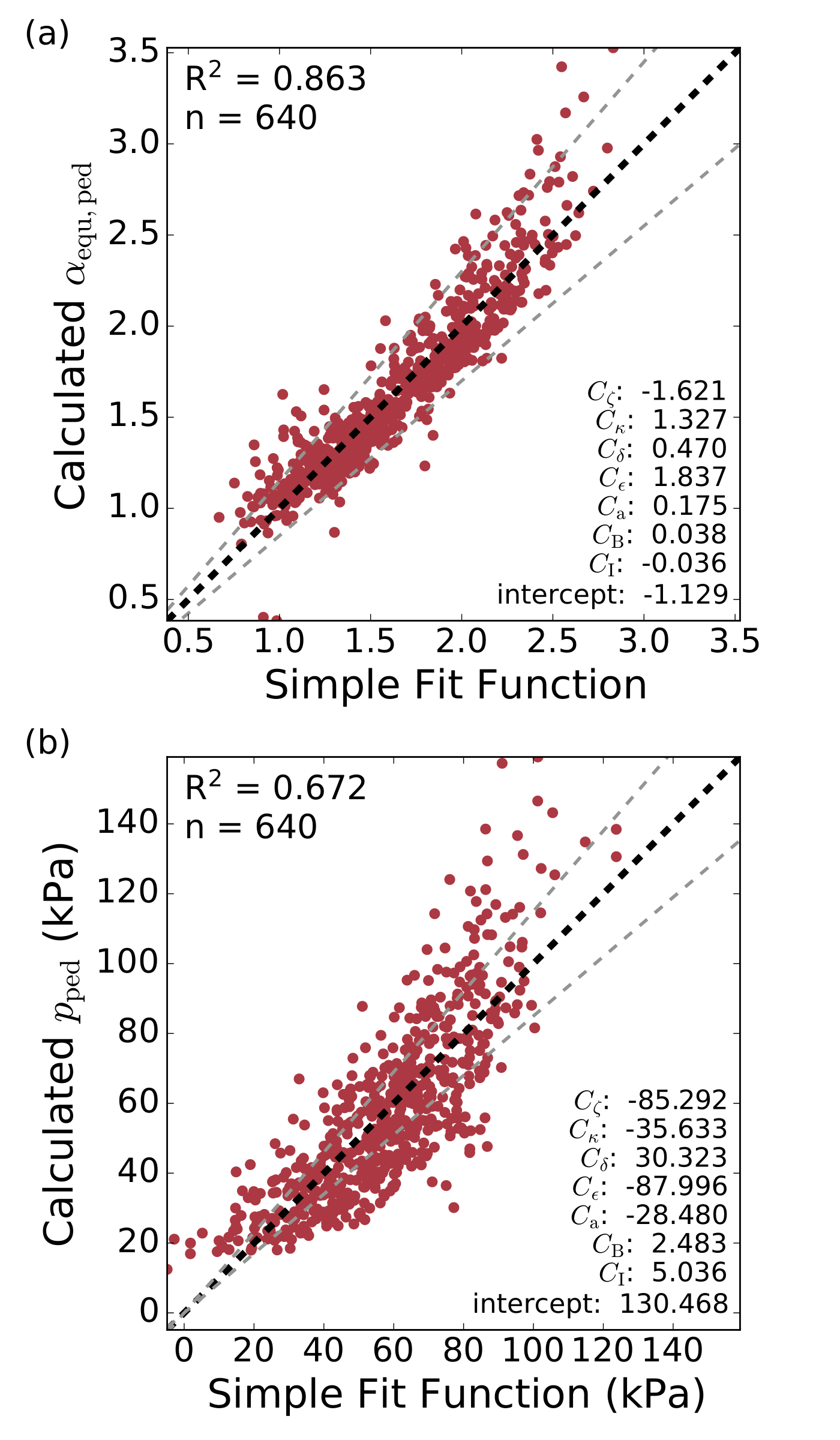}
	\labelphantom{fig:fits-a}
	\labelphantom{fig:fits-b}
	\caption{Closed database equilibria with $\delta<0$ are used to produce scalings of (a) the maximum normalized pressure gradient max($\alpha_\mathrm{ped}$) and (b) the pedestal pressure as a function of machine parameters. Grey dashed lines show $15\%$ error bars.}
	\label{fig:fits}
\end{figure}

To develop a scaling for $\alpha_\mathrm{equ,ped}$ for equilibria limited by the $1^\mathrm{st}$ stability boundary, we fit the modeled $\alpha_\mathrm{equ,ped}$ from the database to the machine parameters $\kappa$, $\epsilon$, $\delta$, $\zeta$, $a_\mathrm{minor}$, $B_\mathrm{T}$ and $I_\mathrm{p}$ in figure~\ref{fig:fits-a}. Almost 700 equilibria covering a large range of $\alpha_\mathrm{equ,ped}$ are selected from the larger database for this analysis by the requiring that access to $2^\mathrm{nd}$ stability is closed that $\delta<0$. A simple linear regression of the form
\begin{equation} \label{eq:fits}
\begin{split}
    \alpha_\mathrm{equ,ped} =\,&C_\mathrm{0} +  C_\mathrm{\kappa}\kappa + C_\mathrm{\epsilon}\epsilon + C_\mathrm{\delta}\delta + C_\mathrm{\zeta}\zeta \\
    +\,&C_\mathrm{a}a_\mathrm{minor} \mathrm{[m]}
    + C_\mathrm{B}B_\mathrm{T} \mathrm{[T]} + C_\mathrm{I}I_\mathrm{p}  \mathrm{[MA]}
\end{split}
\end{equation}
is then applied to this data set, where $a_\mathrm{minor}$, $B_\mathrm{T}$ and $I_\mathrm{p}$ are given in m, T and MA, respectively, and the rest of the parameters are unit-less. All equilibria included in this fit have Greenwald density fractions less than unity. The coefficient of determination $R^2$ is used as a scoring metric for the fit, yielding a reasonable $R^2\sim85\%$, and the majority of tested cases lie within $\pm15\%$ from the expected value. The relative importance of each term in the regression can be considered by comparing the respective regression coefficients after normalizing the data set to account for differing absolute magnitudes of the various parameters. With this procedure, it is evident that the most important shaping contributions to $\alpha_\mathrm{equ,ped}$ come from $\kappa$ and $\epsilon$, which were already shown to have a large impact on $\alpha_\mathrm{equ,ped}$ in section~\ref{sec:shape} due to their role in changing the plasma volume and the bad curvature connection length.

In order to disentangle the volumetric contribution of $\epsilon$ and $\kappa$ to $\alpha_\mathrm{equ,ped}$ from the pressure profile itself, a linear regression directly to the ballooning-limited pedestal pressure ($p_\mathrm{ped}$) is provided in figure~\ref{fig:fits-b} for the same data set. This regression performs significantly worse than the regression for $\alpha_\mathrm{equ,ped}$, with an $R^2$ score of $\sim65\%$, indicating that the critical stabilizing factors are associated with the pressure gradient, rather than the pressure itself. However, general trends of the database are still well described by the model. Comparing coefficients for an identical fit to normalized parameters, it is evident that increased $I_\mathrm{p}$ has the strongest correlation with increased $p_\mathrm{ped}$, whereas increases in both $a_\mathrm{minor}$ and $\epsilon$ correlate with strong decreases in the $p_\mathrm{ped}$. Notably, increased (less negative) $\delta$ is also correlated with an increased pedestal pressure, as was seen previously in the case studies presented in section~\ref{sec:shape}. 

Future work to improve the accuracy of this model should include additional transport calculations to help separate the various effects contributing to gradient production in the L-mode edge, which is outside the scope of this work. Alternatively, since infinite-$n$ ballooning calculations are relatively straightforward and fast to run with codes like BALOO, it should be possible to extend this work in the future to produce a full database of equilibria with arbitrary profiles in order to construct a neural network for ballooning-limited profile prediction similar to EPEDNN. Such a task should be informed by experimental profiles as much as possible, and will be attempted after the additional experiments are performed on DIII-D. As part of the present engagement, a computational loop has been added to OMFIT's STEP framework \cite{Meneghini2015, Meneghini2021} in order to calculate the ballooning boundary and iteratively adjust modeled profiles in the edge when relevant, allowing for profile predictions at $\delta<0$.

\section{Conclusion}

From these results, we predict that robust operation of a NT tokamak reactor with an L-mode-like edge should be possible without significant risk of an unwanted transition to a strong H-mode regime. Through a systematic analysis of over 2000 model equilibria, we have shown that sufficiently negative triangularity robustly destabilizes the infinite-$n$ ballooning mode in the tokamak edge such that the $2^\mathrm{nd}$ stability region becomes inaccessible, and the critical pressure gradient is low compared to positive triangularity cases. This prevents growth of the large pressure gradients needed to access and sustain high pedestal H-mode operation. While week (grassy) H-mode operation may still be possible in the $1^\mathrm{st}$ stability region, high power discharges with $\delta\ll0$ are predicted to operate near the $1^\mathrm{st}$ instability boundary, which will limit the maximum pedestal gradient to significantly depressed values compared to traditional H-mode operation. Further experimental and modeling studies are needed to understand if the prominent infinite-$n$ ballooning modes predicted by this work are benign at reactor scales, as has been suggested by previous work \cite{Kikuchi2019, Austin2019}.

In all cases described here, the transition between an L-mode-like ballooning-limited regime (lower critical pressure gradient) and strong H-mode operation through $2^\mathrm{nd}$ stability access occurs abruptly at a critical $\delta_\mathrm{crit}$, which is a function of various configuration parameters. The transition to $2^\mathrm{nd}$ stability access can be understood by considering the magnitude and sign of the local shear on the bad curvature side of the equilibrium. To achieve stability against ballooning modes, the local shear must be strong and negative across the entire bad curvature region. Decreasing the bad curvature connection length, decreasing the pedestal collisionality or increasing the good curvature stabilization can also help to open access to the $2^\mathrm{nd}$ stability region, though the local shear plays a dominant role. These considerations help explain previously-observed experimental results which document sudden changes in ballooning stability with changes in the plasma shape \cite{ferron_modification_2000, saarelma_ballooning_2021}.

Looking forward towards development of tokamak reactors, these results suggest that NT could be an important and robust mechanism with which to reduce the height of the pressure pedestal and completely eliminate or severely mitigate dangerous instabilities like ELMs \cite{Kikuchi2019,Medvedev2015,Austin2019}. Additionally, if L-mode operation is desired, these results suggest that discharges with exceptionally high squareness (which also maximizes plasma volume) could also provide an attractive alternative to the traditional ``Dee" shape.  Even so, further experimental validation of these results is required, and should be conducted on capable machines as soon as possible. 

\appendix

\section{Appendix A: Variation of pedestal width and justification of the EPED profile model}
\label{app:pedwidth}


\begin{figure}
	\includegraphics[width=1\linewidth]{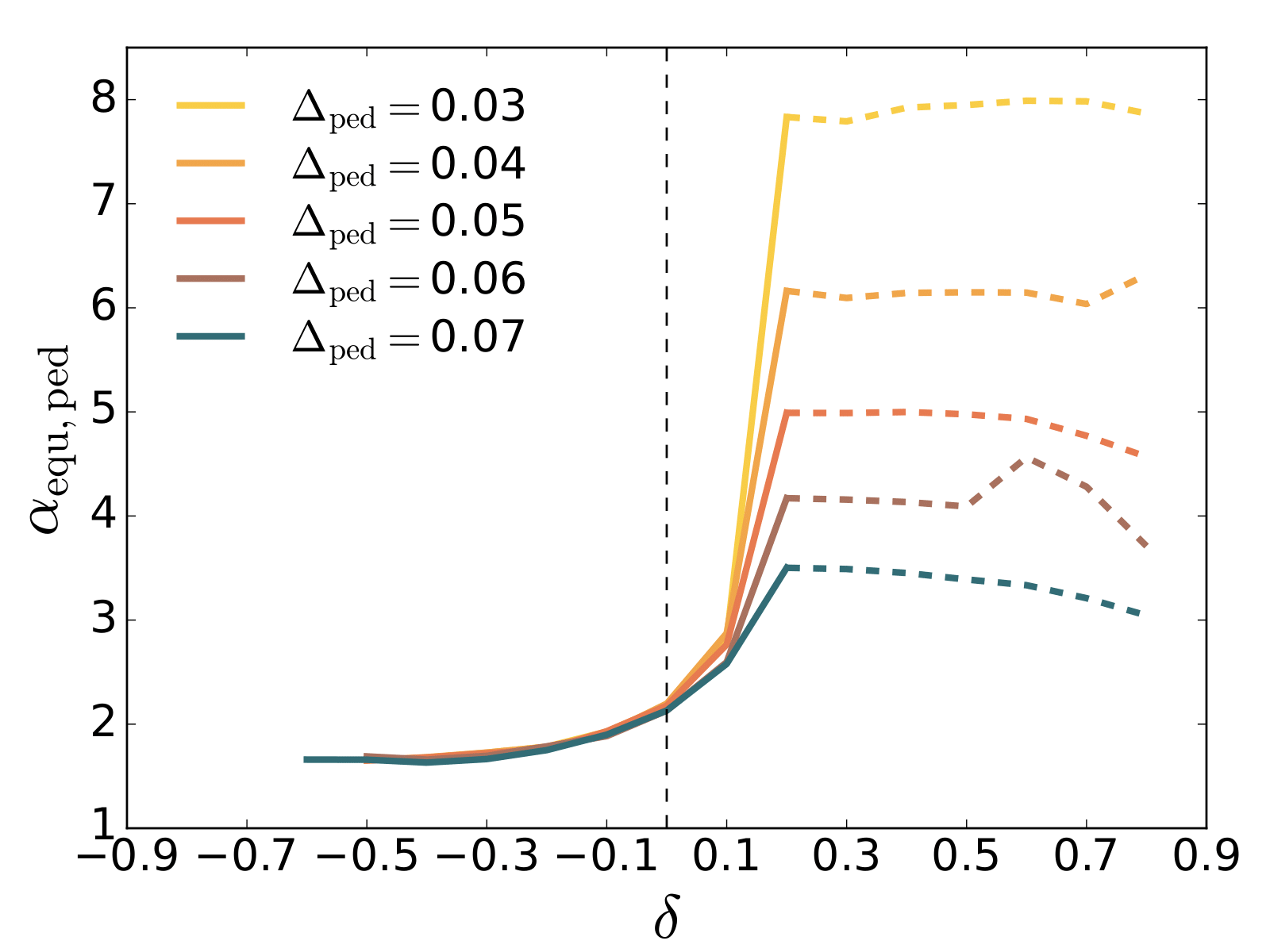}
	\caption{The maximum stable normalized gradient is presented as a function of pedestal width $\Delta_\mathrm{ped}$ and triangularity $\delta$. When the $2^\mathrm{nd}$ stability region is closed at $\delta\lesssim0.2$, the pedestal width has little effect on the maximum achievable gradient. In this and following figures, locations where $2^\mathrm{nd}$ stability is open (and the profiles are thus unconstrained) are marked with a dashed line.}
	\label{fig:pedwidth}
\end{figure}

During actual operation of a reactor, the edge pressure profiles may not be constrained a simple $\texttt{mtanh}$ form with fixed pedestal width and may instead vary in complicated ways to maximize turbulent transport \cite{Nelson_NFjog}. To develop some intuition for how arbitrary profile changes may influence access to the $2^\mathrm{nd}$ stability region, a pedestal width scaling is presented as function of $\delta$ in figure~\ref{fig:pedwidth}. The initial equilibrium $\alpha$ increases with decreasing pedestal width ($\Delta_\mathrm{ped}$) for a fixed pedestal height, as is evident in the unconstrained curves plotted with dashed lines at $\delta>0$. However, at low triangularities $\delta\lesssim0.2$, where the gradient is limited by infinite-$n$ stability, $\Delta_\mathrm{ped}$ has little effect on the maximum achievable $\alpha_\mathrm{ped}$. This stems from the fact that the $1^\mathrm{st}$ stability boundary is relatively flat as a function of $\psi_\mathrm{N}$ for NT equilibria (see, for example, figure~\ref{fig:deltaEx-c}.) As such, the maximum achievable $\alpha_\mathrm{ped}$ remains the same regardless of pedestal width, suggesting that shape plays a much stronger role in determining the onset of $2^\mathrm{nd}$ stability closure and the ballooning-limited $\alpha_\mathrm{ped}$ than the exact shape of the pressure profile, which is consistent with experimental observations reporting sudden shape-triggered transitions between H-mode and L-mode \cite{saarelma_ballooning_2021}.

It should be noted, however, that ballooning-optimized profiles can be artificially constructed to slightly adjust the critical $\delta$ for closing of the $2^\mathrm{nd}$ stability regime. For example, one could imagine squeezing the pedestal for the $\delta=0$ case in figure~\ref{fig:deltaEx-b} to occur only between $0.95<\psi_\mathrm{N}<0.97$. In this case, the $2^\mathrm{nd}$ stability remains open and the profile can be tailored to no longer cross the ballooning instability boundary without reducing $\alpha_\mathrm{ped}$. However, achievement of this profile in experiment remains uncertain due to the large required shift of the entire pedestal away from the separatrix location. Without experimental realization of such a profile, we choose in this work to ignore such potential ``squeezing" effects, instead noting that all reported values of the critical $\delta$ for $2^\mathrm{nd}$ stability suppression should be considered to be slightly over-optimistic to account for additional profile optimization factor. 

\begin{acknowledgments}
\textit{Acknowledgments.} The authors would like to acknowledge helpful discussions with P. B. Snyder which helped to improve this manuscript through development of the BALOO code, the identification of first-stability-limited H-mode discharges and through improvement of the manuscript text. We also thank F. M. Laggner for reviewing the initial manuscript. C. Paz-Soldan acknowledges consulting for General Atomics. All calculations performed in this study were completed through the OMFIT framework \cite{Meneghini2015}, and the generated equilibria are available upon request. This material is based upon work supported by the U.S. Department of Energy, Office of Science, Office of Fusion Energy Sciences under Awards DE-SC0022270 and DE-SC0022272.
\end{acknowledgments}


\begin{thebibliography}{45}%
\makeatletter
\providecommand \@ifxundefined [1]{%
 \@ifx{#1\undefined}
}%
\providecommand \@ifnum [1]{%
 \ifnum #1\expandafter \@firstoftwo
 \else \expandafter \@secondoftwo
 \fi
}%
\providecommand \@ifx [1]{%
 \ifx #1\expandafter \@firstoftwo
 \else \expandafter \@secondoftwo
 \fi
}%
\providecommand \natexlab [1]{#1}%
\providecommand \enquote  [1]{``#1''}%
\providecommand \bibnamefont  [1]{#1}%
\providecommand \bibfnamefont [1]{#1}%
\providecommand \citenamefont [1]{#1}%
\providecommand \href@noop [0]{\@secondoftwo}%
\providecommand \href [0]{\begingroup \@sanitize@url \@href}%
\providecommand \@href[1]{\@@startlink{#1}\@@href}%
\providecommand \@@href[1]{\endgroup#1\@@endlink}%
\providecommand \@sanitize@url [0]{\catcode `\\12\catcode `\$12\catcode
  `\&12\catcode `\#12\catcode `\^12\catcode `\_12\catcode `\%12\relax}%
\providecommand \@@startlink[1]{}%
\providecommand \@@endlink[0]{}%
\providecommand \url  [0]{\begingroup\@sanitize@url \@url }%
\providecommand \@url [1]{\endgroup\@href {#1}{\urlprefix }}%
\providecommand \urlprefix  [0]{URL }%
\providecommand \Eprint [0]{\href }%
\providecommand \doibase [0]{https://doi.org/}%
\providecommand \selectlanguage [0]{\@gobble}%
\providecommand \bibinfo  [0]{\@secondoftwo}%
\providecommand \bibfield  [0]{\@secondoftwo}%
\providecommand \translation [1]{[#1]}%
\providecommand \BibitemOpen [0]{}%
\providecommand \bibitemStop [0]{}%
\providecommand \bibitemNoStop [0]{.\EOS\space}%
\providecommand \EOS [0]{\spacefactor3000\relax}%
\providecommand \BibitemShut  [1]{\csname bibitem#1\endcsname}%
\let\auto@bib@innerbib\@empty
\bibitem [{\citenamefont {Wagner}\ \emph {et~al.}(1984)\citenamefont {Wagner},
  \citenamefont {Fussmann}, \citenamefont {Grave}, \citenamefont {Keilhacker},
  \citenamefont {Kornherr}, \citenamefont {Lackner}, \citenamefont {Mccormick},
  \citenamefont {Muller}, \citenamefont {Stabler}, \citenamefont {Becker},
  \citenamefont {Bernhardi}, \citenamefont {Ditte}, \citenamefont {Eberhagen},
  \citenamefont {Gehre}, \citenamefont {Gernhardt}, \citenamefont {Gierke},
  \citenamefont {Glock}, \citenamefont {Gruber}, \citenamefont {Haas},
  \citenamefont {Hesse}, \citenamefont {Janeschitz}, \citenamefont {Karger},
  \citenamefont {Kissel}, \citenamefont {K1qber}, \citenamefont {Lisitano},
  \citenamefont {Mayer}, \citenamefont {Meisel}, \citenamefont {Mertens},
  \citenamefont {Murrnann}, \citenamefont {Poschenrieder}, \citenamefont
  {Rapp}, \citenamefont {Rohr}, \citenamefont {Ryter}, \citenamefont
  {Schneider}, \citenamefont {Siller}, \citenamefont {Smeulders}, \citenamefont
  {Soldner}, \citenamefont {Speth}, \citenamefont {Steuer}, \citenamefont
  {Szymanski},\ and\ \citenamefont {Vollmer}}]{Wagner1984}%
  \BibitemOpen
  \bibfield  {author} {\bibinfo {author} {\bibfnamefont {F.}~\bibnamefont
  {Wagner}}, \bibinfo {author} {\bibfnamefont {G.}~\bibnamefont {Fussmann}},
  \bibinfo {author} {\bibfnamefont {T.}~\bibnamefont {Grave}}, \bibinfo
  {author} {\bibfnamefont {M.}~\bibnamefont {Keilhacker}}, \bibinfo {author}
  {\bibfnamefont {M.}~\bibnamefont {Kornherr}}, \bibinfo {author}
  {\bibfnamefont {K.}~\bibnamefont {Lackner}}, \bibinfo {author} {\bibfnamefont
  {K.}~\bibnamefont {Mccormick}}, \bibinfo {author} {\bibfnamefont {E.~R.}\
  \bibnamefont {Muller}}, \bibinfo {author} {\bibfnamefont {A.}~\bibnamefont
  {Stabler}}, \bibinfo {author} {\bibfnamefont {G.}~\bibnamefont {Becker}},
  \bibinfo {author} {\bibfnamefont {K.}~\bibnamefont {Bernhardi}}, \bibinfo
  {author} {\bibfnamefont {U.}~\bibnamefont {Ditte}}, \bibinfo {author}
  {\bibfnamefont {A.}~\bibnamefont {Eberhagen}}, \bibinfo {author}
  {\bibfnamefont {O.}~\bibnamefont {Gehre}}, \bibinfo {author} {\bibfnamefont
  {J.}~\bibnamefont {Gernhardt}}, \bibinfo {author} {\bibfnamefont {G.~V.}\
  \bibnamefont {Gierke}}, \bibinfo {author} {\bibfnamefont {E.}~\bibnamefont
  {Glock}}, \bibinfo {author} {\bibfnamefont {O.}~\bibnamefont {Gruber}},
  \bibinfo {author} {\bibfnamefont {G.}~\bibnamefont {Haas}}, \bibinfo {author}
  {\bibfnamefont {M.}~\bibnamefont {Hesse}}, \bibinfo {author} {\bibfnamefont
  {G.}~\bibnamefont {Janeschitz}}, \bibinfo {author} {\bibfnamefont
  {F.}~\bibnamefont {Karger}}, \bibinfo {author} {\bibfnamefont
  {S.}~\bibnamefont {Kissel}}, \bibinfo {author} {\bibfnamefont
  {O.}~\bibnamefont {K1qber}}, \bibinfo {author} {\bibfnamefont
  {G.}~\bibnamefont {Lisitano}}, \bibinfo {author} {\bibfnamefont {H.~M.}\
  \bibnamefont {Mayer}}, \bibinfo {author} {\bibfnamefont {D.}~\bibnamefont
  {Meisel}}, \bibinfo {author} {\bibfnamefont {V.}~\bibnamefont {Mertens}},
  \bibinfo {author} {\bibfnamefont {H.}~\bibnamefont {Murrnann}}, \bibinfo
  {author} {\bibfnamefont {W.}~\bibnamefont {Poschenrieder}}, \bibinfo {author}
  {\bibfnamefont {H.}~\bibnamefont {Rapp}}, \bibinfo {author} {\bibfnamefont
  {H.}~\bibnamefont {Rohr}}, \bibinfo {author} {\bibfnamefont {F.}~\bibnamefont
  {Ryter}}, \bibinfo {author} {\bibfnamefont {F.}~\bibnamefont {Schneider}},
  \bibinfo {author} {\bibfnamefont {G.}~\bibnamefont {Siller}}, \bibinfo
  {author} {\bibfnamefont {P.}~\bibnamefont {Smeulders}}, \bibinfo {author}
  {\bibfnamefont {F.}~\bibnamefont {Soldner}}, \bibinfo {author} {\bibfnamefont
  {E.}~\bibnamefont {Speth}}, \bibinfo {author} {\bibfnamefont {K.-H.}\
  \bibnamefont {Steuer}}, \bibinfo {author} {\bibfnamefont {Z.}~\bibnamefont
  {Szymanski}},\ and\ \bibinfo {author} {\bibfnamefont {O.}~\bibnamefont
  {Vollmer}},\ }\href {https://doi.org/10.1103/PhysRevLett.53.1453} {\bibfield
  {journal} {\bibinfo  {journal} {Physical Review Letters}\ }\textbf {\bibinfo
  {volume} {53}},\ \bibinfo {pages} {1453} (\bibinfo {year}
  {1984})}\BibitemShut {NoStop}%
\bibitem [{\citenamefont {Burrell}(1997)}]{Burrell1997}%
  \BibitemOpen
  \bibfield  {author} {\bibinfo {author} {\bibfnamefont {K.~H.}\ \bibnamefont
  {Burrell}},\ }\href {https://doi.org/10.1063/1.872367} {\bibfield  {journal}
  {\bibinfo  {journal} {Physics of Plasmas}\ }\textbf {\bibinfo {volume} {4}},\
  \bibinfo {pages} {1499} (\bibinfo {year} {1997})}\BibitemShut {NoStop}%
\bibitem [{\citenamefont {Snyder}\ \emph {et~al.}(2002)\citenamefont {Snyder},
  \citenamefont {Wilson}, \citenamefont {Ferron}, \citenamefont {Lao},
  \citenamefont {Leonard}, \citenamefont {Osborne}, \citenamefont {Turnbull},
  \citenamefont {Mossessian}, \citenamefont {Murakami},\ and\ \citenamefont
  {Xu}}]{Snyder2002}%
  \BibitemOpen
  \bibfield  {author} {\bibinfo {author} {\bibfnamefont {P.~B.}\ \bibnamefont
  {Snyder}}, \bibinfo {author} {\bibfnamefont {H.~R.}\ \bibnamefont {Wilson}},
  \bibinfo {author} {\bibfnamefont {J.~R.}\ \bibnamefont {Ferron}}, \bibinfo
  {author} {\bibfnamefont {L.~L.}\ \bibnamefont {Lao}}, \bibinfo {author}
  {\bibfnamefont {A.~W.}\ \bibnamefont {Leonard}}, \bibinfo {author}
  {\bibfnamefont {T.~H.}\ \bibnamefont {Osborne}}, \bibinfo {author}
  {\bibfnamefont {A.~D.}\ \bibnamefont {Turnbull}}, \bibinfo {author}
  {\bibfnamefont {D.}~\bibnamefont {Mossessian}}, \bibinfo {author}
  {\bibfnamefont {M.}~\bibnamefont {Murakami}},\ and\ \bibinfo {author}
  {\bibfnamefont {X.~Q.}\ \bibnamefont {Xu}},\ }\href
  {https://doi.org/10.1063/1.1449463} {\bibfield  {journal} {\bibinfo
  {journal} {Physics of Plasmas}\ }\textbf {\bibinfo {volume} {9}},\ \bibinfo
  {pages} {2037} (\bibinfo {year} {2002})},\ \Eprint
  {https://arxiv.org/abs/https://doi.org/10.1063/1.1449463}
  {https://doi.org/10.1063/1.1449463} \BibitemShut {NoStop}%
\bibitem [{\citenamefont {{ITER Physics Basis Editors}}(1999)}]{ITER1999}%
  \BibitemOpen
  \bibfield  {author} {\bibinfo {author} {\bibnamefont {{ITER Physics Basis
  Editors}}},\ }\href {https://doi.org/10.1088/0029-5515/39/12/301} {\bibfield
  {journal} {\bibinfo  {journal} {Nucl. Fusion}\ }\textbf {\bibinfo {volume}
  {39}},\ \bibinfo {pages} {2137} (\bibinfo {year} {1999})}\BibitemShut
  {NoStop}%
\bibitem [{\citenamefont {Creely}\ \emph {et~al.}(2020)\citenamefont {Creely},
  \citenamefont {Greenwald}, \citenamefont {Ballinger}, \citenamefont
  {Brunner}, \citenamefont {Canik}, \citenamefont {Doody}, \citenamefont
  {Fülöp}, \citenamefont {Garnier}, \citenamefont {Granetz}, \citenamefont
  {Gray}, \citenamefont {Holland}, \citenamefont {Howard}, \citenamefont
  {Hughes}, \citenamefont {Irby}, \citenamefont {Izzo}, \citenamefont {Kramer},
  \citenamefont {Kuang}, \citenamefont {LaBombard}, \citenamefont {Lin},
  \citenamefont {Lipschultz}, \citenamefont {Logan}, \citenamefont {Lore},
  \citenamefont {Marmar}, \citenamefont {Montes}, \citenamefont {Mumgaard},
  \citenamefont {Paz-Soldan}, \citenamefont {Rea}, \citenamefont {Reinke},
  \citenamefont {Rodriguez-Fernandez}, \citenamefont {Särkimäki},
  \citenamefont {Sciortino}, \citenamefont {Scott}, \citenamefont {Snicker},
  \citenamefont {Snyder}, \citenamefont {Sorbom}, \citenamefont {Sweeney},
  \citenamefont {Tinguely}, \citenamefont {Tolman}, \citenamefont {Umansky},
  \citenamefont {Vallhagen}, \citenamefont {Varje}, \citenamefont {Whyte},
  \citenamefont {Wright}, \citenamefont {Wukitch},\ and\ \citenamefont
  {Zhu}}]{Creely2020}%
  \BibitemOpen
  \bibfield  {author} {\bibinfo {author} {\bibfnamefont {A.~J.}\ \bibnamefont
  {Creely}}, \bibinfo {author} {\bibfnamefont {M.~J.}\ \bibnamefont
  {Greenwald}}, \bibinfo {author} {\bibfnamefont {S.~B.}\ \bibnamefont
  {Ballinger}}, \bibinfo {author} {\bibfnamefont {D.}~\bibnamefont {Brunner}},
  \bibinfo {author} {\bibfnamefont {J.}~\bibnamefont {Canik}}, \bibinfo
  {author} {\bibfnamefont {J.}~\bibnamefont {Doody}}, \bibinfo {author}
  {\bibfnamefont {T.}~\bibnamefont {Fülöp}}, \bibinfo {author} {\bibfnamefont
  {D.~T.}\ \bibnamefont {Garnier}}, \bibinfo {author} {\bibfnamefont
  {R.}~\bibnamefont {Granetz}}, \bibinfo {author} {\bibfnamefont {T.~K.}\
  \bibnamefont {Gray}}, \bibinfo {author} {\bibfnamefont {C.}~\bibnamefont
  {Holland}}, \bibinfo {author} {\bibfnamefont {N.~T.}\ \bibnamefont {Howard}},
  \bibinfo {author} {\bibfnamefont {J.~W.}\ \bibnamefont {Hughes}}, \bibinfo
  {author} {\bibfnamefont {J.~H.}\ \bibnamefont {Irby}}, \bibinfo {author}
  {\bibfnamefont {V.~A.}\ \bibnamefont {Izzo}}, \bibinfo {author}
  {\bibfnamefont {G.~J.}\ \bibnamefont {Kramer}}, \bibinfo {author}
  {\bibfnamefont {A.~Q.}\ \bibnamefont {Kuang}}, \bibinfo {author}
  {\bibfnamefont {B.}~\bibnamefont {LaBombard}}, \bibinfo {author}
  {\bibfnamefont {Y.}~\bibnamefont {Lin}}, \bibinfo {author} {\bibfnamefont
  {B.}~\bibnamefont {Lipschultz}}, \bibinfo {author} {\bibfnamefont {N.~C.}\
  \bibnamefont {Logan}}, \bibinfo {author} {\bibfnamefont {J.~D.}\ \bibnamefont
  {Lore}}, \bibinfo {author} {\bibfnamefont {E.~S.}\ \bibnamefont {Marmar}},
  \bibinfo {author} {\bibfnamefont {K.}~\bibnamefont {Montes}}, \bibinfo
  {author} {\bibfnamefont {R.~T.}\ \bibnamefont {Mumgaard}}, \bibinfo {author}
  {\bibfnamefont {C.}~\bibnamefont {Paz-Soldan}}, \bibinfo {author}
  {\bibfnamefont {C.}~\bibnamefont {Rea}}, \bibinfo {author} {\bibfnamefont
  {M.~L.}\ \bibnamefont {Reinke}}, \bibinfo {author} {\bibfnamefont
  {P.}~\bibnamefont {Rodriguez-Fernandez}}, \bibinfo {author} {\bibfnamefont
  {K.}~\bibnamefont {Särkimäki}}, \bibinfo {author} {\bibfnamefont
  {F.}~\bibnamefont {Sciortino}}, \bibinfo {author} {\bibfnamefont {S.~D.}\
  \bibnamefont {Scott}}, \bibinfo {author} {\bibfnamefont {A.}~\bibnamefont
  {Snicker}}, \bibinfo {author} {\bibfnamefont {P.~B.}\ \bibnamefont {Snyder}},
  \bibinfo {author} {\bibfnamefont {B.~N.}\ \bibnamefont {Sorbom}}, \bibinfo
  {author} {\bibfnamefont {R.}~\bibnamefont {Sweeney}}, \bibinfo {author}
  {\bibfnamefont {R.~A.}\ \bibnamefont {Tinguely}}, \bibinfo {author}
  {\bibfnamefont {E.~A.}\ \bibnamefont {Tolman}}, \bibinfo {author}
  {\bibfnamefont {M.}~\bibnamefont {Umansky}}, \bibinfo {author} {\bibfnamefont
  {O.}~\bibnamefont {Vallhagen}}, \bibinfo {author} {\bibfnamefont
  {J.}~\bibnamefont {Varje}}, \bibinfo {author} {\bibfnamefont {D.~G.}\
  \bibnamefont {Whyte}}, \bibinfo {author} {\bibfnamefont {J.~C.}\ \bibnamefont
  {Wright}}, \bibinfo {author} {\bibfnamefont {S.~J.}\ \bibnamefont
  {Wukitch}},\ and\ \bibinfo {author} {\bibfnamefont {J.}~\bibnamefont {Zhu}},\
  }\href {https://doi.org/10.1017/S0022377820001257} {\bibfield  {journal}
  {\bibinfo  {journal} {Journal of Plasma Physics}\ }\textbf {\bibinfo {volume}
  {86}},\ \bibinfo {pages} {865860502} (\bibinfo {year} {2020})}\BibitemShut
  {NoStop}%
\bibitem [{\citenamefont {Nelson}\ \emph {et~al.}(2020)\citenamefont {Nelson},
  \citenamefont {Laggner}, \citenamefont {Groebner}, \citenamefont {Grierson},
  \citenamefont {Izacard}, \citenamefont {Eldon}, \citenamefont {Shafer},
  \citenamefont {Leonard}, \citenamefont {Shiraki}, \citenamefont {Sontag},\
  and\ \citenamefont {Kolemen}}]{Nelson2020}%
  \BibitemOpen
  \bibfield  {author} {\bibinfo {author} {\bibfnamefont {A.~O.}\ \bibnamefont
  {Nelson}}, \bibinfo {author} {\bibfnamefont {F.~M.}\ \bibnamefont {Laggner}},
  \bibinfo {author} {\bibfnamefont {R.~J.}\ \bibnamefont {Groebner}}, \bibinfo
  {author} {\bibfnamefont {B.~A.}\ \bibnamefont {Grierson}}, \bibinfo {author}
  {\bibfnamefont {O.}~\bibnamefont {Izacard}}, \bibinfo {author} {\bibfnamefont
  {D.}~\bibnamefont {Eldon}}, \bibinfo {author} {\bibfnamefont
  {M.}~\bibnamefont {Shafer}}, \bibinfo {author} {\bibfnamefont {A.~W.}\
  \bibnamefont {Leonard}}, \bibinfo {author} {\bibfnamefont {D.}~\bibnamefont
  {Shiraki}}, \bibinfo {author} {\bibfnamefont {A.~C.}\ \bibnamefont
  {Sontag}},\ and\ \bibinfo {author} {\bibfnamefont {E.}~\bibnamefont
  {Kolemen}},\ }\href {https://doi.org/10.1088/1741-4326/ab5e65} {\bibfield
  {journal} {\bibinfo  {journal} {Nuclear Fusion}\ }\textbf {\bibinfo {volume}
  {60}},\ \bibinfo {pages} {046003} (\bibinfo {year} {2020})}\BibitemShut
  {NoStop}%
\bibitem [{\citenamefont {Leonard}(2014)}]{Leonard2014}%
  \BibitemOpen
  \bibfield  {author} {\bibinfo {author} {\bibfnamefont {A.~W.}\ \bibnamefont
  {Leonard}},\ }\href {https://doi.org/10.1063/1.4918359} {\bibfield  {journal}
  {\bibinfo  {journal} {Physics of Plasmas}\ }\textbf {\bibinfo {volume}
  {21}},\ \bibinfo {pages} {090501} (\bibinfo {year} {2014})}\BibitemShut
  {NoStop}%
\bibitem [{\citenamefont {Gunn}\ \emph {et~al.}(2017)\citenamefont {Gunn},
  \citenamefont {Carpentier-Chouchana}, \citenamefont {Escourbiac},
  \citenamefont {Hirai}, \citenamefont {Panayotis}, \citenamefont {Pitts},
  \citenamefont {Corre}, \citenamefont {Dejarnac}, \citenamefont {Firdaouss},
  \citenamefont {Kočan}, \citenamefont {Komm}, \citenamefont {Kukushkin},
  \citenamefont {Languille}, \citenamefont {Missirlian}, \citenamefont {Zhao},\
  and\ \citenamefont {Zhong}}]{Gunn2017}%
  \BibitemOpen
  \bibfield  {author} {\bibinfo {author} {\bibfnamefont {J.~P.}\ \bibnamefont
  {Gunn}}, \bibinfo {author} {\bibfnamefont {S.}~\bibnamefont
  {Carpentier-Chouchana}}, \bibinfo {author} {\bibfnamefont {F.}~\bibnamefont
  {Escourbiac}}, \bibinfo {author} {\bibfnamefont {T.}~\bibnamefont {Hirai}},
  \bibinfo {author} {\bibfnamefont {S.}~\bibnamefont {Panayotis}}, \bibinfo
  {author} {\bibfnamefont {R.~A.}\ \bibnamefont {Pitts}}, \bibinfo {author}
  {\bibfnamefont {Y.}~\bibnamefont {Corre}}, \bibinfo {author} {\bibfnamefont
  {R.}~\bibnamefont {Dejarnac}}, \bibinfo {author} {\bibfnamefont
  {M.}~\bibnamefont {Firdaouss}}, \bibinfo {author} {\bibfnamefont
  {M.}~\bibnamefont {Kočan}}, \bibinfo {author} {\bibfnamefont
  {M.}~\bibnamefont {Komm}}, \bibinfo {author} {\bibfnamefont {A.}~\bibnamefont
  {Kukushkin}}, \bibinfo {author} {\bibfnamefont {P.}~\bibnamefont
  {Languille}}, \bibinfo {author} {\bibfnamefont {M.}~\bibnamefont
  {Missirlian}}, \bibinfo {author} {\bibfnamefont {W.}~\bibnamefont {Zhao}},\
  and\ \bibinfo {author} {\bibfnamefont {G.}~\bibnamefont {Zhong}},\ }\href
  {https://doi.org/10.1088/1741-4326/aa5e2a} {\bibfield  {journal} {\bibinfo
  {journal} {Nuclear Fusion}\ }\textbf {\bibinfo {volume} {57}},\ \bibinfo
  {pages} {046025} (\bibinfo {year} {2017})}\BibitemShut {NoStop}%
\bibitem [{\citenamefont {Austin}\ \emph {et~al.}(2019)\citenamefont {Austin},
  \citenamefont {Marinoni}, \citenamefont {Walker}, \citenamefont {Brookman},
  \citenamefont {Degrassie}, \citenamefont {Hyatt}, \citenamefont {McKee},
  \citenamefont {Petty}, \citenamefont {Rhodes}, \citenamefont {Smith},
  \citenamefont {Sung}, \citenamefont {Thome},\ and\ \citenamefont
  {Turnbull}}]{Austin2019}%
  \BibitemOpen
  \bibfield  {author} {\bibinfo {author} {\bibfnamefont {M.~E.}\ \bibnamefont
  {Austin}}, \bibinfo {author} {\bibfnamefont {A.}~\bibnamefont {Marinoni}},
  \bibinfo {author} {\bibfnamefont {M.~L.}\ \bibnamefont {Walker}}, \bibinfo
  {author} {\bibfnamefont {M.~W.}\ \bibnamefont {Brookman}}, \bibinfo {author}
  {\bibfnamefont {J.~S.}\ \bibnamefont {Degrassie}}, \bibinfo {author}
  {\bibfnamefont {A.~W.}\ \bibnamefont {Hyatt}}, \bibinfo {author}
  {\bibfnamefont {G.~R.}\ \bibnamefont {McKee}}, \bibinfo {author}
  {\bibfnamefont {C.~C.}\ \bibnamefont {Petty}}, \bibinfo {author}
  {\bibfnamefont {T.~L.}\ \bibnamefont {Rhodes}}, \bibinfo {author}
  {\bibfnamefont {S.~P.}\ \bibnamefont {Smith}}, \bibinfo {author}
  {\bibfnamefont {C.}~\bibnamefont {Sung}}, \bibinfo {author} {\bibfnamefont
  {K.~E.}\ \bibnamefont {Thome}},\ and\ \bibinfo {author} {\bibfnamefont
  {A.~D.}\ \bibnamefont {Turnbull}},\ }\href
  {https://doi.org/10.1103/PhysRevLett.122.115001} {\bibfield  {journal}
  {\bibinfo  {journal} {Physical Review Letters}\ }\textbf {\bibinfo {volume}
  {122}},\ \bibinfo {pages} {115001} (\bibinfo {year} {2019})}\BibitemShut
  {NoStop}%
\bibitem [{\citenamefont {Kikuchi}\ \emph {et~al.}(2014)\citenamefont
  {Kikuchi}, \citenamefont {Takizuka},\ and\ \citenamefont
  {Furukawa}}]{Kikuchi}%
  \BibitemOpen
  \bibfield  {author} {\bibinfo {author} {\bibfnamefont {M.}~\bibnamefont
  {Kikuchi}}, \bibinfo {author} {\bibfnamefont {T.}~\bibnamefont {Takizuka}},\
  and\ \bibinfo {author} {\bibfnamefont {M.}~\bibnamefont {Furukawa}},\ }\href
  {https://doi.org/10.7566/JPSCP.1.015014} {\bibfield  {journal} {\bibinfo
  {journal} {JPS Conf. Proc.}\ }\textbf {\bibinfo {volume} {1}},\ \bibinfo
  {pages} {015014} (\bibinfo {year} {2014})}\BibitemShut {NoStop}%
\bibitem [{\citenamefont {Kikuchi}\ \emph {et~al.}(2019)\citenamefont
  {Kikuchi}, \citenamefont {Takizuka}, \citenamefont {Medvedev}, \citenamefont
  {Ando}, \citenamefont {Chen}, \citenamefont {Li}, \citenamefont {Austin},
  \citenamefont {Sauter}, \citenamefont {Villard}, \citenamefont {Merle},
  \citenamefont {Fontana}, \citenamefont {Kishimoto},\ and\ \citenamefont
  {Imadera}}]{Kikuchi2019}%
  \BibitemOpen
  \bibfield  {author} {\bibinfo {author} {\bibfnamefont {M.}~\bibnamefont
  {Kikuchi}}, \bibinfo {author} {\bibfnamefont {T.}~\bibnamefont {Takizuka}},
  \bibinfo {author} {\bibfnamefont {S.}~\bibnamefont {Medvedev}}, \bibinfo
  {author} {\bibfnamefont {T.}~\bibnamefont {Ando}}, \bibinfo {author}
  {\bibfnamefont {D.}~\bibnamefont {Chen}}, \bibinfo {author} {\bibfnamefont
  {J.~X.}\ \bibnamefont {Li}}, \bibinfo {author} {\bibfnamefont
  {M.}~\bibnamefont {Austin}}, \bibinfo {author} {\bibfnamefont
  {O.}~\bibnamefont {Sauter}}, \bibinfo {author} {\bibfnamefont
  {L.}~\bibnamefont {Villard}}, \bibinfo {author} {\bibfnamefont
  {A.}~\bibnamefont {Merle}}, \bibinfo {author} {\bibfnamefont
  {M.}~\bibnamefont {Fontana}}, \bibinfo {author} {\bibfnamefont
  {Y.}~\bibnamefont {Kishimoto}},\ and\ \bibinfo {author} {\bibfnamefont
  {K.}~\bibnamefont {Imadera}},\ }\href
  {https://doi.org/10.1088/1741-4326/ab076d} {\bibfield  {journal} {\bibinfo
  {journal} {Nuclear Fusion}\ }\textbf {\bibinfo {volume} {59}},\ \bibinfo
  {pages} {056017} (\bibinfo {year} {2019})}\BibitemShut {NoStop}%
\bibitem [{\citenamefont {Medvedev}\ \emph {et~al.}(2016)\citenamefont
  {Medvedev}, \citenamefont {Kikuchi}, \citenamefont {Takizuka}, \citenamefont
  {Ivanov}, \citenamefont {Martynov}, \citenamefont {Poshekhonov},
  \citenamefont {Merle}, \citenamefont {Sauter}, \citenamefont {Villard},
  \citenamefont {Chen}, \citenamefont {Jiang}, \citenamefont {Li},
  \citenamefont {Zheng},\ and\ \citenamefont {Ando}}]{medvedev_single_2016}%
  \BibitemOpen
  \bibfield  {author} {\bibinfo {author} {\bibfnamefont {S.~Y.}\ \bibnamefont
  {Medvedev}}, \bibinfo {author} {\bibfnamefont {M.}~\bibnamefont {Kikuchi}},
  \bibinfo {author} {\bibfnamefont {T.}~\bibnamefont {Takizuka}}, \bibinfo
  {author} {\bibfnamefont {A.~A.}\ \bibnamefont {Ivanov}}, \bibinfo {author}
  {\bibfnamefont {A.~A.}\ \bibnamefont {Martynov}}, \bibinfo {author}
  {\bibfnamefont {Y.~Y.}\ \bibnamefont {Poshekhonov}}, \bibinfo {author}
  {\bibfnamefont {A.}~\bibnamefont {Merle}}, \bibinfo {author} {\bibfnamefont
  {O.}~\bibnamefont {Sauter}}, \bibinfo {author} {\bibfnamefont
  {L.}~\bibnamefont {Villard}}, \bibinfo {author} {\bibfnamefont
  {D.}~\bibnamefont {Chen}}, \bibinfo {author} {\bibfnamefont {J.}~\bibnamefont
  {Jiang}}, \bibinfo {author} {\bibfnamefont {J.~X.}\ \bibnamefont {Li}},
  \bibinfo {author} {\bibfnamefont {J.}~\bibnamefont {Zheng}},\ and\ \bibinfo
  {author} {\bibfnamefont {T.}~\bibnamefont {Ando}},\ }\href
  {https://conferences.iaea.org/event/98/contributions/12077/attachments/6334/7718/Medvedev-IAEA-2016-ICC-P3-47.pdf}
  {\bibfield  {journal} {\bibinfo  {journal} {FEC}\ ,\ \bibinfo {pages} {FEC}}
  (\bibinfo {year} {2016})}\BibitemShut {NoStop}%
\bibitem [{\citenamefont {Coda}\ \emph {et~al.}(2022)\citenamefont {Coda},
  \citenamefont {Merle}, \citenamefont {Sauter}, \citenamefont {Porte},
  \citenamefont {Bagnato}, \citenamefont {Boedo}, \citenamefont {Bolzonella},
  \citenamefont {Février}, \citenamefont {Labit}, \citenamefont {Marinoni},
  \citenamefont {Pau}, \citenamefont {Pigatto}, \citenamefont {Sheikh},
  \citenamefont {Tsui}, \citenamefont {Vallar},\ and\ \citenamefont
  {Vu}}]{coda_enhanced_2022}%
  \BibitemOpen
  \bibfield  {author} {\bibinfo {author} {\bibfnamefont {S.}~\bibnamefont
  {Coda}}, \bibinfo {author} {\bibfnamefont {A.}~\bibnamefont {Merle}},
  \bibinfo {author} {\bibfnamefont {O.}~\bibnamefont {Sauter}}, \bibinfo
  {author} {\bibfnamefont {L.}~\bibnamefont {Porte}}, \bibinfo {author}
  {\bibfnamefont {F.}~\bibnamefont {Bagnato}}, \bibinfo {author} {\bibfnamefont
  {J.}~\bibnamefont {Boedo}}, \bibinfo {author} {\bibfnamefont
  {T.}~\bibnamefont {Bolzonella}}, \bibinfo {author} {\bibfnamefont
  {O.}~\bibnamefont {Février}}, \bibinfo {author} {\bibfnamefont
  {B.}~\bibnamefont {Labit}}, \bibinfo {author} {\bibfnamefont
  {A.}~\bibnamefont {Marinoni}}, \bibinfo {author} {\bibfnamefont
  {A.}~\bibnamefont {Pau}}, \bibinfo {author} {\bibfnamefont {L.}~\bibnamefont
  {Pigatto}}, \bibinfo {author} {\bibfnamefont {U.}~\bibnamefont {Sheikh}},
  \bibinfo {author} {\bibfnamefont {C.}~\bibnamefont {Tsui}}, \bibinfo {author}
  {\bibfnamefont {M.}~\bibnamefont {Vallar}},\ and\ \bibinfo {author}
  {\bibfnamefont {T.}~\bibnamefont {Vu}},\ }\href
  {https://doi.org/10.1088/1361-6587/ac3fec} {\bibfield  {journal} {\bibinfo
  {journal} {Plasma Physics and Controlled Fusion}\ }\textbf {\bibinfo {volume}
  {64}},\ \bibinfo {pages} {014004} (\bibinfo {year} {2022})}\BibitemShut
  {NoStop}%
\bibitem [{\citenamefont {Medvedev}\ \emph {et~al.}(2008)\citenamefont
  {Medvedev}, \citenamefont {Ivanov}, \citenamefont {Martynov}, \citenamefont
  {Poshekhonov}, \citenamefont {Behn}, \citenamefont {Martin}, \citenamefont
  {Pochelon}, \citenamefont {Sauter},\ and\ \citenamefont
  {Villard}}]{medvedev_beta_2008}%
  \BibitemOpen
  \bibfield  {author} {\bibinfo {author} {\bibfnamefont {S.~Y.}\ \bibnamefont
  {Medvedev}}, \bibinfo {author} {\bibfnamefont {A.~A.}\ \bibnamefont
  {Ivanov}}, \bibinfo {author} {\bibfnamefont {A.~A.}\ \bibnamefont
  {Martynov}}, \bibinfo {author} {\bibfnamefont {Y.~Y.}\ \bibnamefont
  {Poshekhonov}}, \bibinfo {author} {\bibfnamefont {R.}~\bibnamefont {Behn}},
  \bibinfo {author} {\bibfnamefont {Y.~R.}\ \bibnamefont {Martin}}, \bibinfo
  {author} {\bibfnamefont {A.}~\bibnamefont {Pochelon}}, \bibinfo {author}
  {\bibfnamefont {O.}~\bibnamefont {Sauter}},\ and\ \bibinfo {author}
  {\bibfnamefont {L.}~\bibnamefont {Villard}},\ }\href
  {https://www.researchgate.net/profile/Antoine-Pochelon/publication/37468729_Beta_limits_and_Edge_Stability_for_Negative_Triangularity_Plasma_in_TCV_Tokamak/links/5fd9af1645851553a0bd74f0/Beta-limits-and-Edge-Stability-for-Negative-Triangularity-Plasma-in-TCV-Tokamak.pdf}
  {\bibfield  {journal} {\bibinfo  {journal} {35th EPS Conference on Plasma
  Physics}\ }\textbf {\bibinfo {volume} {32D}},\ \bibinfo {pages} {1.072}
  (\bibinfo {year} {2008})}\BibitemShut {NoStop}%
\bibitem [{\citenamefont {Medvedev}\ \emph {et~al.}(2015)\citenamefont
  {Medvedev}, \citenamefont {Kikuchi}, \citenamefont {Villard}, \citenamefont
  {Takizuka}, \citenamefont {Diamond}, \citenamefont {Zushi}, \citenamefont
  {Nagasaki}, \citenamefont {Duan}, \citenamefont {Wu}, \citenamefont {Ivanov},
  \citenamefont {Martynov}, \citenamefont {Poshekhonov}, \citenamefont
  {Fasoli},\ and\ \citenamefont {Sauter}}]{Medvedev2015}%
  \BibitemOpen
  \bibfield  {author} {\bibinfo {author} {\bibfnamefont {S.~Y.}\ \bibnamefont
  {Medvedev}}, \bibinfo {author} {\bibfnamefont {M.}~\bibnamefont {Kikuchi}},
  \bibinfo {author} {\bibfnamefont {L.}~\bibnamefont {Villard}}, \bibinfo
  {author} {\bibfnamefont {T.}~\bibnamefont {Takizuka}}, \bibinfo {author}
  {\bibfnamefont {P.}~\bibnamefont {Diamond}}, \bibinfo {author} {\bibfnamefont
  {H.}~\bibnamefont {Zushi}}, \bibinfo {author} {\bibfnamefont
  {K.}~\bibnamefont {Nagasaki}}, \bibinfo {author} {\bibfnamefont
  {X.}~\bibnamefont {Duan}}, \bibinfo {author} {\bibfnamefont {Y.}~\bibnamefont
  {Wu}}, \bibinfo {author} {\bibfnamefont {A.~A.}\ \bibnamefont {Ivanov}},
  \bibinfo {author} {\bibfnamefont {A.~A.}\ \bibnamefont {Martynov}}, \bibinfo
  {author} {\bibfnamefont {Y.~Y.}\ \bibnamefont {Poshekhonov}}, \bibinfo
  {author} {\bibfnamefont {A.}~\bibnamefont {Fasoli}},\ and\ \bibinfo {author}
  {\bibfnamefont {O.}~\bibnamefont {Sauter}},\ }\href
  {https://doi.org/10.1088/0029-5515/55/6/063013} {\bibfield  {journal}
  {\bibinfo  {journal} {Nuclear Fusion}\ }\textbf {\bibinfo {volume} {55}},\
  \bibinfo {pages} {063013} (\bibinfo {year} {2015})}\BibitemShut {NoStop}%
\bibitem [{\citenamefont {Merle}\ \emph {et~al.}(2017)\citenamefont {Merle},
  \citenamefont {Sauter},\ and\ \citenamefont
  {Yu~Medvedev}}]{merle_pedestal_2017}%
  \BibitemOpen
  \bibfield  {author} {\bibinfo {author} {\bibfnamefont {A.}~\bibnamefont
  {Merle}}, \bibinfo {author} {\bibfnamefont {O.}~\bibnamefont {Sauter}},\ and\
  \bibinfo {author} {\bibfnamefont {S.}~\bibnamefont {Yu~Medvedev}},\
  }\bibfield  {journal} {\bibinfo  {journal} {Plasma Physics and Controlled
  Fusion}\ }\textbf {\bibinfo {volume} {59}},\ \href
  {https://doi.org/10.1088/1361-6587/aa7ac0} {10.1088/1361-6587/aa7ac0}
  (\bibinfo {year} {2017})\BibitemShut {NoStop}%
\bibitem [{\citenamefont {Saarelma}\ \emph {et~al.}(2021)\citenamefont
  {Saarelma}, \citenamefont {Austin}, \citenamefont {Knolker}, \citenamefont
  {Marinoni}, \citenamefont {Paz-Soldan}, \citenamefont {Schmitz},\ and\
  \citenamefont {Snyder}}]{saarelma_ballooning_2021}%
  \BibitemOpen
  \bibfield  {author} {\bibinfo {author} {\bibfnamefont {S.}~\bibnamefont
  {Saarelma}}, \bibinfo {author} {\bibfnamefont {M.~E.}\ \bibnamefont
  {Austin}}, \bibinfo {author} {\bibfnamefont {M.}~\bibnamefont {Knolker}},
  \bibinfo {author} {\bibfnamefont {A.}~\bibnamefont {Marinoni}}, \bibinfo
  {author} {\bibfnamefont {C.}~\bibnamefont {Paz-Soldan}}, \bibinfo {author}
  {\bibfnamefont {L.}~\bibnamefont {Schmitz}},\ and\ \bibinfo {author}
  {\bibfnamefont {P.~B.}\ \bibnamefont {Snyder}},\ }\bibfield  {journal}
  {\bibinfo  {journal} {Plasma Physics and Controlled Fusion}\ }\textbf
  {\bibinfo {volume} {63}},\ \href {https://doi.org/10.1088/1361}
  {10.1088/1361} (\bibinfo {year} {2021})\BibitemShut {NoStop}%
\bibitem [{\citenamefont {Marinoni}\ \emph {et~al.}(2021)\citenamefont
  {Marinoni}, \citenamefont {Sauter},\ and\ \citenamefont
  {Coda}}]{marinoni_brief_2021}%
  \BibitemOpen
  \bibfield  {author} {\bibinfo {author} {\bibfnamefont {A.}~\bibnamefont
  {Marinoni}}, \bibinfo {author} {\bibfnamefont {O.}~\bibnamefont {Sauter}},\
  and\ \bibinfo {author} {\bibfnamefont {S.}~\bibnamefont {Coda}},\ }\href
  {https://doi.org/10.1007/s41614-021-00054-0} {\bibfield  {journal} {\bibinfo
  {journal} {Reviews of Modern Plasma Physics}\ }\textbf {\bibinfo {volume}
  {5}},\ \bibinfo {pages} {6} (\bibinfo {year} {2021})}\BibitemShut {NoStop}%
\bibitem [{\citenamefont {Freidberg}(2014)}]{Freidberg2014}%
  \BibitemOpen
  \bibfield  {author} {\bibinfo {author} {\bibfnamefont {J.~P.}\ \bibnamefont
  {Freidberg}},\ }\href@noop {} {\emph {\bibinfo {title} {Ideal {MHD}}}}\
  (\bibinfo {year} {2014})\BibitemShut {NoStop}%
\bibitem [{\citenamefont {Mercier}(1960)}]{mercier_necessary_1960}%
  \BibitemOpen
  \bibfield  {author} {\bibinfo {author} {\bibfnamefont {C.}~\bibnamefont
  {Mercier}},\ }\href {https://doi.org/10.1088/0029-5515/1/1/004} {\bibfield
  {journal} {\bibinfo  {journal} {Nuclear Fusion}\ }\textbf {\bibinfo {volume}
  {1}},\ \bibinfo {pages} {47} (\bibinfo {year} {1960})}\BibitemShut {NoStop}%
\bibitem [{\citenamefont {Connor}\ \emph {et~al.}(1978)\citenamefont {Connor},
  \citenamefont {Hastie},\ and\ \citenamefont {Taylor}}]{connor_shear_1978}%
  \BibitemOpen
  \bibfield  {author} {\bibinfo {author} {\bibfnamefont {J.~W.}\ \bibnamefont
  {Connor}}, \bibinfo {author} {\bibfnamefont {R.~J.}\ \bibnamefont {Hastie}},\
  and\ \bibinfo {author} {\bibfnamefont {J.~B.}\ \bibnamefont {Taylor}},\
  }\href {https://doi.org/10.1103/PhysRevLett.40.396} {\bibfield  {journal}
  {\bibinfo  {journal} {Physical Review Letters}\ }\textbf {\bibinfo {volume}
  {40}},\ \bibinfo {pages} {396} (\bibinfo {year} {1978})}\BibitemShut
  {NoStop}%
\bibitem [{\citenamefont {Greene}\ and\ \citenamefont
  {Chance}(1981)}]{greene_second_1981}%
  \BibitemOpen
  \bibfield  {author} {\bibinfo {author} {\bibfnamefont {J.}~\bibnamefont
  {Greene}}\ and\ \bibinfo {author} {\bibfnamefont {M.}~\bibnamefont
  {Chance}},\ }\href {https://doi.org/10.1088/0029-5515/21/4/002} {\bibfield
  {journal} {\bibinfo  {journal} {Nuclear Fusion}\ }\textbf {\bibinfo {volume}
  {21}},\ \bibinfo {pages} {453} (\bibinfo {year} {1981})}\BibitemShut
  {NoStop}%
\bibitem [{\citenamefont {Gerver}\ \emph {et~al.}(1988)\citenamefont {Gerver},
  \citenamefont {Kesner},\ and\ \citenamefont {Ramos}}]{gerver_access_1988}%
  \BibitemOpen
  \bibfield  {author} {\bibinfo {author} {\bibfnamefont {M.~J.}\ \bibnamefont
  {Gerver}}, \bibinfo {author} {\bibfnamefont {J.}~\bibnamefont {Kesner}},\
  and\ \bibinfo {author} {\bibfnamefont {J.~J.}\ \bibnamefont {Ramos}},\ }\href
  {https://doi.org/10.1063/1.866545} {\bibfield  {journal} {\bibinfo  {journal}
  {Physics of Fluids}\ }\textbf {\bibinfo {volume} {31}},\ \bibinfo {pages}
  {2674} (\bibinfo {year} {1988})}\BibitemShut {NoStop}%
\bibitem [{\citenamefont {Wilson}\ and\ \citenamefont
  {Miller}(1999)}]{wilson_access_1999}%
  \BibitemOpen
  \bibfield  {author} {\bibinfo {author} {\bibfnamefont {H.~R.}\ \bibnamefont
  {Wilson}}\ and\ \bibinfo {author} {\bibfnamefont {R.~L.}\ \bibnamefont
  {Miller}},\ }\href {https://doi.org/10.1063/1.873326} {\bibfield  {journal}
  {\bibinfo  {journal} {Physics of Plasmas}\ }\textbf {\bibinfo {volume} {6}},\
  \bibinfo {pages} {873} (\bibinfo {year} {1999})}\BibitemShut {NoStop}%
\bibitem [{\citenamefont {Ozeki}\ \emph {et~al.}(1990)\citenamefont {Ozeki},
  \citenamefont {Chu}, \citenamefont {Lao}, \citenamefont {Taylor},
  \citenamefont {Chance}, \citenamefont {Kinoshita}, \citenamefont {Burrell},\
  and\ \citenamefont {Stambaugh}}]{ozeki_plasma_1990}%
  \BibitemOpen
  \bibfield  {author} {\bibinfo {author} {\bibfnamefont {T.}~\bibnamefont
  {Ozeki}}, \bibinfo {author} {\bibfnamefont {M.}~\bibnamefont {Chu}}, \bibinfo
  {author} {\bibfnamefont {L.}~\bibnamefont {Lao}}, \bibinfo {author}
  {\bibfnamefont {T.}~\bibnamefont {Taylor}}, \bibinfo {author} {\bibfnamefont
  {M.}~\bibnamefont {Chance}}, \bibinfo {author} {\bibfnamefont
  {S.}~\bibnamefont {Kinoshita}}, \bibinfo {author} {\bibfnamefont
  {K.}~\bibnamefont {Burrell}},\ and\ \bibinfo {author} {\bibfnamefont
  {R.}~\bibnamefont {Stambaugh}},\ }\href
  {https://doi.org/10.1088/0029-5515/30/8/003} {\bibfield  {journal} {\bibinfo
  {journal} {Nuclear Fusion}\ }\textbf {\bibinfo {volume} {30}},\ \bibinfo
  {pages} {1425} (\bibinfo {year} {1990})}\BibitemShut {NoStop}%
\bibitem [{\citenamefont {Lao}\ \emph {et~al.}(1999)\citenamefont {Lao},
  \citenamefont {Ferron}, \citenamefont {Miller}, \citenamefont {Osborne},
  \citenamefont {Chan}, \citenamefont {Groebner}, \citenamefont {Jackson},
  \citenamefont {Haye}, \citenamefont {Strait}, \citenamefont {Taylor},
  \citenamefont {Turnbull}, \citenamefont {Doyle}, \citenamefont {Lazarus},
  \citenamefont {Murakami}, \citenamefont {McKee}, \citenamefont {Rice},
  \citenamefont {Zhang},\ and\ \citenamefont {Chen}}]{lao_effects_1999}%
  \BibitemOpen
  \bibfield  {author} {\bibinfo {author} {\bibfnamefont {L.}~\bibnamefont
  {Lao}}, \bibinfo {author} {\bibfnamefont {J.}~\bibnamefont {Ferron}},
  \bibinfo {author} {\bibfnamefont {R.}~\bibnamefont {Miller}}, \bibinfo
  {author} {\bibfnamefont {T.}~\bibnamefont {Osborne}}, \bibinfo {author}
  {\bibfnamefont {V.}~\bibnamefont {Chan}}, \bibinfo {author} {\bibfnamefont
  {R.}~\bibnamefont {Groebner}}, \bibinfo {author} {\bibfnamefont
  {G.}~\bibnamefont {Jackson}}, \bibinfo {author} {\bibfnamefont {R.~L.}\
  \bibnamefont {Haye}}, \bibinfo {author} {\bibfnamefont {E.}~\bibnamefont
  {Strait}}, \bibinfo {author} {\bibfnamefont {T.}~\bibnamefont {Taylor}},
  \bibinfo {author} {\bibfnamefont {A.}~\bibnamefont {Turnbull}}, \bibinfo
  {author} {\bibfnamefont {E.}~\bibnamefont {Doyle}}, \bibinfo {author}
  {\bibfnamefont {E.}~\bibnamefont {Lazarus}}, \bibinfo {author} {\bibfnamefont
  {M.}~\bibnamefont {Murakami}}, \bibinfo {author} {\bibfnamefont
  {G.}~\bibnamefont {McKee}}, \bibinfo {author} {\bibfnamefont
  {B.}~\bibnamefont {Rice}}, \bibinfo {author} {\bibfnamefont {C.}~\bibnamefont
  {Zhang}},\ and\ \bibinfo {author} {\bibfnamefont {L.}~\bibnamefont {Chen}},\
  }\href {https://doi.org/10.1088/0029-5515/39/11Y/319} {\bibfield  {journal}
  {\bibinfo  {journal} {Nuclear Fusion}\ }\textbf {\bibinfo {volume} {39}},\
  \bibinfo {pages} {1785} (\bibinfo {year} {1999})}\BibitemShut {NoStop}%
\bibitem [{\citenamefont {Ferron}\ \emph {et~al.}(2000)\citenamefont {Ferron},
  \citenamefont {Lao}, \citenamefont {Luce}, \citenamefont {Miller},
  \citenamefont {Osborne}, \citenamefont {Rice}, \citenamefont {Strait},\ and\
  \citenamefont {Taylor}}]{ferron_modification_2000}%
  \BibitemOpen
  \bibfield  {author} {\bibinfo {author} {\bibfnamefont {J.}~\bibnamefont
  {Ferron}}, \bibinfo {author} {\bibfnamefont {L.}~\bibnamefont {Lao}},
  \bibinfo {author} {\bibfnamefont {T.}~\bibnamefont {Luce}}, \bibinfo {author}
  {\bibfnamefont {R.}~\bibnamefont {Miller}}, \bibinfo {author} {\bibfnamefont
  {T.}~\bibnamefont {Osborne}}, \bibinfo {author} {\bibfnamefont
  {B.}~\bibnamefont {Rice}}, \bibinfo {author} {\bibfnamefont {E.}~\bibnamefont
  {Strait}},\ and\ \bibinfo {author} {\bibfnamefont {T.}~\bibnamefont
  {Taylor}},\ }\href {https://doi.org/10.1088/0029-5515/40/7/310} {\bibfield
  {journal} {\bibinfo  {journal} {Nuclear Fusion}\ }\textbf {\bibinfo {volume}
  {40}},\ \bibinfo {pages} {1411} (\bibinfo {year} {2000})}\BibitemShut
  {NoStop}%
\bibitem [{\citenamefont {Korotkov}\ \emph {et~al.}(2000)\citenamefont
  {Korotkov}, \citenamefont {Summers}, \citenamefont {Hender},\ and\
  \citenamefont {Pietrzyk}}]{korotkov_edge_2000}%
  \BibitemOpen
  \bibfield  {author} {\bibinfo {author} {\bibfnamefont {A.}~\bibnamefont
  {Korotkov}}, \bibinfo {author} {\bibfnamefont {D.}~\bibnamefont {Summers}},
  \bibinfo {author} {\bibfnamefont {T.}~\bibnamefont {Hender}},\ and\ \bibinfo
  {author} {\bibfnamefont {Z.}~\bibnamefont {Pietrzyk}},\ }in\ \href
  {http://infoscience.epfl.ch/record/123952} {\emph {\bibinfo {booktitle}
  {Proc. 27th {EPS} {Conference} on {Controlled} {Fusion} and {Plasma}
  {Physics}}}},\ Vol.\ \bibinfo {volume} {24B}\ (\bibinfo {year}
  {2000})\BibitemShut {NoStop}%
\bibitem [{\citenamefont {Saibene}\ \emph {et~al.}(2002)\citenamefont
  {Saibene}, \citenamefont {Sartori}, \citenamefont {Loarte}, \citenamefont
  {Campbell}, \citenamefont {Lomas}, \citenamefont {Parail}, \citenamefont
  {Zastrow}, \citenamefont {Andrew}, \citenamefont {Sharapov}, \citenamefont
  {Korotkov}, \citenamefont {Becoulet}, \citenamefont {Huysmans}, \citenamefont
  {Koslowski}, \citenamefont {Budny}, \citenamefont {Conway}, \citenamefont
  {Stober}, \citenamefont {Suttrop}, \citenamefont {Kallenbach}, \citenamefont
  {Hellermann},\ and\ \citenamefont {Beurskens}}]{saibene_improved_2002}%
  \BibitemOpen
  \bibfield  {author} {\bibinfo {author} {\bibfnamefont {G.}~\bibnamefont
  {Saibene}}, \bibinfo {author} {\bibfnamefont {R.}~\bibnamefont {Sartori}},
  \bibinfo {author} {\bibfnamefont {A.}~\bibnamefont {Loarte}}, \bibinfo
  {author} {\bibfnamefont {D.~J.}\ \bibnamefont {Campbell}}, \bibinfo {author}
  {\bibfnamefont {P.~J.}\ \bibnamefont {Lomas}}, \bibinfo {author}
  {\bibfnamefont {V.}~\bibnamefont {Parail}}, \bibinfo {author} {\bibfnamefont
  {K.~D.}\ \bibnamefont {Zastrow}}, \bibinfo {author} {\bibfnamefont
  {Y.}~\bibnamefont {Andrew}}, \bibinfo {author} {\bibfnamefont
  {S.}~\bibnamefont {Sharapov}}, \bibinfo {author} {\bibfnamefont
  {A.}~\bibnamefont {Korotkov}}, \bibinfo {author} {\bibfnamefont
  {M.}~\bibnamefont {Becoulet}}, \bibinfo {author} {\bibfnamefont {G.~T.~A.}\
  \bibnamefont {Huysmans}}, \bibinfo {author} {\bibfnamefont {H.~R.}\
  \bibnamefont {Koslowski}}, \bibinfo {author} {\bibfnamefont {R.}~\bibnamefont
  {Budny}}, \bibinfo {author} {\bibfnamefont {G.~D.}\ \bibnamefont {Conway}},
  \bibinfo {author} {\bibfnamefont {J.}~\bibnamefont {Stober}}, \bibinfo
  {author} {\bibfnamefont {W.}~\bibnamefont {Suttrop}}, \bibinfo {author}
  {\bibfnamefont {A.}~\bibnamefont {Kallenbach}}, \bibinfo {author}
  {\bibfnamefont {M.~v.}\ \bibnamefont {Hellermann}},\ and\ \bibinfo {author}
  {\bibfnamefont {M.}~\bibnamefont {Beurskens}},\ }\href
  {https://doi.org/10.1088/0741-3335/44/9/301} {\bibfield  {journal} {\bibinfo
  {journal} {Plasma Physics and Controlled Fusion}\ }\textbf {\bibinfo {volume}
  {44}},\ \bibinfo {pages} {1769} (\bibinfo {year} {2002})}\BibitemShut
  {NoStop}%
\bibitem [{\citenamefont {Lao}\ \emph {et~al.}(2001)\citenamefont {Lao},
  \citenamefont {Kamada}, \citenamefont {Oikawa}, \citenamefont {Baylor},
  \citenamefont {Burrell}, \citenamefont {Chan}, \citenamefont {Chance},
  \citenamefont {Chu}, \citenamefont {Ferron}, \citenamefont {Fukuda},
  \citenamefont {Hatae}, \citenamefont {Isayama}, \citenamefont {Jackson},
  \citenamefont {Leonard}, \citenamefont {Makowski}, \citenamefont {Manickam},
  \citenamefont {Murakami}, \citenamefont {Okabayashi}, \citenamefont
  {Osborne}, \citenamefont {Snyder}, \citenamefont {Strait}, \citenamefont
  {Takeji}, \citenamefont {Takizuka}, \citenamefont {Taylor}, \citenamefont
  {Turnbull}, \citenamefont {Tsuchiya},\ and\ \citenamefont
  {Wade}}]{lao_dependence_2001}%
  \BibitemOpen
  \bibfield  {author} {\bibinfo {author} {\bibfnamefont {L.}~\bibnamefont
  {Lao}}, \bibinfo {author} {\bibfnamefont {Y.}~\bibnamefont {Kamada}},
  \bibinfo {author} {\bibfnamefont {T.}~\bibnamefont {Oikawa}}, \bibinfo
  {author} {\bibfnamefont {L.}~\bibnamefont {Baylor}}, \bibinfo {author}
  {\bibfnamefont {K.}~\bibnamefont {Burrell}}, \bibinfo {author} {\bibfnamefont
  {V.}~\bibnamefont {Chan}}, \bibinfo {author} {\bibfnamefont {M.}~\bibnamefont
  {Chance}}, \bibinfo {author} {\bibfnamefont {M.}~\bibnamefont {Chu}},
  \bibinfo {author} {\bibfnamefont {J.}~\bibnamefont {Ferron}}, \bibinfo
  {author} {\bibfnamefont {T.}~\bibnamefont {Fukuda}}, \bibinfo {author}
  {\bibfnamefont {T.}~\bibnamefont {Hatae}}, \bibinfo {author} {\bibfnamefont
  {A.}~\bibnamefont {Isayama}}, \bibinfo {author} {\bibfnamefont
  {G.}~\bibnamefont {Jackson}}, \bibinfo {author} {\bibfnamefont
  {A.}~\bibnamefont {Leonard}}, \bibinfo {author} {\bibfnamefont
  {M.}~\bibnamefont {Makowski}}, \bibinfo {author} {\bibfnamefont
  {J.}~\bibnamefont {Manickam}}, \bibinfo {author} {\bibfnamefont
  {M.}~\bibnamefont {Murakami}}, \bibinfo {author} {\bibfnamefont
  {M.}~\bibnamefont {Okabayashi}}, \bibinfo {author} {\bibfnamefont
  {T.}~\bibnamefont {Osborne}}, \bibinfo {author} {\bibfnamefont
  {P.}~\bibnamefont {Snyder}}, \bibinfo {author} {\bibfnamefont
  {E.}~\bibnamefont {Strait}}, \bibinfo {author} {\bibfnamefont
  {S.}~\bibnamefont {Takeji}}, \bibinfo {author} {\bibfnamefont
  {T.}~\bibnamefont {Takizuka}}, \bibinfo {author} {\bibfnamefont
  {T.}~\bibnamefont {Taylor}}, \bibinfo {author} {\bibfnamefont
  {A.}~\bibnamefont {Turnbull}}, \bibinfo {author} {\bibfnamefont
  {K.}~\bibnamefont {Tsuchiya}},\ and\ \bibinfo {author} {\bibfnamefont
  {M.}~\bibnamefont {Wade}},\ }\href
  {https://doi.org/10.1088/0029-5515/41/3/306} {\bibfield  {journal} {\bibinfo
  {journal} {Nuclear Fusion}\ }\textbf {\bibinfo {volume} {41}},\ \bibinfo
  {pages} {295} (\bibinfo {year} {2001})}\BibitemShut {NoStop}%
\bibitem [{\citenamefont {Canik}\ \emph {et~al.}(2013)\citenamefont {Canik},
  \citenamefont {Guttenfelder}, \citenamefont {Maingi}, \citenamefont
  {Osborne}, \citenamefont {Kubota}, \citenamefont {Ren}, \citenamefont {Bell},
  \citenamefont {Kugel}, \citenamefont {Leblanc},\ and\ \citenamefont
  {Souhkanovskii}}]{canik_edge_2013}%
  \BibitemOpen
  \bibfield  {author} {\bibinfo {author} {\bibfnamefont {J.~M.}\ \bibnamefont
  {Canik}}, \bibinfo {author} {\bibfnamefont {W.}~\bibnamefont {Guttenfelder}},
  \bibinfo {author} {\bibfnamefont {R.}~\bibnamefont {Maingi}}, \bibinfo
  {author} {\bibfnamefont {T.~H.}\ \bibnamefont {Osborne}}, \bibinfo {author}
  {\bibfnamefont {S.}~\bibnamefont {Kubota}}, \bibinfo {author} {\bibfnamefont
  {Y.}~\bibnamefont {Ren}}, \bibinfo {author} {\bibfnamefont {R.~E.}\
  \bibnamefont {Bell}}, \bibinfo {author} {\bibfnamefont {H.~W.}\ \bibnamefont
  {Kugel}}, \bibinfo {author} {\bibfnamefont {B.~P.}\ \bibnamefont {Leblanc}},\
  and\ \bibinfo {author} {\bibfnamefont {V.~A.}\ \bibnamefont
  {Souhkanovskii}},\ }\bibfield  {journal} {\bibinfo  {journal} {Nuclear
  Fusion}\ }\textbf {\bibinfo {volume} {53}},\ \href
  {https://doi.org/10.1088/0029-5515/53/11/113016}
  {10.1088/0029-5515/53/11/113016} (\bibinfo {year} {2013})\BibitemShut
  {NoStop}%
\bibitem [{\citenamefont {Mossessian}\ \emph {et~al.}(2002)\citenamefont
  {Mossessian}, \citenamefont {Snyder}, \citenamefont {Greenwald},
  \citenamefont {Hughes}, \citenamefont {Lin}, \citenamefont {Mazurenko},
  \citenamefont {Medvedev}, \citenamefont {Wilson},\ and\ \citenamefont
  {Wolfe}}]{mossessian_h-mode_2002}%
  \BibitemOpen
  \bibfield  {author} {\bibinfo {author} {\bibfnamefont {D.~A.}\ \bibnamefont
  {Mossessian}}, \bibinfo {author} {\bibfnamefont {P.~B.}\ \bibnamefont
  {Snyder}}, \bibinfo {author} {\bibfnamefont {M.}~\bibnamefont {Greenwald}},
  \bibinfo {author} {\bibfnamefont {J.~W.}\ \bibnamefont {Hughes}}, \bibinfo
  {author} {\bibfnamefont {Y.}~\bibnamefont {Lin}}, \bibinfo {author}
  {\bibfnamefont {A.}~\bibnamefont {Mazurenko}}, \bibinfo {author}
  {\bibfnamefont {S.}~\bibnamefont {Medvedev}}, \bibinfo {author}
  {\bibfnamefont {H.~R.}\ \bibnamefont {Wilson}},\ and\ \bibinfo {author}
  {\bibfnamefont {S.}~\bibnamefont {Wolfe}},\ }\href
  {https://doi.org/10.1088/0741-3335/44/4/303} {\bibfield  {journal} {\bibinfo
  {journal} {Plasma Physics and Controlled Fusion}\ }\textbf {\bibinfo {volume}
  {44}},\ \bibinfo {pages} {423} (\bibinfo {year} {2002})}\BibitemShut
  {NoStop}%
\bibitem [{\citenamefont {Chance}\ \emph {et~al.}(1983)\citenamefont {Chance},
  \citenamefont {Jardin},\ and\ \citenamefont {Stix}}]{chance_ballooning_1983}%
  \BibitemOpen
  \bibfield  {author} {\bibinfo {author} {\bibfnamefont {M.~S.}\ \bibnamefont
  {Chance}}, \bibinfo {author} {\bibfnamefont {S.~C.}\ \bibnamefont {Jardin}},\
  and\ \bibinfo {author} {\bibfnamefont {T.~H.}\ \bibnamefont {Stix}},\ }\href
  {https://doi.org/10.1103/PhysRevLett.51.1963} {\bibfield  {journal} {\bibinfo
   {journal} {Physical Review Letters}\ }\textbf {\bibinfo {volume} {51}},\
  \bibinfo {pages} {1963} (\bibinfo {year} {1983})}\BibitemShut {NoStop}%
\bibitem [{\citenamefont {Grimm}\ \emph {et~al.}(1985)\citenamefont {Grimm},
  \citenamefont {Chance}, \citenamefont {Todd}, \citenamefont {Manickam},
  \citenamefont {Okabayashi}, \citenamefont {Tang}, \citenamefont {Dewar},
  \citenamefont {Fishman}, \citenamefont {Mendelsohn}, \citenamefont
  {Monticello}, \citenamefont {Phillips},\ and\ \citenamefont
  {Reusch}}]{grimm_mhd_1985}%
  \BibitemOpen
  \bibfield  {author} {\bibinfo {author} {\bibfnamefont {R.~C.}\ \bibnamefont
  {Grimm}}, \bibinfo {author} {\bibfnamefont {M.~S.}\ \bibnamefont {Chance}},
  \bibinfo {author} {\bibfnamefont {A.~M.~M.}\ \bibnamefont {Todd}}, \bibinfo
  {author} {\bibfnamefont {J.}~\bibnamefont {Manickam}}, \bibinfo {author}
  {\bibfnamefont {M.}~\bibnamefont {Okabayashi}}, \bibinfo {author}
  {\bibfnamefont {W.~M.}\ \bibnamefont {Tang}}, \bibinfo {author}
  {\bibfnamefont {R.~L.}\ \bibnamefont {Dewar}}, \bibinfo {author}
  {\bibfnamefont {H.}~\bibnamefont {Fishman}}, \bibinfo {author} {\bibfnamefont
  {S.~L.}\ \bibnamefont {Mendelsohn}}, \bibinfo {author} {\bibfnamefont
  {D.~A.}\ \bibnamefont {Monticello}}, \bibinfo {author} {\bibfnamefont
  {M.~W.}\ \bibnamefont {Phillips}},\ and\ \bibinfo {author} {\bibfnamefont
  {M.}~\bibnamefont {Reusch}},\ }\href
  {https://doi.org/10.1088/0029-5515/25/7/005} {\bibfield  {journal} {\bibinfo
  {journal} {Nuclear Fusion}\ }\textbf {\bibinfo {volume} {25}},\ \bibinfo
  {pages} {805} (\bibinfo {year} {1985})}\BibitemShut {NoStop}%
\bibitem [{\citenamefont {Lütjens}\ \emph {et~al.}(1996)\citenamefont
  {Lütjens}, \citenamefont {Bondeson},\ and\ \citenamefont
  {Sauter}}]{lutjens_chease_1996}%
  \BibitemOpen
  \bibfield  {author} {\bibinfo {author} {\bibfnamefont {H.}~\bibnamefont
  {Lütjens}}, \bibinfo {author} {\bibfnamefont {A.}~\bibnamefont {Bondeson}},\
  and\ \bibinfo {author} {\bibfnamefont {O.}~\bibnamefont {Sauter}},\ }\href
  {https://doi.org/10.1016/0010-4655(96)00046-X} {\bibfield  {journal}
  {\bibinfo  {journal} {Computer Physics Communications}\ }\textbf {\bibinfo
  {volume} {97}},\ \bibinfo {pages} {219} (\bibinfo {year} {1996})}\BibitemShut
  {NoStop}%
\bibitem [{\citenamefont {Miller}\ \emph {et~al.}(1997)\citenamefont {Miller},
  \citenamefont {Lin-Liu}, \citenamefont {Turnbull}, \citenamefont {Chan},
  \citenamefont {Pearlstein}, \citenamefont {Sauter},\ and\ \citenamefont
  {Villard}}]{miller_stable_1997}%
  \BibitemOpen
  \bibfield  {author} {\bibinfo {author} {\bibfnamefont {R.~L.}\ \bibnamefont
  {Miller}}, \bibinfo {author} {\bibfnamefont {Y.~R.}\ \bibnamefont {Lin-Liu}},
  \bibinfo {author} {\bibfnamefont {A.~D.}\ \bibnamefont {Turnbull}}, \bibinfo
  {author} {\bibfnamefont {V.~S.}\ \bibnamefont {Chan}}, \bibinfo {author}
  {\bibfnamefont {L.~D.}\ \bibnamefont {Pearlstein}}, \bibinfo {author}
  {\bibfnamefont {O.}~\bibnamefont {Sauter}},\ and\ \bibinfo {author}
  {\bibfnamefont {L.}~\bibnamefont {Villard}},\ }\href
  {https://doi.org/10.1063/1.872193} {\bibfield  {journal} {\bibinfo  {journal}
  {Physics of Plasmas}\ }\textbf {\bibinfo {volume} {4}},\ \bibinfo {pages}
  {1062} (\bibinfo {year} {1997})}\BibitemShut {NoStop}%
\bibitem [{\citenamefont {Snyder}\ \emph {et~al.}(2009)\citenamefont {Snyder},
  \citenamefont {Groebner}, \citenamefont {Leonard}, \citenamefont {Osborne},\
  and\ \citenamefont {Wilson}}]{Snyder2009}%
  \BibitemOpen
  \bibfield  {author} {\bibinfo {author} {\bibfnamefont {P.~B.}\ \bibnamefont
  {Snyder}}, \bibinfo {author} {\bibfnamefont {R.~J.}\ \bibnamefont
  {Groebner}}, \bibinfo {author} {\bibfnamefont {A.~W.}\ \bibnamefont
  {Leonard}}, \bibinfo {author} {\bibfnamefont {T.~H.}\ \bibnamefont
  {Osborne}},\ and\ \bibinfo {author} {\bibfnamefont {H.~R.}\ \bibnamefont
  {Wilson}},\ }\href {https://doi.org/10.1063/1.3122146} {\bibfield
  {journal} {\bibinfo  {journal} {Physics of Plasmas}\ }\textbf {\bibinfo
  {volume} {16}},\ \bibinfo {pages} {056118} (\bibinfo {year}
  {2009})}\BibitemShut {NoStop}%
  
\bibitem [{\citenamefont {Snyder}\ \emph {et~al.}(2011)\citenamefont {Snyder},
  \citenamefont {Groebner}, \citenamefont {Hughes}, \citenamefont {Osborne},
  \citenamefont {Beurskens}, \citenamefont {Leonard}, \citenamefont {Wilson},\
  and\ \citenamefont {Xu}}]{Snyder2011}%
  \BibitemOpen
  \bibfield  {author} {\bibinfo {author} {\bibfnamefont {P.}~\bibnamefont
  {Snyder}}, \bibinfo {author} {\bibfnamefont {R.}~\bibnamefont {Groebner}},
  \bibinfo {author} {\bibfnamefont {J.}~\bibnamefont {Hughes}}, \bibinfo
  {author} {\bibfnamefont {T.}~\bibnamefont {Osborne}}, \bibinfo {author}
  {\bibfnamefont {M.}~\bibnamefont {Beurskens}}, \bibinfo {author}
  {\bibfnamefont {A.}~\bibnamefont {Leonard}}, \bibinfo {author} {\bibfnamefont
  {H.}~\bibnamefont {Wilson}},\ and\ \bibinfo {author} {\bibfnamefont
  {X.}~\bibnamefont {Xu}},\ }\href
  {https://doi.org/10.1088/0029-5515/51/10/103016} {\bibfield  {journal}
  {\bibinfo  {journal} {Nuclear Fusion}\ }\textbf {\bibinfo {volume} {51}},\
  \bibinfo {pages} {103016} (\bibinfo {year} {2011})}\BibitemShut {NoStop}%
  
\bibitem [{\citenamefont {Sauter}\ \emph {et~al.}(2014)\citenamefont {Sauter},
  \citenamefont {Brunner}, \citenamefont {Kim}, \citenamefont {Merlo},
  \citenamefont {Behn}, \citenamefont {Camenen}, \citenamefont {Coda},
  \citenamefont {Duval}, \citenamefont {Federspiel}, \citenamefont {Goodman},
  \citenamefont {Karpushov}, \citenamefont {Merle},\ and\ \citenamefont
  {Team}}]{Sauter2014}%
  \BibitemOpen
  \bibfield  {author} {\bibinfo {author} {\bibfnamefont {O.}~\bibnamefont
  {Sauter}}, \bibinfo {author} {\bibfnamefont {S.}~\bibnamefont {Brunner}},
  \bibinfo {author} {\bibfnamefont {D.}~\bibnamefont {Kim}}, \bibinfo {author}
  {\bibfnamefont {G.}~\bibnamefont {Merlo}}, \bibinfo {author} {\bibfnamefont
  {R.}~\bibnamefont {Behn}}, \bibinfo {author} {\bibfnamefont {Y.}~\bibnamefont
  {Camenen}}, \bibinfo {author} {\bibfnamefont {S.}~\bibnamefont {Coda}},
  \bibinfo {author} {\bibfnamefont {B.~P.}\ \bibnamefont {Duval}}, \bibinfo
  {author} {\bibfnamefont {L.}~\bibnamefont {Federspiel}}, \bibinfo {author}
  {\bibfnamefont {T.~P.}\ \bibnamefont {Goodman}}, \bibinfo {author}
  {\bibfnamefont {A.}~\bibnamefont {Karpushov}}, \bibinfo {author}
  {\bibfnamefont {A.}~\bibnamefont {Merle}},\ and\ \bibinfo {author}
  {\bibfnamefont {T.}~\bibnamefont {Team}},\ }\href
  {https://doi.org/10.1063/1.4876612} {\bibfield  {journal} {\bibinfo
  {journal} {Physics of Plasmas}\ }\textbf {\bibinfo {volume} {21}},\ \bibinfo
  {pages} {055906} (\bibinfo {year} {2014})}\BibitemShut {NoStop}%
\bibitem [{\citenamefont {Meneghini}\ \emph {et~al.}(2021)\citenamefont
  {Meneghini}, \citenamefont {Snoep}, \citenamefont {Lyons}, \citenamefont
  {McClenaghan}, \citenamefont {Imai}, \citenamefont {Grierson}, \citenamefont
  {Smith}, \citenamefont {Staebler}, \citenamefont {Snyder}, \citenamefont
  {Candy}, \citenamefont {Belli}, \citenamefont {Lao}, \citenamefont {Park},
  \citenamefont {Citrin}, \citenamefont {Luda~di Cortemiglia}, \citenamefont
  {Tema~Biwole},\ and\ \citenamefont {Mordijck}}]{Meneghini2021}%
  \BibitemOpen
  \bibfield  {author} {\bibinfo {author} {\bibfnamefont {O.}~\bibnamefont
  {Meneghini}}, \bibinfo {author} {\bibfnamefont {G.}~\bibnamefont {Snoep}},
  \bibinfo {author} {\bibfnamefont {B.~C.}\ \bibnamefont {Lyons}}, \bibinfo
  {author} {\bibfnamefont {J.}~\bibnamefont {McClenaghan}}, \bibinfo {author}
  {\bibfnamefont {C.~S.}\ \bibnamefont {Imai}}, \bibinfo {author}
  {\bibfnamefont {B.~A.}\ \bibnamefont {Grierson}}, \bibinfo {author}
  {\bibfnamefont {S.~P.}\ \bibnamefont {Smith}}, \bibinfo {author}
  {\bibfnamefont {G.~M.}\ \bibnamefont {Staebler}}, \bibinfo {author}
  {\bibfnamefont {P.~B.}\ \bibnamefont {Snyder}}, \bibinfo {author}
  {\bibfnamefont {J.}~\bibnamefont {Candy}}, \bibinfo {author} {\bibfnamefont
  {E.~A.}\ \bibnamefont {Belli}}, \bibinfo {author} {\bibfnamefont {L.~L.}\
  \bibnamefont {Lao}}, \bibinfo {author} {\bibfnamefont {J.~M.}\ \bibnamefont
  {Park}}, \bibinfo {author} {\bibfnamefont {J.}~\bibnamefont {Citrin}},
  \bibinfo {author} {\bibfnamefont {T.}~\bibnamefont {Luda~di Cortemiglia}},
  \bibinfo {author} {\bibfnamefont {A.~S.}\ \bibnamefont {Tema~Biwole}},\ and\
  \bibinfo {author} {\bibfnamefont {S.}~\bibnamefont {Mordijck}},\ }\href
  {https://doi.org/10.1088/1741-4326/abb918} {\bibfield  {journal} {\bibinfo
  {journal} {Nuclear Fusion}\ }\textbf {\bibinfo {volume} {61}},\ \bibinfo
  {pages} {026006} (\bibinfo {year} {2021})}\BibitemShut {NoStop}%
\bibitem [{\citenamefont {Sauter}\ \emph {et~al.}(1999)\citenamefont {Sauter},
  \citenamefont {Angioni},\ and\ \citenamefont {Lin-Liu}}]{Sauter1999}%
  \BibitemOpen
  \bibfield  {author} {\bibinfo {author} {\bibfnamefont {O.}~\bibnamefont
  {Sauter}}, \bibinfo {author} {\bibfnamefont {C.}~\bibnamefont {Angioni}},\
  and\ \bibinfo {author} {\bibfnamefont {Y.~R.}\ \bibnamefont {Lin-Liu}},\
  }\href {https://doi.org/10.1063/1.873240} {\bibfield  {journal} {\bibinfo
  {journal} {Physics of Plasmas}\ }\textbf {\bibinfo {volume} {6}},\ \bibinfo
  {pages} {2834} (\bibinfo {year} {1999})}\BibitemShut {NoStop}%
\bibitem [{\citenamefont {Sauter}\ \emph {et~al.}(2002)\citenamefont {Sauter},
  \citenamefont {Angioni},\ and\ \citenamefont {Lin-Liu}}]{Sauter2002}%
  \BibitemOpen
  \bibfield  {author} {\bibinfo {author} {\bibfnamefont {O.}~\bibnamefont
  {Sauter}}, \bibinfo {author} {\bibfnamefont {C.}~\bibnamefont {Angioni}},\
  and\ \bibinfo {author} {\bibfnamefont {Y.~R.}\ \bibnamefont {Lin-Liu}},\
  }\href {https://doi.org/10.1063/1.1517052} {\bibfield  {journal} {\bibinfo
  {journal} {Physics of Plasmas}\ }\textbf {\bibinfo {volume} {9}},\ \bibinfo
  {pages} {5140} (\bibinfo {year} {2002})}\BibitemShut {NoStop}%
\bibitem [{\citenamefont {Schwartz}\ \emph {et~al.}(2022)\citenamefont
  {Schwartz}, \citenamefont {Nelson},\ and\ \citenamefont
  {Kolemen}}]{Schwartz2022}%
  \BibitemOpen
  \bibfield  {author} {\bibinfo {author} {\bibfnamefont {J.~A.}\ \bibnamefont
  {Schwartz}}, \bibinfo {author} {\bibfnamefont {A.~O.}\ \bibnamefont
  {Nelson}},\ and\ \bibinfo {author} {\bibfnamefont {E.}~\bibnamefont
  {Kolemen}},\ }\href
  {https://doi.org/https://doi.org/10.1088/1741-4326/ac62f6} {\bibfield
  {journal} {\bibinfo  {journal} {Nuclear Fusion}\ ,\ \bibinfo {pages} {in
  press}} (\bibinfo {year} {2022})}\BibitemShut {NoStop}%
\bibitem [{\citenamefont {Meneghini}\ \emph {et~al.}(2015)\citenamefont
  {Meneghini}, \citenamefont {Smith}, \citenamefont {Lao}, \citenamefont
  {Izacard}, \citenamefont {Ren}, \citenamefont {Park}, \citenamefont {Candy},
  \citenamefont {Wang}, \citenamefont {Luna}, \citenamefont {Izzo},
  \citenamefont {Grierson}, \citenamefont {Snyder}, \citenamefont {Holland},
  \citenamefont {Penna}, \citenamefont {Lu}, \citenamefont {Raum},
  \citenamefont {McCubbin}, \citenamefont {Orlov}, \citenamefont {Belli},
  \citenamefont {Ferraro}, \citenamefont {Prater}, \citenamefont {Osborne},
  \citenamefont {Turnbull},\ and\ \citenamefont {Staebler}}]{Meneghini2015}%
  \BibitemOpen
  \bibfield  {author} {\bibinfo {author} {\bibfnamefont {O.}~\bibnamefont
  {Meneghini}}, \bibinfo {author} {\bibfnamefont {S.~P.}\ \bibnamefont
  {Smith}}, \bibinfo {author} {\bibfnamefont {L.~L.}\ \bibnamefont {Lao}},
  \bibinfo {author} {\bibfnamefont {O.}~\bibnamefont {Izacard}}, \bibinfo
  {author} {\bibfnamefont {Q.}~\bibnamefont {Ren}}, \bibinfo {author}
  {\bibfnamefont {J.~M.}\ \bibnamefont {Park}}, \bibinfo {author}
  {\bibfnamefont {J.}~\bibnamefont {Candy}}, \bibinfo {author} {\bibfnamefont
  {Z.}~\bibnamefont {Wang}}, \bibinfo {author} {\bibfnamefont {C.~J.}\
  \bibnamefont {Luna}}, \bibinfo {author} {\bibfnamefont {V.~A.}\ \bibnamefont
  {Izzo}}, \bibinfo {author} {\bibfnamefont {B.~A.}\ \bibnamefont {Grierson}},
  \bibinfo {author} {\bibfnamefont {P.~B.}\ \bibnamefont {Snyder}}, \bibinfo
  {author} {\bibfnamefont {C.}~\bibnamefont {Holland}}, \bibinfo {author}
  {\bibfnamefont {J.}~\bibnamefont {Penna}}, \bibinfo {author} {\bibfnamefont
  {G.}~\bibnamefont {Lu}}, \bibinfo {author} {\bibfnamefont {P.}~\bibnamefont
  {Raum}}, \bibinfo {author} {\bibfnamefont {A.}~\bibnamefont {McCubbin}},
  \bibinfo {author} {\bibfnamefont {D.~M.}\ \bibnamefont {Orlov}}, \bibinfo
  {author} {\bibfnamefont {E.~A.}\ \bibnamefont {Belli}}, \bibinfo {author}
  {\bibfnamefont {N.~M.}\ \bibnamefont {Ferraro}}, \bibinfo {author}
  {\bibfnamefont {R.}~\bibnamefont {Prater}}, \bibinfo {author} {\bibfnamefont
  {T.~H.}\ \bibnamefont {Osborne}}, \bibinfo {author} {\bibfnamefont {A.~D.}\
  \bibnamefont {Turnbull}},\ and\ \bibinfo {author} {\bibfnamefont {G.~M.}\
  \bibnamefont {Staebler}},\ }\href
  {https://doi.org/10.1088/0029-5515/55/8/083008} {\bibfield  {journal}
  {\bibinfo  {journal} {Nuclear Fusion}\ }\textbf {\bibinfo {volume} {55}},\
  \bibinfo {pages} {083008} (\bibinfo {year} {2015})}\BibitemShut {NoStop}%
\bibitem [{\citenamefont {Nelson}\ \emph {et~al.}(2021)\citenamefont {Nelson},
  \citenamefont {Laggner}, \citenamefont {Diallo}, \citenamefont {Smith},
  \citenamefont {Xing}, \citenamefont {Shousha},\ and\ \citenamefont
  {Kolemen}}]{Nelson_NFjog}%
  \BibitemOpen
  \bibfield  {author} {\bibinfo {author} {\bibfnamefont {A.~O.}\ \bibnamefont
  {Nelson}}, \bibinfo {author} {\bibfnamefont {F.~M.}\ \bibnamefont {Laggner}},
  \bibinfo {author} {\bibfnamefont {A.}~\bibnamefont {Diallo}}, \bibinfo
  {author} {\bibfnamefont {D.~R.}\ \bibnamefont {Smith}}, \bibinfo {author}
  {\bibfnamefont {Z.~A.}\ \bibnamefont {Xing}}, \bibinfo {author}
  {\bibfnamefont {R.}~\bibnamefont {Shousha}},\ and\ \bibinfo {author}
  {\bibfnamefont {E.}~\bibnamefont {Kolemen}},\ }\href
  {https://doi.org/10.1088/1741-4326/ac27ca} {\bibfield  {journal} {\bibinfo
  {journal} {Nuclear Fusion}\ }\textbf {\bibinfo {volume} {61}},\ \bibinfo
  {pages} {116083} (\bibinfo {year} {2021})}\BibitemShut {NoStop}%
\end{thebibliography}


\end{document}